\documentclass[iop]{emulateapj}

\usepackage{graphicx}
\usepackage{gensymb}
\usepackage{amsfonts,amsmath,amssymb}
\usepackage{natbib}
\usepackage{amsmath}
\usepackage{comment}
\usepackage{hyperref}
\usepackage{hypcap}
\shorttitle{Quasar Classification Using Color and Variability}
\shortauthors{C. M. Peters et al.}

\begin{document}

\title{Quasar Classification Using Color and Variability}
\author{Christina M. Peters\altaffilmark{1, $\star$}} \author{Gordon T. Richards\altaffilmark{1}} \author{Adam D. Myers\altaffilmark{2}} \author{Michael A. Strauss\altaffilmark{3}} \author{Kasper B. Schmidt\altaffilmark{4}} \author{\v{Z}eljko Ivezi\'{c}\altaffilmark{5}} \author{Nicholas P. Ross\altaffilmark{6}} \author{Chelsea L. MacLeod\altaffilmark{6}} \author{Ryan Riegel\altaffilmark{7}}

\altaffiltext{$\star$}{For correspondence regarding this article, please write to C.~M. Peters: \email{Christina.M.Peters@drexel.edu}}
\altaffiltext{1}{Department of Physics, Drexel University, Philadelphia, PA 19104, USA.}
\altaffiltext{2}{Department of Physics and Astronomy, University of Wyoming, Laramie, WY 82071, USA.}
\altaffiltext{3}{Department of Astrophysical Sciences, Princeton University, Princeton, NJ  08544, USA.}
\altaffiltext{4}{Department of Physics, University of California, Santa Barbara, CA 93106, USA.}
\altaffiltext{5}{Astronomy Department, University of Washington, Seattle, WA 98195, USA.}
\altaffiltext{6}{Institute for Astronomy, The University of Edinburgh, Edinburgh EH9 3HJ, U.K.}
\altaffiltext{7}{Skytree, Inc., 1731 Technology Drive, Suite 700, San Jose, CA 95110, USA.}

\slugcomment{Draft Version \today}

\begin{abstract}

We conduct a pilot investigation to determine the optimal combination of color and variability information to identify quasars in current and future multi-epoch optical surveys.  We use a Bayesian quasar selection algorithm (\citealt{Richards:2004}) to identify 35,820 type~1 quasar candidates in a 239 $\mathrm{deg}^2$ field of the Sloan Digital Sky Survey (SDSS) Stripe 82, using a combination of optical photometry and variability. Color analysis is performed on 5-band single- and multi-epoch SDSS optical photometry to a depth of $r \sim 22.4$.  From these data, variability parameters are calculated by fitting the structure function of each object in each band with a power law model using $10$ to $>100$ observations over timescales from $\sim 1$ day to $\sim 8$ years. Selection was based on a training sample of 13,221 spectroscopically-confirmed type-1 quasars, largely from the SDSS. Using variability alone, colors alone, and combining variability and colors we achieve 91\%, 93\%, and 97\% quasar completeness and 98\%, 98\%, and 97\% efficiency respectively, with particular improvement in the selection of quasars at $2.7<z<3.5$ where quasars and stars have similar optical colors. The 22,867 quasar candidates that are not spectroscopically confirmed reach a depth of $i \sim 22.0$; 21,876 (95.7\%) are dimmer than coadded $i$-band magnitude of 19.9, the cut off for spectroscopic follow-up for SDSS on Stripe 82. Brighter than 19.9, we find 5.7\% more quasar candidates without confirming spectra in sky regions otherwise considered complete. The resulting quasar sample has sufficient purity (and statistically correctable incompleteness) to produce a luminosity function comparable to those determined by spectroscopic investigations. We discuss improvements that can be made to the process in preparation for performing similar photometric selection and science on data from post-SDSS sky surveys.

\end{abstract}


\section{Introduction}\label{sec:Introduction}
Identification of large numbers of quasars/active galactic nuclei (AGN) over a broad range of redshift and luminosity is crucial for many science projects. Work that requires object densities higher than have been provided to date by spectroscopic surveys includes cross-correlating the catalogs with the cosmic microwave background \citep{Giannantonio:2008} to constrain dark energy; using quasars to measure cosmic magnification \citep{Scranton:2005}; finding binary quasars which can be used to test the merger hypothesis of quasars \citep{Hennawi:2010}; finding gravitationally lensed quasars \citep{Oguri:2006}; constraining quasar evolution \citep{Myers:2006}; studying dust in galaxies \citep{Menard:2010}; and broader cosmological studies \citep{Leistedt:2013}.

Historically, quasar candidates have been identified by virtue of their colors, variability, and (lack of) proper motion---but generally not through all of these methods combined.  The standard way of identifying large numbers of candidate quasars is to make ``color cuts'' using optical (or infrared) photometry (e.g., \citealt{Richards:2002}: \citealt{Croom:2004}; \citealt{Warren:2000}; \citealt{Lacy:2004}; \citealt{Stern:2005}; \citealt{Maddox:2012}; \citealt{Assef:2013}).  This is because the majority of unobscured quasars at $z < 2.5$ are much bluer than the majority of stars in the optical and are much redder in the infrared.  However, this process is neither complete (identifying all true quasars) nor efficient (minimizing false positives).  Such methods do an effective job of identifying a large number of interesting objects with relatively little effort; however, better methods are needed to scale to future surveys in a way that allows scientific analysis without the need for spectroscopic confirmation.

In addition to classification by color, time-domain data make variability a promising way for classifying objects. For examples of such work, see \cite{Koo:1986}, \cite{Hughes:1992}, \cite{Vanden-Berk:2004}, \cite{de-Vries:2005}, \cite{Sesar:2007}, \cite{Kelly:2009}, \cite{Kozlowski:2010}, \cite{Schmidt:2010}, \cite{Butler:2011}, MacLeod et al. (\citeyear{MacLeod:2010}, \citeyear{MacLeod:2011}, and \citeyear{MacLeod:2012}), and \cite{Graham:2014}. Specifically, quasars exhibit stochastic, aperiodic variability with variations of order 10\% on the timescale of years (\citealt{de-Vries:2003}; \citealt{Vanden-Berk:2004}).  The amplitude and time scale of this variability are sufficiently distinctive to allow one to identify an object as a candidate quasar.

Many current and future astronomical imaging surveys (SkyMapper: \citealt{Keller:2007}; Palomar Transient Factory: \citealt{Law:2009}; Pan-STARRS: \citealt{Kaiser:2010}; DES:  \citealt{DES:2005}; LSST: \citealt{Ivezic:2008}) are focusing on time-domain astronomy and in anticipation it is important to determine the effectiveness of classification using variability information. These surveys will observe areas of sky many times.  There is great hope that variability selection will fill in the gaps in color selection methods (or replace color selection entirely).  Indeed, investigations such as \cite{Schmidt:2010}, \cite{MacLeod:2011}, and \citet{Butler:2011} have been quite successful.  However, variability-only selection suffers from its own set of problems.  For example, high-redshift quasars can be lost when using a fixed observed-frame variability analysis: Ly$\alpha$ absorption reduces the quasar continuum in blue bands and the redder bands have larger photometric errors for fainter objects.  In addition, variability increases with lower luminosity (e.g., \citealt{Vanden-Berk:2004}), but so does the host galaxy contribution---potentially complicating selection of such objects without careful difference imaging to remove the host galaxy contribution.  Thus it is important to investigate how well variability selection works by itself versus being combined with other methods (e.g., colors and astrometry).

The premise of this project is to {\em simultaneously} use the distinctive and quantifiable characteristics of color and variability to distinguish quasars from stars and inactive galaxies. The Sloan Digital Sky Survey (SDSS; \citealt{York:2000}) repeatedly imaged a 2.5$\degree$ equatorial section of the sky referred to as \href{http://www.sdss.org/legacy/stripe82.html}{Stripe 82}\footnote{sdss.org/legacy/stripe82.html} (\citealt{Abazajian:2009}; \citealt{Annis:2014}; \citealt{Jiang:2014}). The light curves of spectroscopically confirmed quasars and stars from Stripe 82 give us the information we need to develop and test classification of quasars. 

The specific goal of this project is to use color, variability, and astrometric data in combination with modern machine learning techniques to uncover previously unidentified quasars in the SDSS Stripe 82 region and to pave the way for improved multi-faceted selection in the future. The goal is not necessarily to produce the most complete or efficient catalog possible, but to test the combined use of colors and variability data in classification. In this pilot investigation we make some simplifications to the process that will be explored in more detail in future work.  Specifically, we concentrate on point sources to avoid the problem of the host galaxy washing out the variable nucleus (reducing our sensitivity to low-redshift quasars), we utilize a simple power-law model of variability as opposed to more sophisticated (but not necessarily ``correct'') models such as the damped random walk, we use variability data from each band separately instead of merging them together, and we take a simplistic approach to combining photometric redshift information from different methods. Each of these simplifications for this pilot study is worthy of their own separate investigation to determine how to best deal with these issues.

A shortcoming of the traditional quasar identification process is that it usually involves selecting quasar candidates by identifying them as outliers using cuts {\em in the observed data space} (e.g., selecting all point sources with $u-g<0.6$).  Our classification instead makes {\em simultaneous} use of all of the data types available and uses modern statistical techniques (based on kernel density estimation; KDE) to make cuts in {\em probability space} (e.g., objects with an expected quasar probability greater than 50\%).  We will extend the methods developed by our group (\citealt{Richards:2004}; \citet{Riegel:2008};\citealt{Richards:2009_DR6catalog}; \citealt{Richards:2009_OpticalIRcatalog}) and others (e.g., \citealt{Suchkov:2005}; \citealt{Ball:2006}; \citealt{Davoodi:2006}; \citealt{Gao:2008}; \citealt{Bailer-Jones:2008}; \citealt{DAbrusco:2009}; \citealt{Guy:2010}; \citealt{Schmidt:2010}; \citealt{Abraham:2012}; \citealt{Bovy:2012}; \citealt{Peng:2012}; \citealt{Gupta:2014}) to create a classification algorithm for time-domain focused sky surveys.  While this approach has been shown to work well in the past \citep[e.g.,][]{Richards:2004,Richards:2009_DR6catalog}, in future work we also intend to explore other modern statistical techniques such as described by \cite{Feigelson:2012} and references therein.

The quasar candidates that result from application of this method are only identified {\em photometrically}; they lack spectroscopy which not only would confirm the type of an object, it crucially also would determine the redshift. There are many sophisticated methods for estimating photometric redshifts (e.g. \citealt{Rowan-Robinson:2008}; \citealt{Salvato:2009}); we use the algorithm described in \cite{Richards:2001} and \cite{Weinstein:2004} which ranks among the most accurate for (luminous) quasar photometric redshift estimates.  We improve this process further by using the effective prismatic effects of the Earth's atmosphere as a low-resolution spectrograph \citep{Kaczmarczik:2009}. In short, the positions of quasars, with their strong emission features, is a function of pass band and redshift.  This behavior of quasars allows us to uniquely incorporate {\em astrometric} information into our photometric redshift estimates.

Our work provides a stepping stone for quasar classification for future surveys such as the \href{http://www.lsst.org}{Large Synoptic Survey Telescope}\footnote{lsst.org} (LSST).  Eventually, each region of LSST will be imaged about 200 times in each filter over the 10 years of the survey, allowing for study of the variability of the object on scales of minutes to a decade. This focus on time-domain astronomy is an exciting new era in surveys, but the lack of spectroscopy creates a problem for confirming the type of an object.  As the number of spectroscopic fibers allocated to quasar identification pales in comparison to the number of photometrically detected objects that merit spectroscopic follow-up, it is only through complete and efficient object classification coupled with accurate redshift estimates that we can overcome the lack of spectroscopy in LSST and other future astronomical surveys and maximize their science output.

\begin{deluxetable*}{llrrr}
\tablecolumns{5} 
\tablewidth{0pc} 
\tabletypesize{\small}
\tablecaption{Master Quasar Catalog \label{table:MQC}}
\tablehead{ \colhead{Source} & \colhead{Description} & \colhead{w/ spectra} & \colhead{w/o spectra} & \colhead{Training Set}}
\startdata
Table 5 from \citet{Schneider:2010} & SDSS I/II & 105472 & 0 & 6082  \\
\citet{Croom:2004} & 2QZ & 9663 & 0 & 0 \\
\citet{Croom:2009} & 2SLAQ & 8881 & 0 & 1576 \\
Croom et al. (in prep.) & AUS & 2200 & 0 & 1706 \\
\citet{Kochanek:2012} & AGES & 2844 & 4 & 0 \\
\citet{Lilly:2007} and \citet{Elvis:2009} & COSMOS & 259 & 0 & 0 \\
\citet{Fan:2006} and \citet{Jiang:2008} & $z > 5.8$ & 27 & 0 & 0 \\
\citet{Paris:2014} & SDSS-III/BOSS & 168820 & 0 & 7383 \\
\citet{Ross:2012} & MMT & 836 & 0 & 278 \\
\citet{Richards:2009_DR6catalog} & NBCKDE Photometric Catalog & 174663 & 965542 & 9061 \\
\citet{Bovy:2011} & XDQSO Photometric Catalog & 142567 & 682831& 7088 \\
Table 5 of \citet{Papovich:2006} & ÒBROADLINEÓ objects & 104 & 0 & 0 \\
Table 5 of \citet{Glikman:2006} &  $z \sim 4$ & 10 & 0 & 0 \\
Tables 4 and 6 of \citet{Maddox:2012} & KX-selected & 3608 &0 & 986 \\
\tableline
Total & & 274329 & 1301846 & 13221
\enddata
\end{deluxetable*}

The layout of this paper is as follows. In Section~\ref{sec:Data} we introduce the SDSS Stripe 82 data that we will use. We then describe how the variability parameters used for classification are calculated. In Section~\ref{sec:Method} we summarize the NBC KDE selection algorithm and describe how it is used in this case. We test the various classification parameters and determine the optimal combination in Section~\ref{sec:TestingParameters}. Then, in Section~\ref{sec:BuildingCatalog}, we build the quasar candidate catalog using these optimal parameters, first using the full quasar training set, then using the training set divided into redshift bins to perform simultaneous classification and redshift estimation. In Section~\ref{sec:ResultsPhotometricRedshifts} we describe how the astrometric parameters are calculated, then estimate photometric and astrometric redshifts for all the candidate quasars. Next, we describe a cut to remove contamination and describe the final catalog of quasar candidates in Section~\ref{sec:Catalog}.  In Section~\ref{sec:Discussion} we compare to cuts in variability space and to color-based quasar selection, and calculate number counts and a luminosity function for the candidates. We discuss possible next steps in Section~\ref{sec:FutureWork} and conclude in Section~\ref{sec:Conclusions}.

Cosmology-dependent parameters are determined using $H_{o}$ = 70 km s$^{-1}$ Mpc$^{-1}$, $\Omega_m$ = 0.3, and $\Omega_\Lambda$ = 0.7 (\citealt{Hinshaw:2013}). Throughout this paper magnitudes will be reported on the AB system of \cite{Oke:1983}.

\section{Data}\label{sec:Data}

In this section, we describe the origin of the data and the parameters used for classification by our algorithm.   Section~\ref{sec:S82} describes the imaging data and \ref{sec:MQC} the spectroscopic data.  Sections~\ref{sec:Colors} and \ref{sec:Variability} discuss derivation of the color and variability classification parameters, respectively.  In principle, we could use astrometric information for classification as well; however, for this pilot study we have limited astrometric data to estimate photometric redshifts as discussed in Section~\ref{sec:ResultsPhotometricRedshifts}.  Machine learning algorithms need both training sets to find patterns in the data and a test set of data to verify that these patterns are useful; these data sets are described in Section~\ref{sec:Test Set and Training Sets}.

\subsection{SDSS Stripe 82}\label{sec:S82}

The SDSS is an optical survey that has used the 2.5-m Sloan telescope (\citealt{Gunn:2006}) at Apache Point Observatory in New Mexico to map 14,500 deg$^2$ of the sky (\citealt{Aihara:2011}). Photometry was performed with a drift-scan CCD camera (\citealt{Gunn:1998}) taking nearly simultaneous 54.1 second exposures in five broad optical bands ($u$, $g$, $r$, $i$, and $z$) between 3,000\AA\ and 10,000\AA\  (\citealt{Fukugita:1996}).

The imaging data used in our analysis consists of objects solely from the SDSS Stripe 82 area, which were made available as part of SDSS Data Release 7 (DR7; \citealt{Abazajian:2009}) and includes observations from October 1999 to November 2007.  The Stripe 82 region covers a 2.5\degree\ wide `stripe' on the celestial equator from right ascension $\sim$300\degree\ to $\sim$60\degree\ in the Southern Galactic Cap. Repeated observations were performed on this region throughout the SDSS I/II, with increasing frequency as part of the SDSS Supernova Survey (\citealt{Frieman:2008}), with $\sim$100 repeat imaging scans by the end of observations. The initial observations were done under optimal seeing, sky brightness, and photometric conditions. The supernova survey runs were done on useable nights, but under less than optimal conditions.  We limit our analysis to those objects detected as point sources.

The multiple observations on Stripe 82 were aligned and stacked into a coadded catalog described in \cite{Annis:2014} (see also \citealt{Jiang:2014} and \citealt{Huff:2014}). This catalog uses 20 to 40 observations on the region, mostly the early runs under optimal conditions. The data were downloaded from the SDSS Stripe 82 Catalog Archive Server (CAS)\footnote{http://cas.sdss.org/stripe82/en}. Database entries having SDSS ``run'' numbers of 106 and 206, representing objects with co-added photometry,  were extracted along with the individual epoch photometry for each of these objects in order to generate light curves\footnote{This process has since been made somewhat easier through the use of a unifying ``thingIndex'' table in Data Release 12 \citep{Alam:2015}: http://skyserver.sdss.org/dr12/en/help/browser/browser.aspx}.  The single epoch images go to a depth of $r \sim 22.4$ (5$\sigma$) with a median seeing of 1.4$\arcsec$. Coaddition of the imaging data reaches $\sim$ 2 magnitudes deeper and improves the median seeing to 1.1$\arcsec$.  The improvement in using coadded magnitudes over single epoch magnitudes for classification is demonstrated in Section~\ref{sec:ResultsTrainingSet}; see also \cite{Ivezic:2007}. 

\subsection{Master Quasar Catalog}\label{sec:MQC}

Definition of our quasar training set requires a subsample with spectroscopic confirmation.  Our primary source of spectroscopy comes from a ``Master'' Quasar Catalog (MQC), described in Section~2.1 of Richards et al.\ (2015, submitted), containing over 1.5 million sources, for which over 250,000 have confirming spectroscopy. This dataset consists of sources within the SDSS survey areas and draws objects from the sources described in Table~\ref{table:MQC}.

\begin{figure*}
\capstart
\centering
\begin{tabular}{cc}
\includegraphics[width=3in,height=3in]{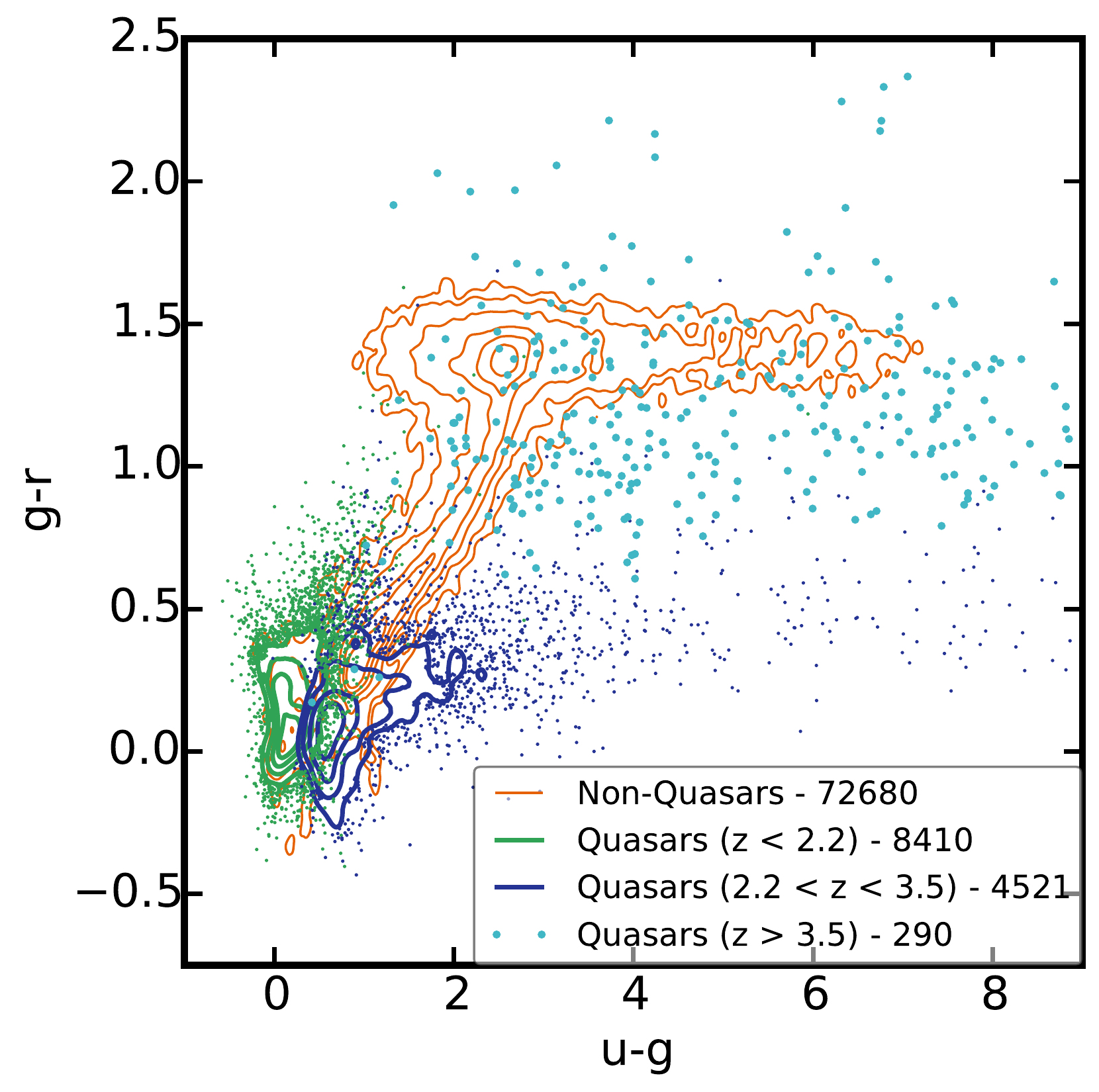} & 
\includegraphics[width=3in,height=3in]{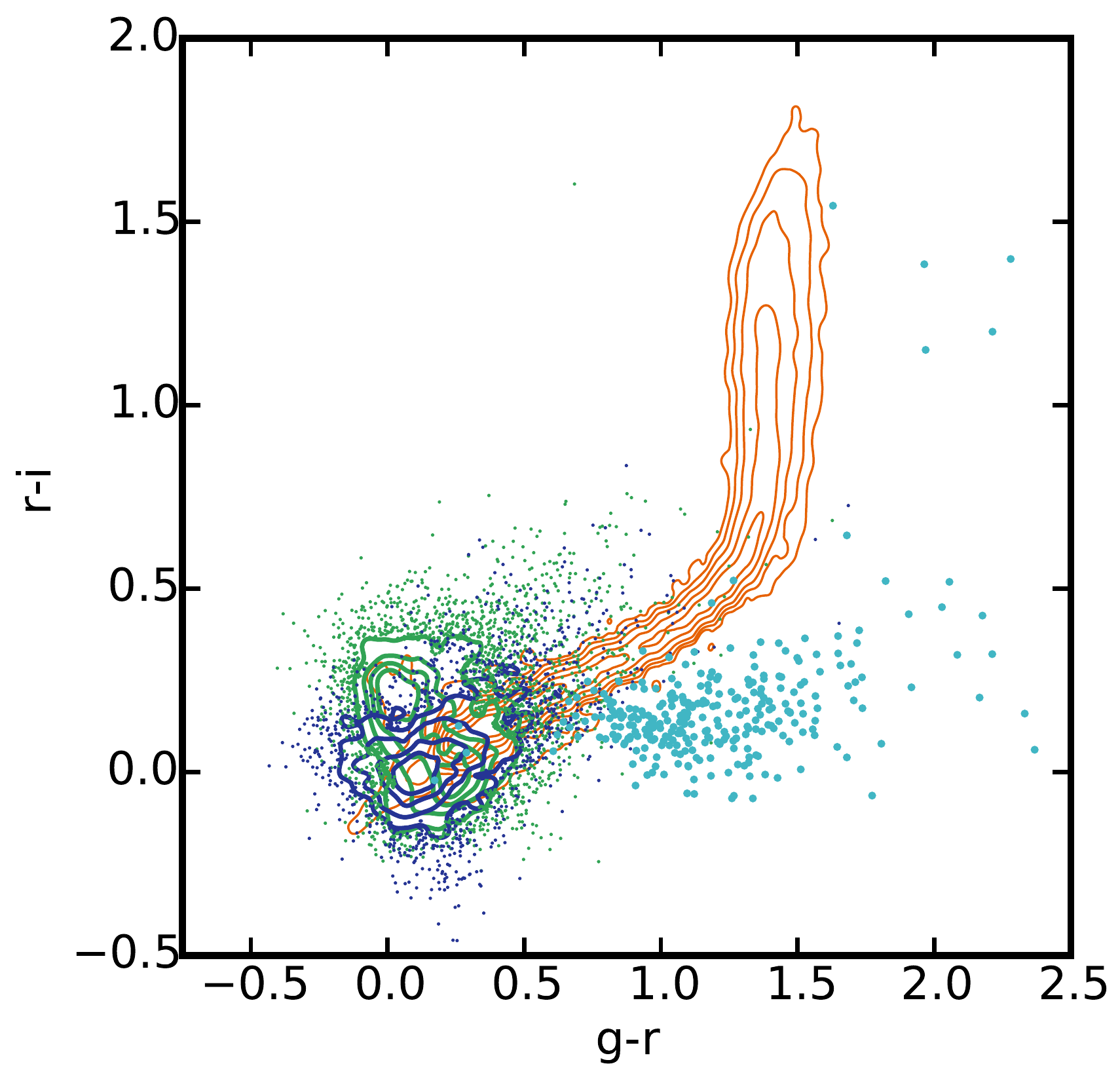}
\end{tabular}
\caption{Quasar and non-quasar training sets in two projections of the SDSS color space using coadded photometry. Non-quasars (shown in orange contours), such as stars and compact galaxies, are considered contaminants when trying to accurately classify quasars (shown in cool colors). Notice the number of non-quasars in the region in which mid-redshift quasars ($2.2 < z < 3.5$; shown as dark blue contours and scatter points for outliers) lie. This overlap makes it difficult to accurately classify an object in this region as a quasar or non-quasar and motivates searches for alternative methods of classification, like variability. Quasars are shown as three redshift regions: low-redshift ($z < 2.2$;  shown as green contours and scatter points for outliers), mid-redshift, and high-redshift ($z > 3.5$; shown as light blue dots). The extension of the non-quasar color space at $g - r \sim 1.4$ is not real, but an artifact of including objects with large $u$-band photometric errors (and thus spilling into the true quasar parameter space).}
\label{fig:colorcolor_redshift}
\end{figure*}

This quasar sample represents nearly every quasar known fainter than $i \sim 16$ (including candidate photometric quasars) at the time of Data Release 10 (DR10; \citealt{Ahn:2014}) of SDSS-III (\citealt{Eisenstein:2011}; \citet{Dawson:2013}). The majority of the confirmed quasars come from the SDSS I/II quasar catalog, which is described in detail by \citet{Richards:2002} and \citet{Schneider:2010} and from the SDSS-III/BOSS quasar catalog, which is described in detail by \citet{Ross:2012} and \citet{Paris:2014}.

The SDSS I/II quasars were primarily color selected (with some radio and X-ray selection) over a broad redshift range ($0<z<5$).  \citet{Richards:2002} describe the quasar target selection of the main quasar survey, which went to $i<19.1$ for quasars with colors consistent with $z<3$ and to $i<20.2$ for quasars expected to be at higher redshifts.  On Stripe 82, deeper targeting was performed (\citealt{Adelman-McCarthy:2006}) going to $i=19.9$ and $i=20.4$, respectively, in targeting ``chunk'' 22; to $i=20.2$ (for low-redshift sources) and $i=20.65$ (for radio sources) in targeting chunk 48; and to $i<21$ for sources more variable (between two epochs) than 3$\sigma$ (and 0.1 mag) in both $g$ and $r$ in targeting chunk 73.  The BOSS quasars (focused on $2.2<z<3.5.$; \citealt{Ross:2012}) were, in addition to color selection, also targeted by variability (on Stripe 82).  This variability selection is described in \citet{Palanque-Delabrouille:2011} and uses an algorithm that was also based on the same parameterization of variability as used herein (see Section~\ref{sec:Variability}).  Thus it is interesting to see if our method finds additional quasars beyond those already spectroscopically confirmed.   Quasar candidates in our catalog that are previously known from SDSS-I/II and SDSS-III spectroscopy are indicated as such in our catalog; see Appendix~\ref{sec:CatalogColumns}.

\vfill

\subsection{Classification Parameters: Colors}\label{sec:Colors}

The optical color information used in our analysis consists of the four adjacent SDSS colors ($u - g$, $g - r$, $r - i$, and $i - z$), which were determined from the cataloged photometry using point-spread-function magnitudes, corrected for Galactic extinction \citep{Schlegel:1998}. We used both single-epoch colors, from a single observation of the object, and the coadded colors, from the \cite{Annis:2014} catalog.

The level of contamination from stars and galaxies varies significantly in various regions of colorspace; see Figure~\ref{fig:colorcolor_redshift}. Optical surveys for quasars often use relatively simple color cuts (drawing lines of demarcation in these color spaces) to select objects that are likely to be quasars.  In SDSS, outliers from the stellar locus in the color space were potential spectroscopic target candidates \citep{Richards:2002}. The $ugri$ bands were used to identify low-redshift quasars and the $griz$ bands for high-redshift quasars. For low- and high-redshift quasars, selecting by colors is effective, but mid-redshift quasars ($2.2 < z < 3.5$) occupy the same region of color space as many stars and contamination becomes a serious problem. Note how the mid-redshift quasars, shown as dark blue contours and scatter points in Figure~\ref{fig:colorcolor_redshift}, overlap with the non-quasars, shown as orange contours. It is most efficient to choose quasars outside of this redshift region for spectroscopic follow-up, but this creates a strong selection effect in the quasar sample. For efficient selection of mid-redshift quasars, it becomes necessary to have another method to distinguish the quasars from non-quasars and this is where the variable nature of quasars becomes particularly useful.

\vfill

\subsection{Classification Parameters: Variability}\label{sec:Variability}

\begin{figure*}
\capstart
\centering
\begin{tabular}{cc}
\includegraphics[width=3in,height=3in]{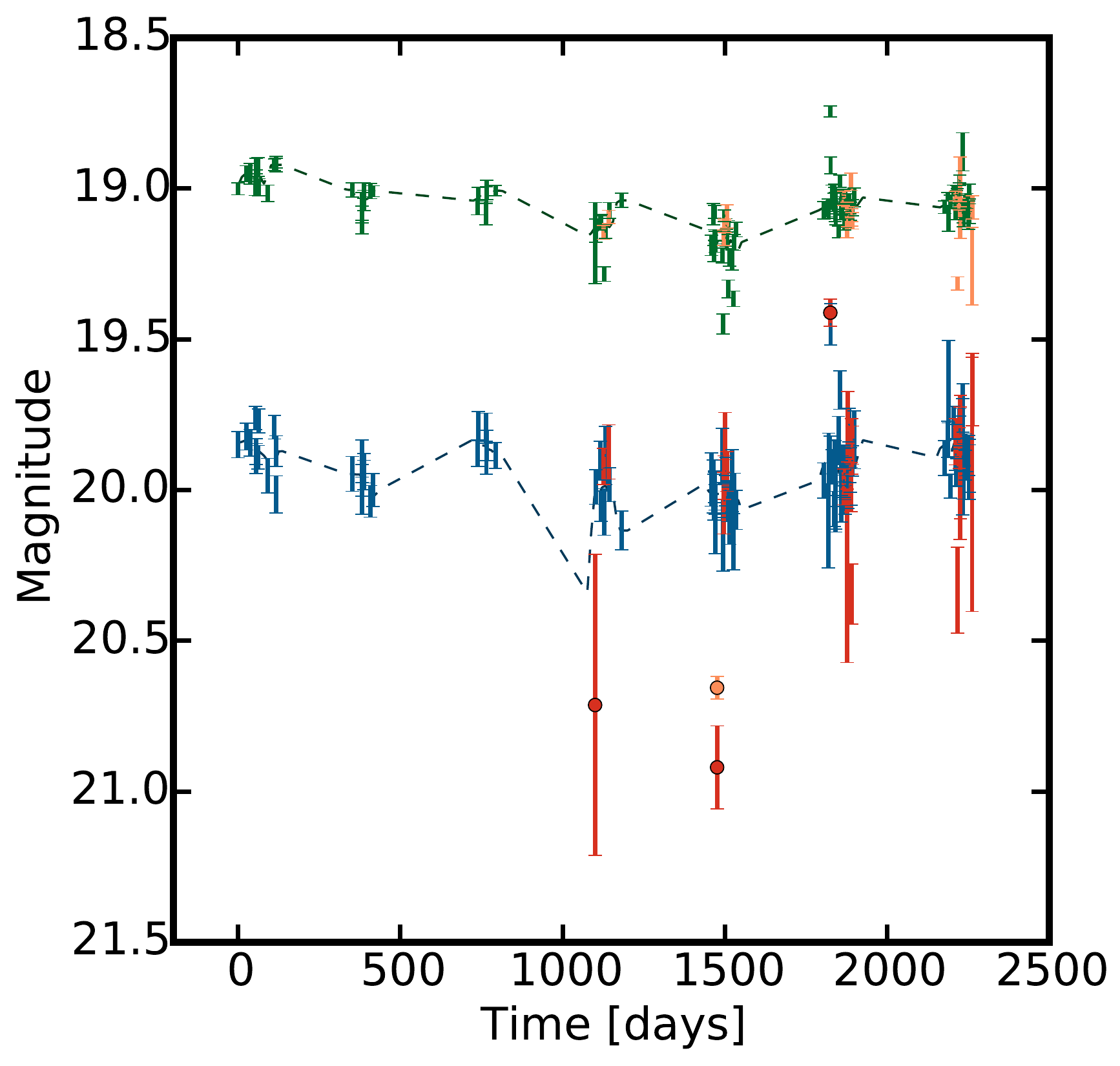} & 
\includegraphics[width=3in,height=3in]{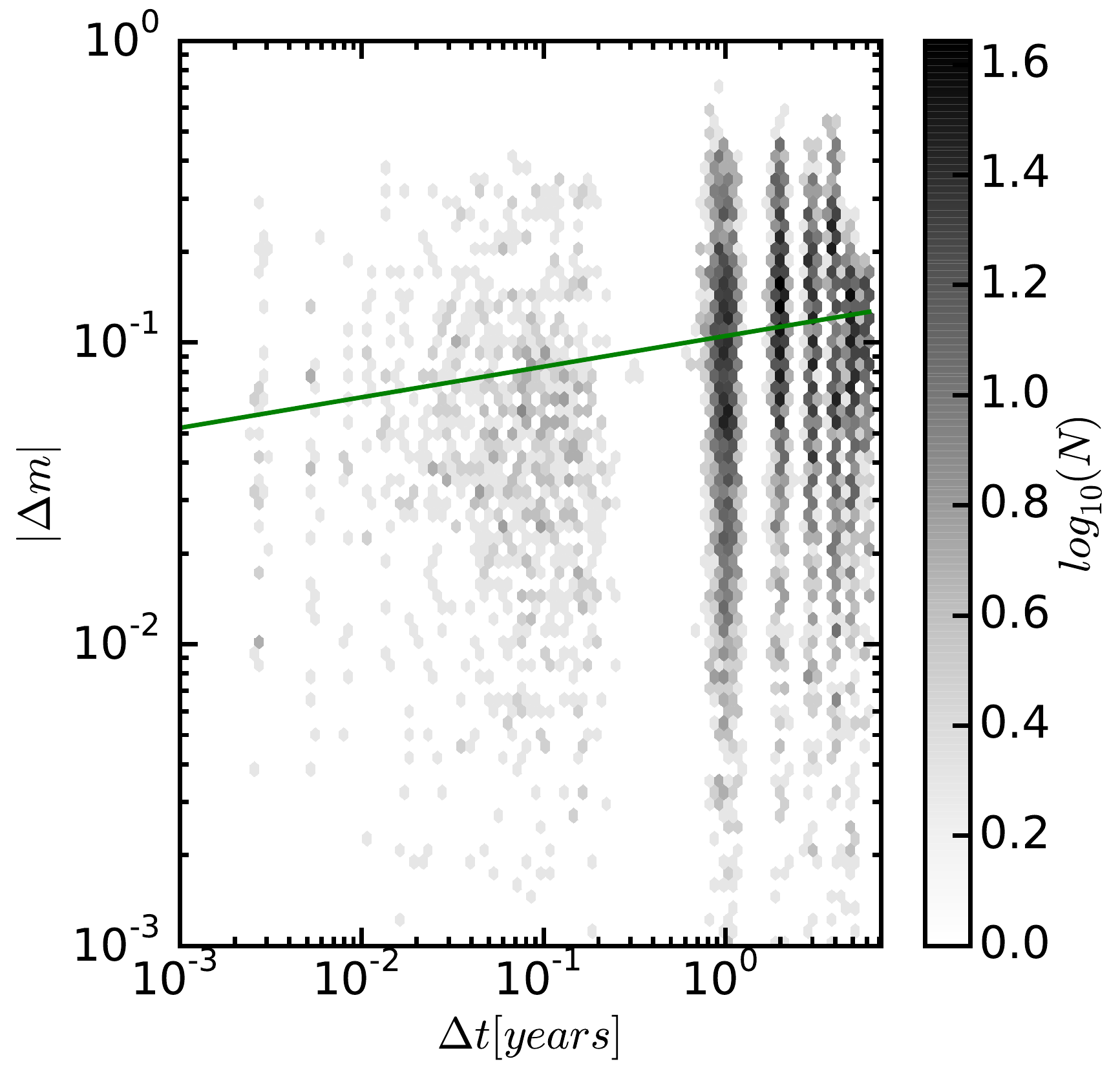}
\end{tabular}
\caption{$g$ and $u$-band light curves ({\it left panel}) and $g$-band structure function fit with a power law model ({\it right panel}) of SDSS J013417.81-005036.2, a redshift 2.26 quasar from SDSS Stripe 82 (also shown in Figure~\ref{fig:DCRprocess}). This quasar is shown as an example representative of the data set. {\it Left panel:} There are 126 observations in the $g$-band. The 106 observations that meet the PSF-width and the airmass requirements are shown as green points with error bars, while those that were removed are shown in orange. The dark green dashed line is the running median (with a window of 50 days and steps of 5 days) calculated from the $g$-band observations. The orange dot was removed from the light curve because it is more than 0.25 magnitude from the median. The $u$-band observations are similarly shown in blue and red. {\it Right panel:} The pairs of photometric points from the $g$-band light curve in the left panel are shown as a hex-bin density plot where the darkness of the hex bin indicates the number of points in that bin. The power law fit is shown as a green line. The method for calculating the structure function and the equation used to fit the structure function are detailed in Section~\ref{sec:Variability}. In the case of this object, the fitting algorithm gives $A_g = 0.105$ and $\gamma_g =  0.102$.  The points removed as outliers in the left panel would only contribute $|\Delta m| > 0.25$ mag values.}
\label{fig:lightcurve}
\end{figure*}

Most quasars vary at optical wavelengths by about 10\% over several years, which distinguishes them from most normal galaxies and stars (\citealt{de-Vries:2003}; \citealt{Vanden-Berk:2004}). Most variable stars vary periodically or quasi-periodically (\citealt{RichardsJ:2012}) and with smaller amplitude, but quasars generally show no periodic variability (\citealt{Bailer-Jones:2012}; \citealt{Andrae:2013}). While the physical causes for the variability in quasars are not well understood (see \citealt{Dexter:2011} for a recent investigation), the nature of the variability enables one to distinguish quasars from non-quasars. 

We use the structure function to characterize variability by quantifying the amplitude of variability as a function of the time difference between paired observations. For this analysis, based on empirical experiment (balancing the number of epochs with the quality of the data), we required that the FWHM of the PSF fit in the $r$ band be less than 2$\arcsec$ and the airmass in the $r$ band be less than 1.575 for the observation to be included. These cuts remove approximately 15\% of observations. After this procedure, we found that a small number of non-astrophysical outliers in the light curve still must be removed; these points are such strong outliers that we are not concerned that removing them is compromising the variability analysis.  Similar to the approach in \cite{Schmidt:2010}, we accomplish this by calculating a running median light curve then removing all measurements with a difference between the median light curve and the observed magnitude greater than 0.25 magnitudes (Figure~\ref{fig:lightcurve} {\it left panel}). The structure function is calculated in all of the SDSS bands where at least 10 observations remain after these cuts.

There are other methods currently being used to characterize the variability of quasars including Slepian wavelet variance (SWV; \citealt{Graham:2014}), AutoRegressive Moving Average, or ARMA, processes \citep{Kasliwal:2015}, or damped random walk (DRW; \citealt{Kelly:2009}; \citealt{Kozlowski:2010}) . Future work could consider using these methods instead of the structure function.

In our work, the structure function is defined as the root mean square magnitude difference as a function of time lag between magnitude measurements:
\begin{equation}
V^{2}(\Delta t) = \langle (m(t) - m(t + \Delta t))^2 \rangle 
\end{equation}
In the above equation, $m(t) - m(t + \Delta t)$ is the measured magnitude difference between two observations in a given band and $\Delta t$ is the time difference between the two observations in the observer's frame. The SDSS has a high cadence of observations during the fall months each year and then gaps of $\sim$9 months before the next set of observations. This irregular sampling in the light curve (Figure~\ref{fig:lightcurve} {\it left panel}) results in a structure function with gaps (Figure~\ref{fig:lightcurve} {\it right panel}).

The structure function can be modeled as a power law (Equation 3 in \citealt{Schmidt:2010}):
\begin{equation}
V_{PowerLaw}(\Delta t | A, \gamma) = A  \left( \cfrac{\Delta t}{\mathrm{1 year}} \right) ^\gamma .
\label{Eq:PowerLaw}
\end{equation}
Such a parameterization is not effective at describing the underlying type of variability or the mechanism for it, but provides a sufficiently robust statistical description for the timescales ($\sim 1$ day to $\sim 8$ years) covered by our data \citep{Schmidt:2010} to distinguish variable sources from non-variable sources, which is our objective. Using this model for the structure function, we find that 93\% of quasars are more variable than non-variable stars on average (using white dwarfs as representative) and show more growth in variability at longer time scales than 80\% of non-quasar point sources.

The variability can also be modeled as a DRW (\citealt{Kelly:2009}, \citealt{Kozlowski:2010}, \citealt{MacLeod:2010}), which predicts the following form of the structure function:
\begin{equation}
V_{DRW}(\Delta t | \sigma, \tau) = \sqrt{2} \sigma  \left( 1 - \mathrm{e}^{-\Delta t / \tau} \right)^\frac{1}{2} .
\end{equation}
To first order in $\Delta t$, the DRW behaves as:
\begin{equation}
V_{DRW}(\Delta t | \sigma, \tau) \sim \sqrt{2} \sigma  \left( \cfrac{\Delta t}{\tau} \right)^\frac{1}{2} ,
\end{equation}
a realization of Equation~\ref{Eq:PowerLaw} where $\gamma = 1/2$. In short, the DRW model is similar to the power-law model except that it truncates the growth of the magnitude differences at some characteristic timescale.  For the sake of this proof of concept, the power law model will suffice and is what we shall use hereafter. In future work we will investigate whether a more sophisticated model, such as the DRW model, improves quasar selection; however, even that model may be too simplistic to describe quasar variability across the range of timescales probed by modern optical monitoring data (\citealt{Mushotzky:2011}; \citealt{Zu:2013}; \citealt{Graham:2014}; \citealt{Kasliwal:2015}).

To fit the power law model to the observational data for each object we used the likelihood function (Equation 4 in \citealt{Schmidt:2010}):
\begin{equation}
\mathcal{L}(A,\gamma) = \prod_{j,k} L_{j,k} ,
\end{equation}
where $L_{j,k}$ is the likelihood of observing one particular magnitude difference $\Delta m_{j,k}$ between two light curve points separated by $\Delta t_{j,k}$. To determine the maximum likelihood of a Gaussian distribution, as in the case of the noise and intrinsic photometric variability, the likelihood function is:
\begin{equation}
\mathcal{L} = \prod_{i}^{N} \cfrac{1}{\sqrt{2 \pi} \sigma_i} \exp \left( - \cfrac{1}{2} \cfrac{(\Delta m_{i})^2}{\sigma_i^2}\right)
\end{equation}
The variance $\sigma^2 = (A(t_j - t_k)^\gamma)^2 + {\sigma_{phot,j}}^2 + {\sigma_{phot,k}}^2$ represents the scatter around the line that we are fitting and includes both intrinsic variability and noise. The $\sigma_{phot,j}$ and $\sigma_{phot,k}$ are the measured photometric errors on the measurements. Both the noise and the intrinsic photometric variability are assumed to have a Gaussian distribution.

If there is no variability or measurement noise, the structure function would be equal to zero for all $\Delta t$. The likelihood function now has the form: 
\begin{equation}
\begin{split}
\mathcal{L} = \prod_{j>k} \cfrac{1}{\sqrt{2 \pi [(A(t_j - t_k)^\gamma)^2 +{\sigma_{phot,j}}^2 + {\sigma_{phot,k}}^2]}} \\ \exp \left( - \cfrac{1}{2} \cfrac{(m_j - m_k)^2}{(A(t_j - t_k)^\gamma)^2 + {\sigma_{phot,j}}^2 + {\sigma_{phot,k}}^2}\right)
\end{split}
\end{equation}

The product only counts those observations where $j>k$, so there is no double counting and there are $\frac{n(n-1)}{2}$ data pairs where $n$ is the number of observations. We require the fitting to return physical values, $A > 0$ and $\gamma > 0$, so that the power law exponent and the average variability on a 1-year timescale are positive. This is because we are fitting $| \Delta m|$ and $|\Delta t|$ and all light curves will have some level of measurement noise, causing $A > 0$. Non-variable stars generally have $\gamma$ approaching 0. The expected increasing deviation from the mean for quasars with increasing $|\Delta t|$ will cause $\gamma > 0$.

We found a strong degeneracy between $A$ and $\gamma$ when maximizing the likelihood. To break this degeneracy, we applied a Gaussian prior to the likelihood on $A$. With a typical observing cadence of $\sim$1 year, the prior is centered on the observed median $|\Delta m|$ value, $\hat{A}$, at $0.5$ years $ < |\Delta t| < 1.5$ years and the standard deviation, $\sigma_{A}$, for those values. We place no explicit prior on $\gamma$ in the likelihood, but the requirement that $\gamma > 0$ functions as a flat prior. In addition to breaking the degeneracy, this prior encourages the minimization routine to converge on a realistic $A$ value more quickly. The cadence of the Stripe 82 data gives sufficient data points over this time difference to support this constraint. We combine the log of the likelihood function and the prior as follows:
\begin{equation}\label{eq:log likelihood function}
\begin{split}
S = -2 \cfrac{1}{N} \log (\mathcal{L}) + P(A) \\ = \cfrac{2}{n(n-1)} \sum_{j>k}\left[ \log ( (A(t_j - t_k)^\gamma)^2 + {\sigma_j}^2 + {\sigma_k}^2) \right. \\  \left. + \cfrac{(m_j - m_k)^2}{(A(t_j - t_k)^\gamma)^2 + {\sigma_j}^2 + {\sigma_k}^2} \right] + \cfrac{(A - \hat{A})^2}{\sigma_{A}^2},
\end{split}
\end{equation}
where $N$ is the number of terms in the sum and $P(A)$ is the prior on $A$.

\begin{figure}
\capstart
\centering
\includegraphics[width=3in,height=3in]{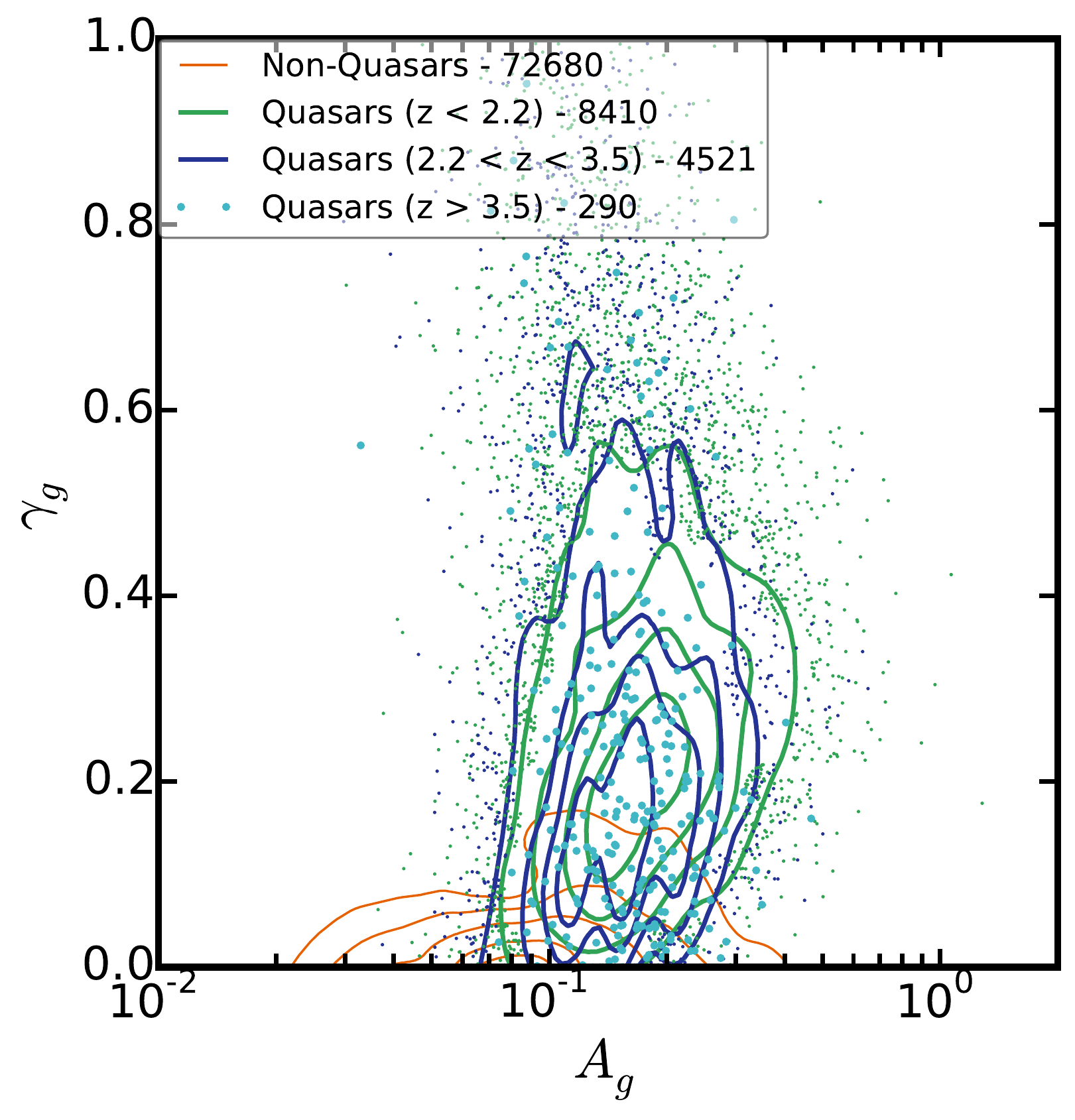}
\caption{Quasar and non-quasar training sets in variability parameter space for the $g$-band observations. Note that, unlike in the color-color plots in Figure~\ref{fig:colorcolor_redshift}, there are no distinct changes in the variability parameters as a function of quasar redshift in this parameter space. This is advantageous because it allows us to separate the quasars from the non-quasars in the variability space without extreme changes in completeness at specific redshifts, as seen with color selection. Non-quasars, such as stars and normal galaxies, are shown in orange contours. Quasars are shown in cool colors as three redshift regions: low-redshift ($z < 2.2$; shown as green contours and scatter points for outliers), mid-redshift ($2.2 < z < 3.5$; shown as dark blue contours and scatter points for outliers), and high-redshift ($z > 3.5$; shown as light blue dots).}
\label{fig:A_gamma}
\end{figure}

The posterior probability is maximized, by minimizing\footnote{Using Scipy's Optimization package, Powell's method: {\tt scipy.optimize.fmin\_powell}.} Equation 8 (the negative of the posterior probability), for each object in each of the five bands, so that for each object there are now ten variability parameters that can be used for classification: $A_u$, $\gamma_u$, $A_g$, $\gamma_g$, $A_r$, $\gamma_r$, $A_i$, $\gamma_i$, $A_z$, and $\gamma_z$.  Figure~\ref{fig:A_gamma} shows an example for the $g$-band variability parameters; note that the different redshift ranges are well mixed (but are largely distinct from non-quasars) in this case. In practice, our implementation of the likelihood method is biased (10 - 20\% in the best-fit values) which becomes relevant when light curves are much better sampled than those discussed here. An approach such as that described in the appendices of \cite{Kozlowski:2010} or \cite{Hernitschek:2015} would be more robust.  However, for the sake of this pilot investigation, our approach is more than sufficient, particularly because any bias in the variability parameters is the same for both selection by variability only, and by combined color and variability selection.

We currently fit the structure function to the multi-epoch data for all bands separately to compare their performance in the NBC KDE selection algorithm (see Section~\ref{sec:Method}). However, there are several ideas on how best to combine the observations in all five bands to obtain one light curve and one structure function to describe the overall variability. These methods are complicated by differences in how quasars vary in the different bands. For example, different bands represent different distances in the accretion disk resulting in a time lag between the bands and different characteristic timescales. 

\begin{figure*}
\capstart
\centering
\includegraphics[width=7in]{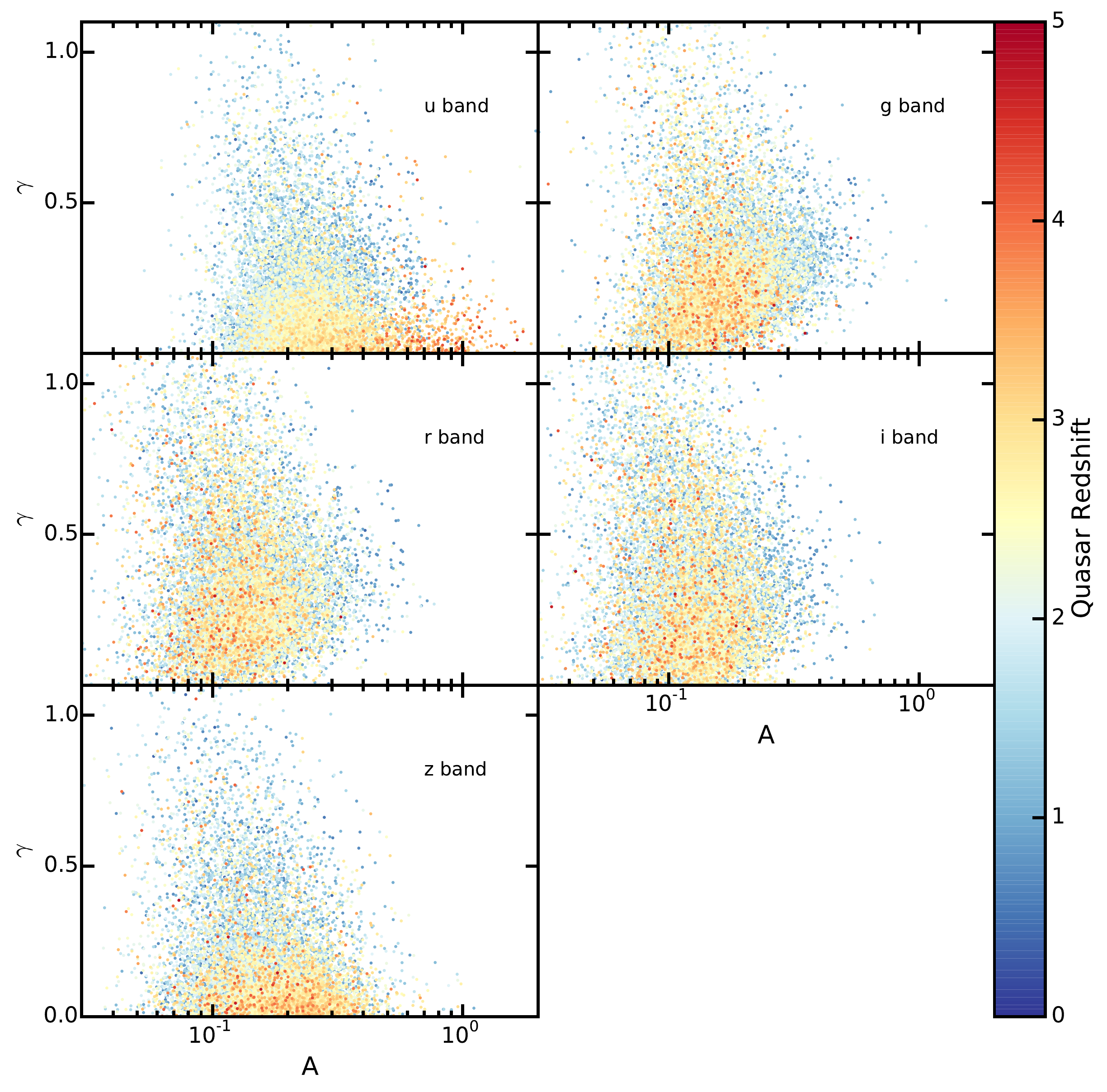}
\caption{All spectroscopically confirmed quasars shown in $A$ vs. $\gamma$ space in each of the SDSS bands, colored by redshift. Shown to demonstrate the difficulty involved in combining the observations in all five bands to obtain one light curve and one structure function in order to describe the overall variability without previously knowing the object's redshift. Note how the distribution of points shifts with band and with redshift. In particular, $A$ and $\gamma$ values agree well in the $g$, $r$, and $i$ band, but the large photometric errors in $u$ and $z$ bands artificially increase the apparent amplitude of the variability.}
\label{fig:A_gamma_redshift}
\end{figure*}

As shown in Figure \ref{fig:A_gamma_redshift}, there are different amplitudes of variability in different bands. Additionally, Ly$\alpha$ absorption obscures the true variability of quasars at high redshift. This is quite apparent in the $u$-band ({\it top left panel}) where the measured variability parameters for high-z quasars are caused by the high photometric errors of the $u$-band dropouts. It is also recognized that quasars become more luminous as they become bluer (Schmidt et al. \citeyear{Schmidt:2010} and \citeyear{Schmidt:2012}) and that bluer quasars in general are more variable (\citealt{Vanden-Berk:2004}; \citealt{MacLeod:2010}). Both of these effects must be taken into account when combining observations to describe the overall variability. A further complication for LSST will be the non-simultaneity of the observations  in different bands. Thus, proper treatment of the combined variability data is complex and beyond the scope of this paper.  For our purposes, describing the variability in each band is sufficient, and we therefore proceed with fitting the structure function for each of the bands separately.

\subsection{Test Set and Training Sets}\label{sec:Test Set and Training Sets}

Now that we have described the data inputs to our algorithm we can formally define the {\em test} and {\em training} sets.  The test set begins with all stellar morphology ({\tt objc\_type} $ == 6$) objects on the SDSS Stripe 82 with observations in DR7.  Restricting our sample to point sources allows us to concentrate on the improvements gained by combining colors and variability without having to worry about the differences in color and variability at redshifts and luminosities where the host galaxy contributes significantly to these properties. This set of observations was then limited by the following criteria: $-40\degree<RA<55\degree$, $g-i<6.0$, $g<23.5$, $i<22$, $\sigma_g<0.5$, and $\sigma_i<0.33$. These cuts are intended to reduce scatter due to high stellar density near the Galactic plane, high dust obscuration, and non-astrophysical colors. Observations with flags indicating poor photometry, such as those discussed in Section 3.2 of \cite{Richards:2002} were also excluded. There are 1,163,174 objects with 49,274,136 observations that meet these cuts.

Only objects where we had sufficient observations to calculate variability parameters in all five bands and astrometric parameters in $u$ and $g$ were included in the test and  training sets. Additionally, we require coadded colors $-1.0 < u-g < 9.0$, $-0.75 < g-r < 2.5$, $-0.5 < r-i < 3.0$, and $-1.5 < i-z < 1.75$, to constrain the parameter space for the NBC KDE to limit the necessary computational time for objects with unusually deviant colors. After these cuts, 916,587 objects remain. These objects compose the {\em cleaned data set}. The test set consists of the 903,366 sources from the cleaned data set that have not been spectroscopically identified as quasars. 

The quasar training set is formed from the 13,221 spectroscopically confirmed quasars in the MQC that have matches in the cleaned data set. To keep computational time reasonable, we select a subsample of 72,680 non-matches for the non-quasar training set.  As with our previous work \citep[e.g.,][]{Richards:2009_DR6catalog}, we note that the vast majority of these non-quasar training set objects are not actually spectroscopically confirmed to be non-quasars and thus there will be some level of contamination as is discussed further in Section~\ref{sec:Method}.   We do not explicitly include or exclude spectroscopically confirmed stars or galaxies in the non-quasar training set as most of these were selected as quasars (and found to be contaminants) and are thus biased in their color-space distribution. In practice, when we run the classification on the test set we include the training set objects so that our catalog of candidate objects includes the known quasars, making it easier to determine our completeness of these sources.

\section{NBC KDE Algorithm}\label{sec:Method}

Using training sets described in Section~\ref{sec:Test Set and Training Sets}, classification of the test set objects (based on parameters described in Sections \ref{sec:Colors} and \ref{sec:Variability}) was performed using Non-parametric Bayesian Classification (NBC) based on applying Kernel Density Estimation (KDE) to select quasars; see \citet{Richards:2004}, \citet{Gray:2005}, and \citet{Riegel:2008}. The algorithm takes training sets of objects divided into quasars and non-quasars. It creates an N\--dimensional probability space for each of the classes, where N is the number of parameters that describe each type of object and the parameter space is normalized to give equal weight to each parameter \citep{Gray:2005}. A probability density function (PDF) is constructed for each class of objects using KDE, by representing each individual object within a class by an N-dimensional Gaussian distribution and summing together the result for each object. Using the NBC KDE selection algorithm it is possible to combine all the classification parameters ($u - g$, $g - r$,  $r - i$, $i - z$, $A_u$, $\gamma_u$, $A_g$, $\gamma_g$, $A_r$, $\gamma_r$, $A_i$, $\gamma_i$, $A_z$, and $\gamma_z$) and perform the classification simultaneously considering all the characteristics to determine if the object is a quasar or a non-quasar.

From this PDF, the probability of an unclassified object being a quasar or non-quasar can be calculated, but first we need an understanding of the real-world ratio of quasars to non-quasars. When a new point is placed in the PDF, the probability of it being a quasar or a non-quasar is weighted by its prior probability. This prior is an expectation of how many of the unknown objects are non-quasars. This weighting is an application of Bayes' Theorem:

\begin{equation}
P(M |  D, I) = \frac{p(D | M, I) P(M | I)}{p(D | I)}.
\label{eq:BayesTheorem}
\end{equation}

In Equation~\ref{eq:BayesTheorem},  Bayes' Theorem (\citealt{Bayes:1763}; \citealt[Chapter 5]{Ivezic:2014}), $D$ stands for data, $M$ for model, and $I$ for prior information. This relates the posterior for the model based on the likelihood given the data and a prior. The pair of multi\--dimensional weighted PDFs measures the probability of an unknown object being a quasar or a non-quasar, while taking into account the expected ratio of quasars to non-quasars, and classifies it accordingly. Throughout this work we use a prior of 0.95, meaning that we expect 95\% of the objects to be non-quasars. The lower limit for the prior is determined by the fraction of known quasars in the test set. In \cite{Richards:2009_DR6catalog} the ratio of quasar candidates to the test set was 2.6\%. We use a slightly lower prior to capture some of the quasars that \citet{Richards:2009_DR6catalog} did not. We assumed the prior to be independent of position on the sky and magnitude. Changing the prior by 1\% does not change the number of quasar candidates by 1\% of the test set, but changes the number by roughly 1\% of the quasar candidates (Richards et al. 2015, submitted).

The algorithm requires a bandwidth for each of the training sets. The bandwidth controls the width of the kernel (a Gaussian distribution in our case) used to build the KDE. It is important to choose an optimal bandwidth when calculating the KDE or the distribution will be too smooth (under-fit) or will be too structured (over-fit)---in the same way as choosing an incorrect bin size for a histogram. The optimal bandwidth was found by performing leave-one-out cross-validation (leaving one object out and using the remainder of the training set to classify) over a range of bandwidths. We also refer to this as a {\it self test}. 

This process was repeated to find the optimal bandwidth based on the product of completeness and efficiency. {\it Completeness} is defined as the number of known quasars correctly classified as quasars divided by the number of known quasars. It is also referred to as sensitivity. {\it Efficiency} is defined as the number of known quasars correctly classified as quasars divided by the number of objects (known quasars and non-quasars) classified as quasars. It is also referred to as purity. Different metrics could be chosen depending on the desired science and whether completeness is needed over efficiency, but we use the product of completeness and efficiency as a middle ground for this proof of concept. That is, an efficiency of 85\% and a completeness of 70\% is considered a better selection than efficiency of 99\% and a completeness of 55\%.

\begin{figure*}
\centering
\begin{tabular}{cc}
\includegraphics[width=3in,height=3in]{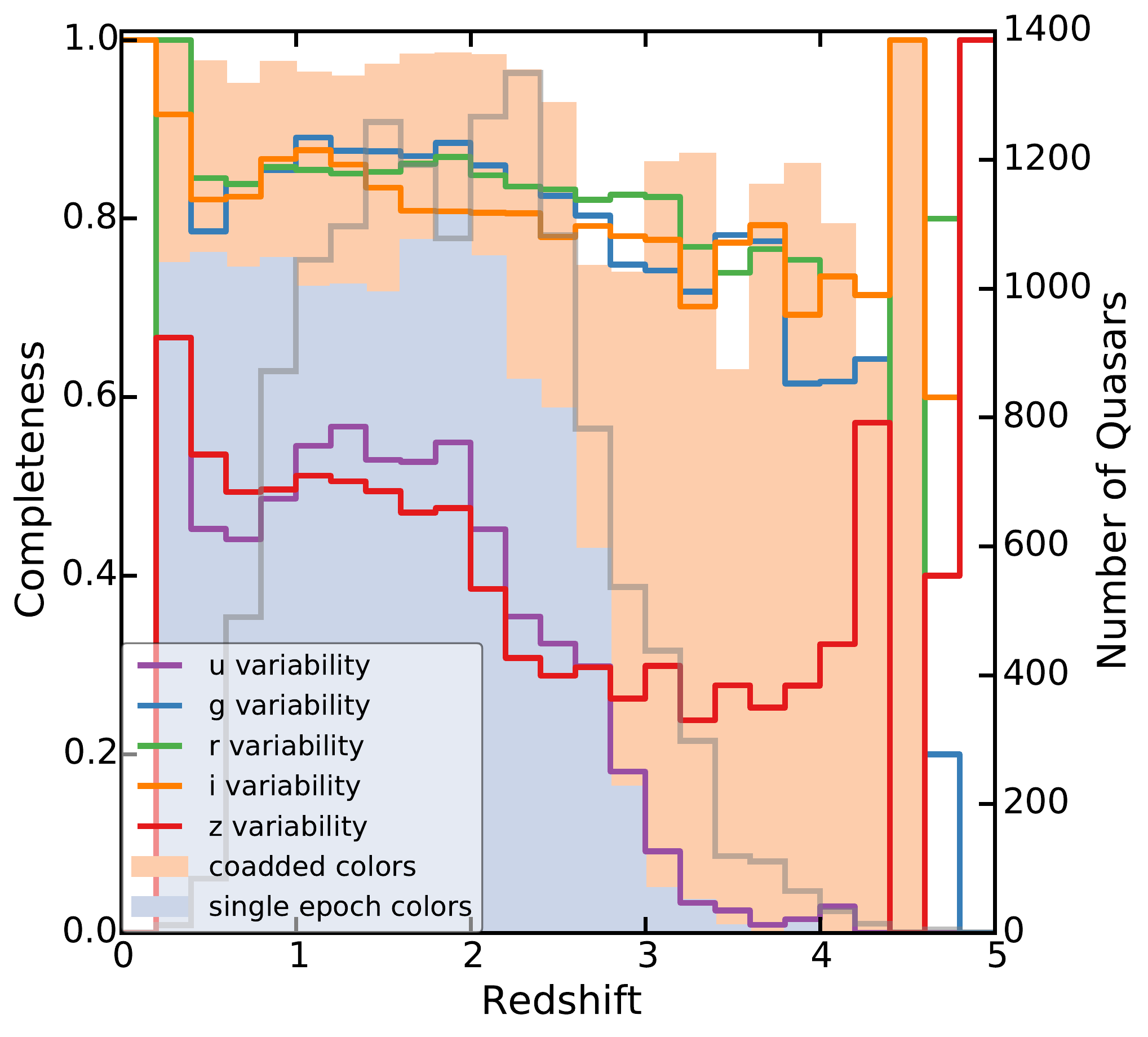} & 
\includegraphics[width=3in,height=3in]{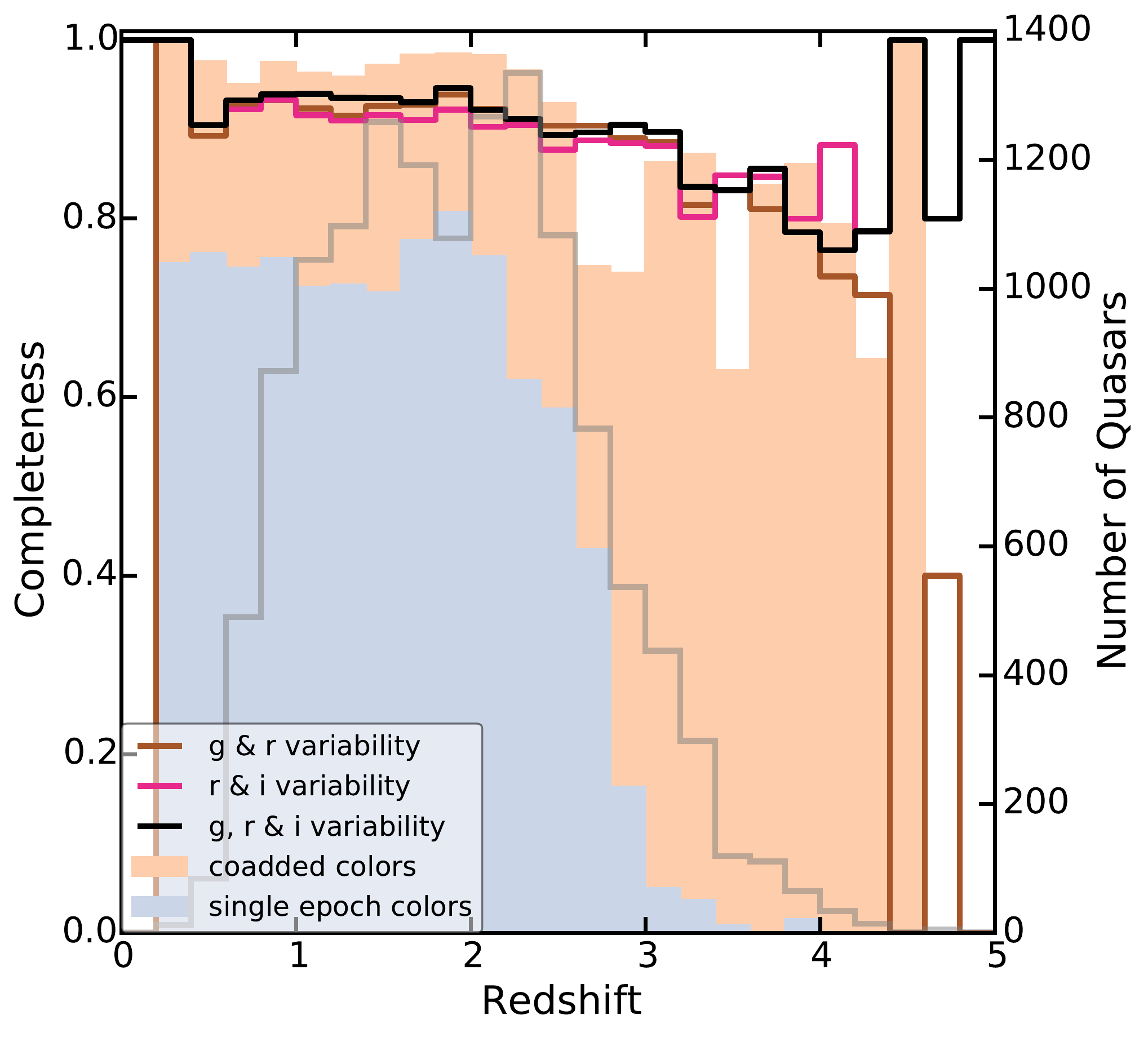}
\end{tabular}
\caption{Fraction of quasars correctly classified as quasars (completeness). These panels demonstrate that we are able to separate the quasars from the non-quasars in the variability space without extreme changes in completeness at specific redshifts. In both panels the gray line shows the number of quasars in each bin (right axis) and light blue (single epoch) and peach (coadded epochs) histograms show the completeness of color-only selection (left axis, Section~\ref{sec:ClassificationUsingColor}). Note the catastrophic loss of high-$z$ quasars from single-epoch colors and the incompleteness at $z\sim2.8$ even for coadded colors.  We also show classification from variability only: single bands ({\it left panel}) and combinations of multiple bands ({\it right panel}). The $g$, $r$, and $i$ bands are shown as blue, green, and orange lines respectively. There are no dramatic drops in the $g-$, $r-$, or $i-$bands variability at distinct redshifts, just a gradual decline with increasing redshift, which is related to observed magnitude, signal to noise ratio, and time scale of variability in the observer's frame. The overall completeness using variability alone is not as high as coadded colors alone at low redshifts, but is more successful than single-epoch colors alone at high redshifts.}
\label{fig:zhistogram_fraction_None}
\end{figure*}

After an initial self-classification of the training set is done, all those objects in the non-quasar training set that were classified as quasars in the self test are removed. This process is expected to remove the majority of quasars that may have contaminated the non-quasar training set due to lack of prior spectroscopic confirmation.  This new ``cleaned" non-quasar training set is used for the final classification. This cleaning process is a single iteration process and is performed separately for each of the classifications that we attempt below.

Having established the quasar prior probability, the quasar training set, a ``cleaned'' non-quasar training set, and the bandwidths for each of the training sets, we can proceed to classification of the unknown sources (i.e., the test set).   Application of the NBC KDE algorithm results in each object receiving a binary quasar vs.\ non-quasar classification, bifurcated at $P(M |  D, I) =0.5$. In the future, it may make more sense to simply output a probability for each object to facilitate combining this information with other data, but for the sake of this pilot study, we have chosen to make a hard cut (but in probability space rather than color space).

We explore which set of parameters (color, variability, or both) produces the best results in Section~\ref{sec:TestingParameters}, then we will apply the algorithm to the test set to obtain a set of quasar candidates in Section~\ref{sec:BuildingCatalog}.

\section{Testing Classification Parameters}\label{sec:TestingParameters}

Our goal is to establish whether combining color and variability information in quasar selection is superior to using just colors or variability alone. To accomplish this goal, the NBC KDE algorithm was used in a series of {\em self tests}, which consists of performing leave-one-out cross-validation on the training sets (rather than on a test set). The object being classified is not included in the training set and the process is repeated for each object in the training sets. The classifications returned by the algorithm are compared to the known classifications of the objects to estimate the completeness and efficiency of selection using those particular input parameters.

Section~\ref{sec:ClassificationUsingColor} uses the NBC KDE algorithm with the above quasar and non-quasar training sets to perform a self test using colors alone. This process serves as our basis of comparison: do other parameters enable more robust quasar selection than colors alone? In Section~\ref{sec:ResultsTrainingSet}, we attempt variability-only classification along with combined color and variability classification.  We then compare the results of these self tests.  This process reveals which variability (and color) parameters yield the most robust classification.

\subsection{Classification Using Color}\label{sec:ClassificationUsingColor}

Our first self test was performed using only the {\it single-epoch} SDSS adjacent colors ($u-g$, $g-r$, $r-i$, $i-z$) as inputs to the algorithm.  In practice, we chose a random epoch (meeting our requirements for good photometric and astrometric data) for each object. Using single epoch data is the most fair comparison for the majority of the objects in the SDSS footprint and we can use this as a control to compare how our method improves selection by adding variability. We could have chosen the `best' epoch for optimal classification by single-epoch colors alone; however, as we are testing the improvement from adding variability to the color classification, any epoch with quality data will serve. 

\begin{deluxetable*}{lrrlrrl}
\tablecolumns{3} 
\tablewidth{0pc} 
\tabletypesize{\small}
\tablecaption{NBC KDE Results - Self Test Non-quasar and Quasar Fraction \label{table:NBCKDEResultsSelfTestFraction}}
\tablehead{ \colhead{Self Test} & \multicolumn{3}{c}{non-quasars as non-quasars} & \multicolumn{3}{c}{quasars as quasars} \\
\cline{2-4} \cline{5-7}
\colhead{} &  \colhead{correct} &  \colhead{total} &  \colhead{fraction} &  \colhead{correct} &  \colhead{total} &  \colhead{fraction} }
\startdata
single epoch colors & 68611 & 69566 & 0.986 & 8232 & 13221 & 0.623 \\
coadded colors & 69474 & 69738 & 0.996 & 12353 & 13221 & 0.934 \\
$u$ variability & 70970 & 71936 & 0.987 & 5550 & 13221 & 0.420 \\
$g$ variability & 69489 & 70040 & 0.992 & 11138 & 13221 & 0.842 \\
$r$ variability & 69998 & 70476 & 0.993 & 11137 & 13221 & 0.842 \\
$i$ variability & 69935 & 70397 & 0.993 & 10782 & 13221 & 0.816 \\
$z$ variability & 70665 & 71372 & 0.990 & 5403 & 13221 & 0.409 \\
$g$ \& $r$ variability & 69777 & 70054 & 0.996 & 12060 & 13221 & 0.912 \\
$r$ \& $i$ variability & 69714 & 70050 & 0.995 & 11933 & 13221 & 0.903 \\
$g$, $r$, \& $i$ variability & 69728 & 70034 & 0.996 & 12150 & 13221 & 0.919 \\
coadded colors; $u$ variability & 69644 & 70077 & 0.994 & 12311 & 13221 & 0.931 \\
coadded colors; $g$ variability & 69822 & 70114 & 0.996 & 12739 & 13221 & 0.964 \\
coadded colors; $r$ variability & 69912 & 70229 & 0.996 & 12741 & 13221 & 0.964 \\
coadded colors; $i$ variability & 69880 & 70157 & 0.996 & 12634 & 13221 & 0.956 \\
coadded colors; $z$ variability & 69682 & 69990 & 0.996 & 12359 & 13221 & 0.935 \\
coadded colors; $g$ \& $r$ variability & 69663 & 70081 & 0.994 & 12816 & 13221 & 0.969 \\
coadded colors; $r$ \& $i$ variability & 69658 & 70096 & 0.994 & 12800 & 13221 & 0.968 \\
coadded colors; $g$, $r$, \& $i$ variability & 69948 & 70108 & 0.998 & 12626 & 13221 & 0.955
\enddata
\tablecomments{Fraction of non-quasars correctly classified as non-quasars and quasars correctly classified as quasars from the leave-one-out cross-validation of the training sets. The non-quasar total is different in the different rows because the non-quasar training set is ``cleaned" before it is used for the final classification, as described in Section~\ref{sec:Method}. The bandwidths are chosen to optimize the product of completeness and efficiency.}
\end{deluxetable*}

The results of the classification are shown in Table~\ref{table:NBCKDEResultsSelfTestFraction}, row 1, which indicates that these parameters are successful at not classifying non-quasars as quasars, at the expense of missing more than 37\% of known quasars. Indicative of the well-known problem of separating high-redshift quasars from the locus of moderate-to-cool temperature stars (e.g., \citealt{Richards:2002}), most of these missing quasars are at high redshift as can be seen from Figure~\ref{fig:zhistogram_fraction_None}. On the other hand, low-redshift quasars, which can be selected robustly by traditional color cuts, are also easily identified using the NBC KDE algorithm as shown in \cite{Richards:2004}.

The completeness of our single-epoch selection is distinctly different from \cite{Richards:2006}: it is seemingly too high at low-z (given our restriction to point sources) and too low at high-z.    For low-z this merely reflects the completeness of point sources.  At high-z it is important to realize that in \cite{Richards:2006} the purpose was to perform as complete a selection as possible, with efficiency as low as 50\%, using hard color cuts.  We will discuss how complete our selection is for all quasars, including extended sources, in Section~\ref{sec:Discussion}.

\begin{figure*}
\centering
\begin{tabular}{cc}
\includegraphics[width=3in,height=3in]{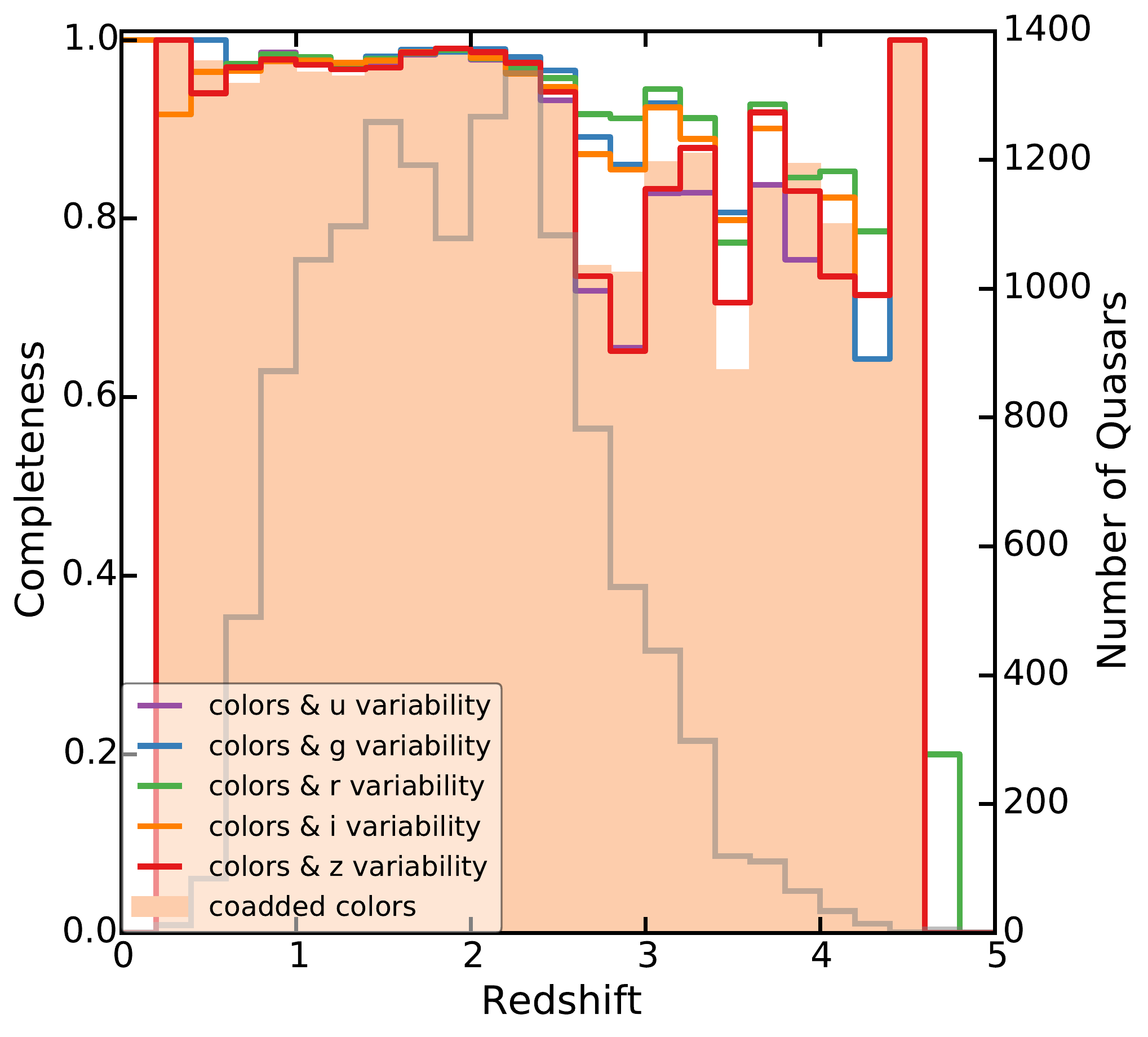} & 
\includegraphics[width=3in,height=3in]{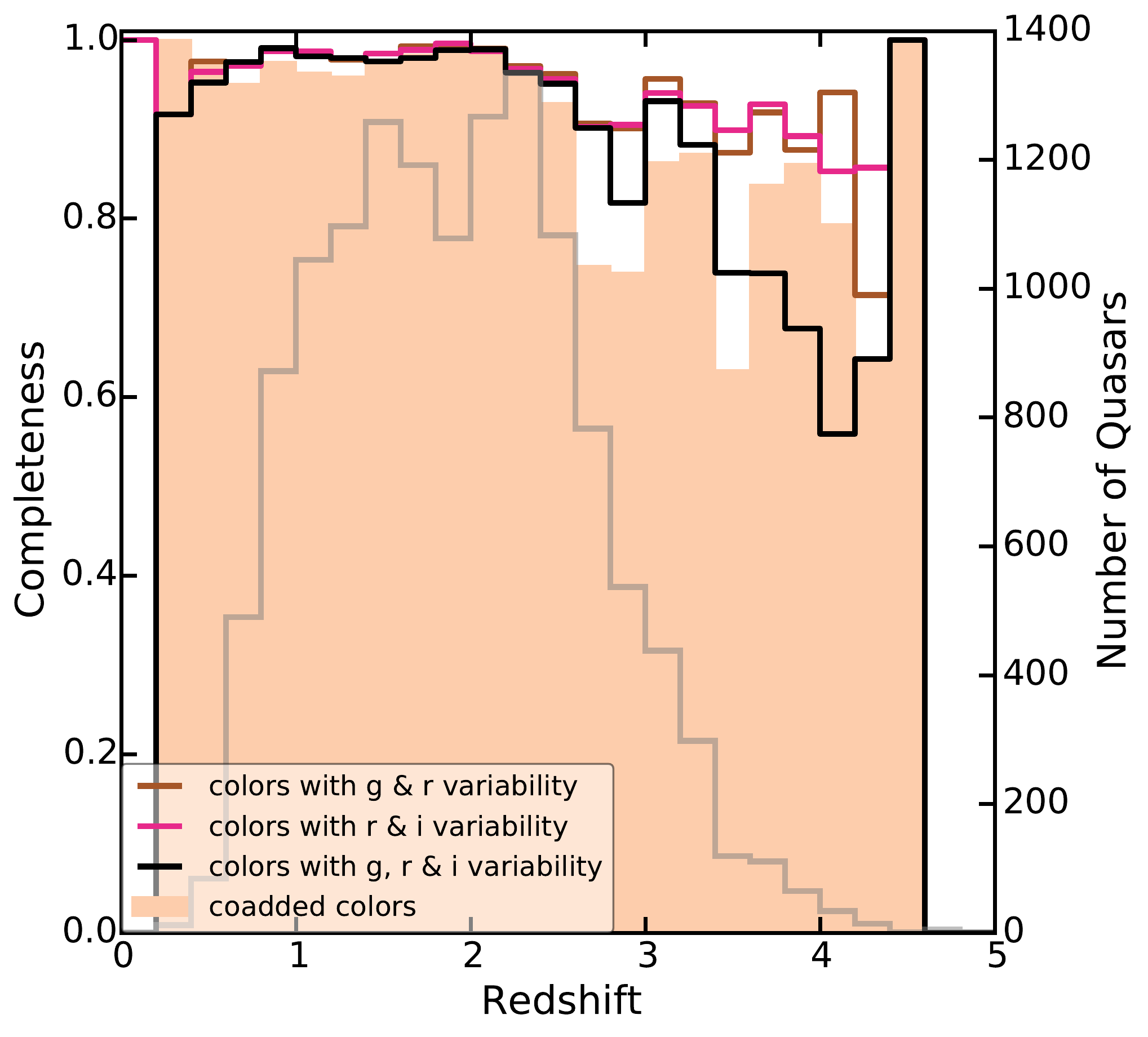}
\end{tabular}
\caption{Fraction of quasars correctly classified as quasars using coadded colors {\it and} variability, as a function of redshift. Notice the improved completeness near redshifts 2.7 and 3.5, where the quasars and non-quasars overlap in color space, with the addition of variability parameters. Shown are single bands of variability combined with coadded colors ({\it left panel}) and combinations of multiple bands of variability combined with coadded colors ({\it right panel}). In both panels the gray line shows the number of quasars in each bin (right axis).}
\label{fig:zhistogram_fraction_coadded}
\end{figure*}

\begin{deluxetable*}{l rl rl rl}
\tablecolumns{7}
\tablewidth{0pc} 
\tabletypesize{\small}
\tablecaption{NBC KDE Results: Self Test Completeness and Efficiency \label{table:Self Test Completeness and Efficiency}}
\tablehead{ \colhead{Self Test}  & \multicolumn{2}{c}{Variability Only} & \multicolumn{2}{c}{Single Epoch Colors w/ Variability} &  \multicolumn{2}{c}{Coadded Colors w/ Variability} \\
\cline{2-3} \cline{4-5} \cline{6-7}
\colhead{} & Completeness & Efficiency & Completeness & Efficiency & Completeness & Efficiency}
\startdata
color only & \nodata & \nodata & 0.6226 & 0.8960 & 0.9343 & 0.9791 \\
$u$ variability & 0.4198 & 0.8517 & 0.6934 & 0.9289 & 0.9312 & 0.9660 \\
$g$ variability & 0.8424 & 0.9529 & 0.8372 & 0.9149 & 0.9635 & 0.9776 \\
$r$ variability & 0.8424 & 0.9588 & 0.8583 & 0.9165 & 0.9637 & 0.9757 \\
$i$ variability & 0.8155 & 0.9589 & 0.8126 & 0.9235 & 0.9556 & 0.9785 \\
$z$ variability & 0.4087 & 0.8843 & 0.7158 & 0.9214 & 0.9348 & 0.9757 \\
$g$ \& $r$ variability & 0.9122 & 0.9775 & 0.8115 & 0.9758 & 0.9694 & 0.9684 \\
$r$ \& $i$ variability & 0.9026 & 0.9726 & 0.8076 & 0.9734 & 0.9682 & 0.9669 \\
$g$, $r$,  \& $i$ variability & 0.9190 & 0.9754 & 0.8573 & 0.9761 & 0.9550 & 0.9875
\enddata
\tablecomments{Completeness (known quasars classified as quasars divided by known quasars) and efficiency (known quasars classified as quasars divided all objects classified as quasars) for each of the self tests described in Section~\ref{sec:ResultsTrainingSet}. This indicates that the most successful option is a combination of coadded colors and variability, but no particular variability bands stood out when in combination with colors.}
\end{deluxetable*}

In the SDSS Stripe 82 region, where we will conduct our experiments on variability selection of quasars, we are able to combine multiple epochs of imaging data to produce more accurate color measurements of the quasars (as discussed in Section~\ref{sec:S82}).  Thus, we perform a second self test using {\em coadded} colors for each object. Table~\ref{table:NBCKDEResultsSelfTestFraction}, row 2 demonstrates that the use of coadded colors yields a small improvement in the efficiency of the sample, but a large improvement in the completeness---now being 93\% complete. Figure~\ref{fig:zhistogram_fraction_None} shows that most of this improvement comes from the recovery of high-redshift quasars; smaller photometric errors make it easier to distinguish the high-redshift quasar distribution from stars.  However, there is still a dip at $z\sim2.8$ where even the coadded colors do not enable better than 75\% completeness.

\subsection{Choosing Optimal Classification Parameters}\label{sec:ResultsTrainingSet}

Variability alone can be the basis for a robust quasar classification (e.g., \citealt{Schmidt:2010}; \citealt{Butler:2011}; \citealt{MacLeod:2011}), so we next perform a self test by applying KDE to the pair of variability parameters for each band (as defined in Section~\ref{sec:Variability}) and then on combinations of variability parameters from the multiple bands. The results are shown in Table~\ref{table:NBCKDEResultsSelfTestFraction} and Figure~\ref{fig:zhistogram_fraction_None}. It is interesting to compare the performance of the bands because each represents different distances from the center of the accretion disk, different characteristic timescales, and different (redshift-dependent) peak amplitudes.

Particularly important is that variability selection has a higher completeness in the range $2.6 < z < 3.0$ than do colors. There are no significant trends with redshift in the $A$--$\gamma$ space in the $g$, $r$, and $i$ bands, so the quasars can be separated out from the non-quasars in the variability space without completeness issues at specific redshifts (unlike the dramatic drops seen for color-only selection). The completeness drops off gradually with higher redshift, which is a result of changes in observed magnitude, signal-to-noise ratio, and time scale of variability in the observer's frame. Combining $g$ and $r$, $r$ and $i$, and $g$, $r$, and $i$, we find similar trends as using just the variability parameters from a single band, with marginally higher completeness (and efficiency) at all redshifts.

Selection by $u$- and $z$-band variability performs much worse than both coadded and single epoch colors. The $u$ band is strongly influenced by Ly$\alpha$ forest absorption of the (variable) quasar continuum at high redshift, thus suppressing the signal-to-noise ratio. This results in discordant variability parameters for quasars that are quite apparent in Figure~\ref{fig:A_gamma_redshift}. The lower performance of the $z$-band is likely due to the lower signal-to-noise ratio of the photometry and thus the larger scatter of the variability parameters as seen in Figure~\ref{fig:A_gamma_redshift}. These discrepant values increase the probability of high-redshift quasars being classified as stars.

While variability selection produces more consistent results with redshift than color selection, we find that, at many redshifts, color selection is still superior. We thus consider coadded colors with combinations of variability parameters from single and multiple bands. The results are shown in Table~\ref{table:NBCKDEResultsSelfTestFraction} and Figure~\ref{fig:zhistogram_fraction_coadded}.  Adding variability parameters from just one band significantly improves the selection, especially the high signal-to-noise ratio bands $g$, $r$, and $i$. The addition of the $u$- and $z$-band variability to colors still fails at z$\sim$2.8 because the variability signal is not strong enough (as demonstrated in Figures~\ref{fig:A_gamma_redshift} and \ref{fig:zhistogram_fraction_None}) to overcome color selection bias.

\begin{figure*}
\capstart
\centering
\begin{tabular}{ccc}
\includegraphics[width=2.3in]{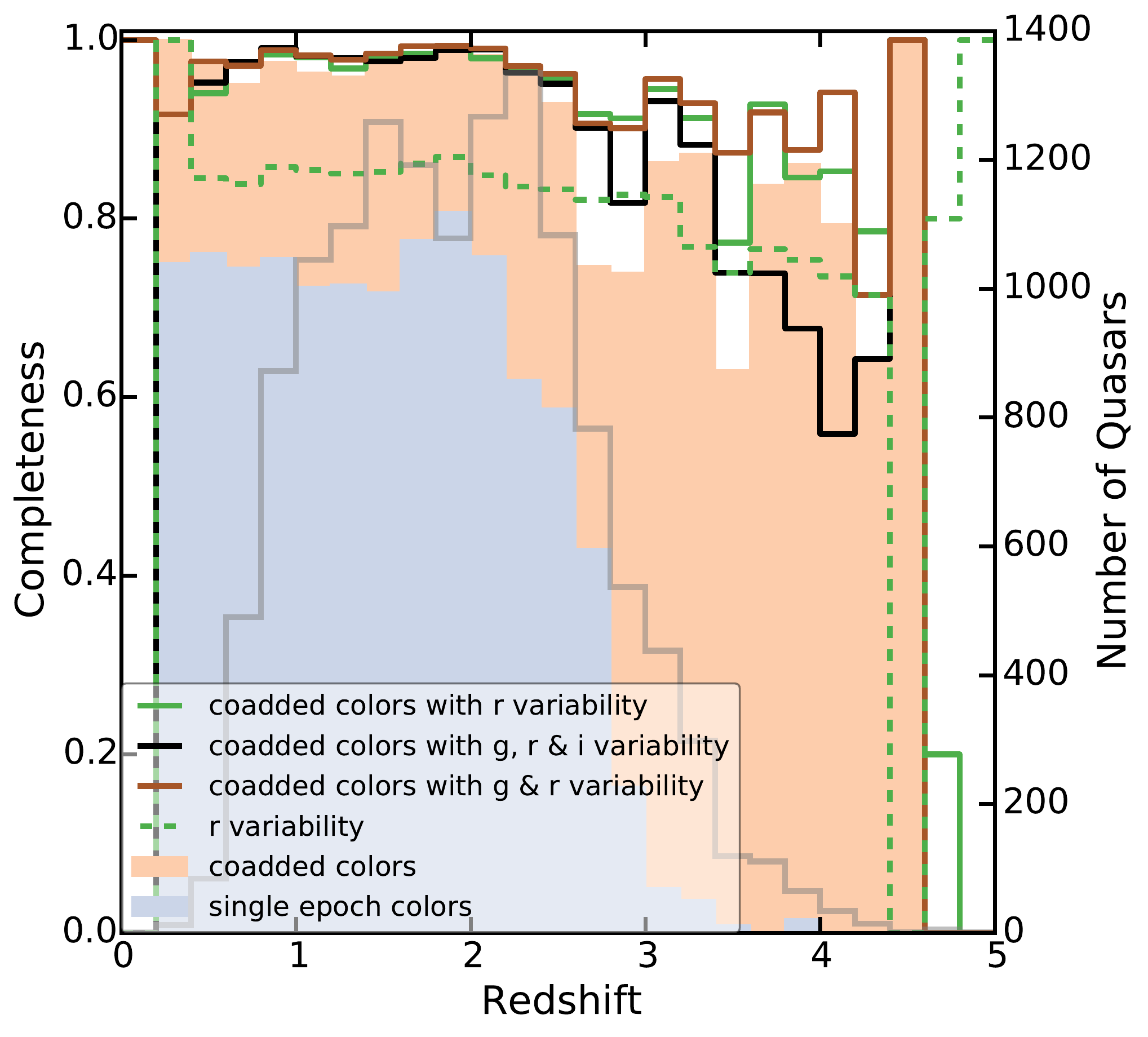} & 
\includegraphics[width=2.3in]{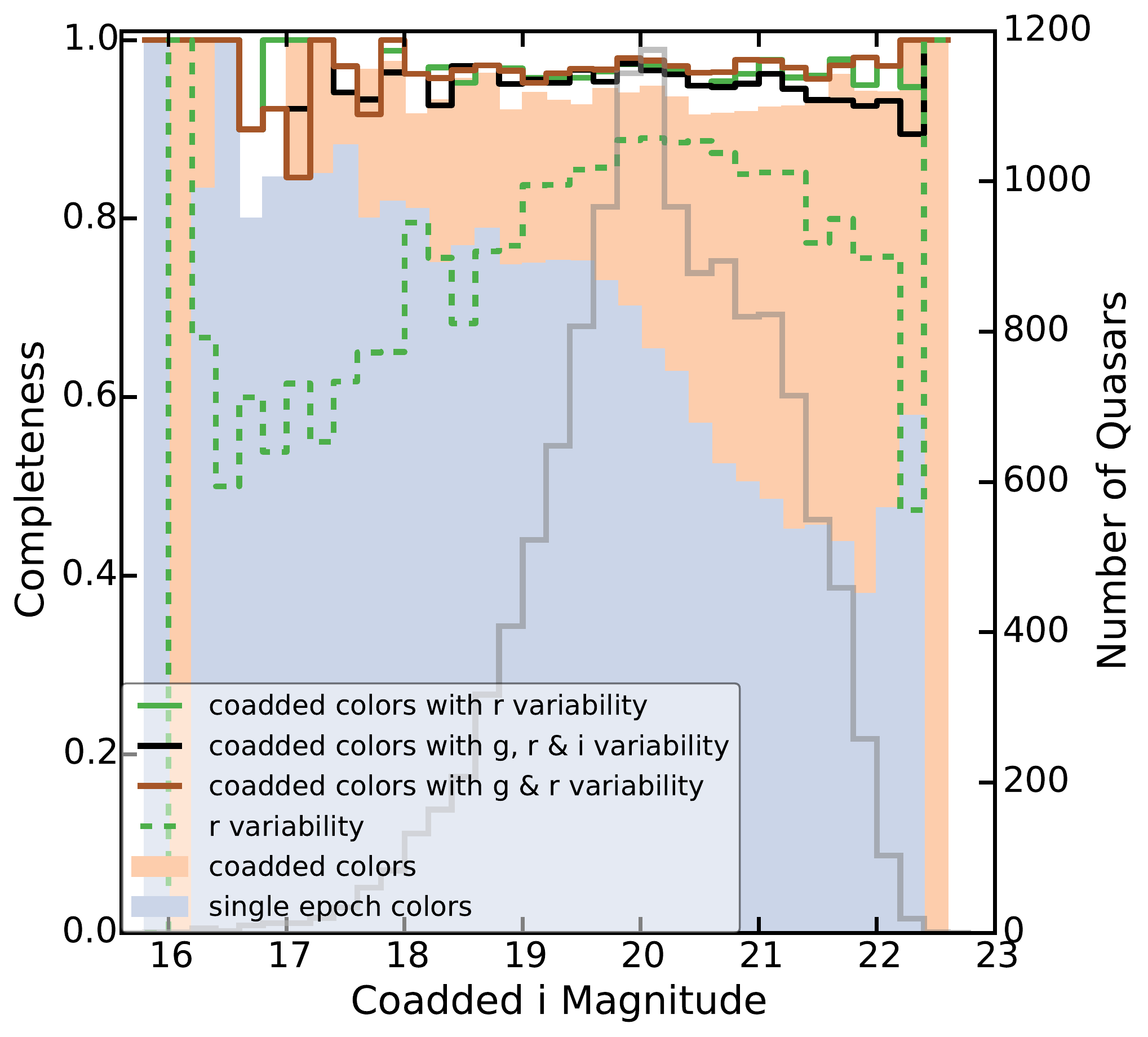} & 
\includegraphics[width=2.3in]{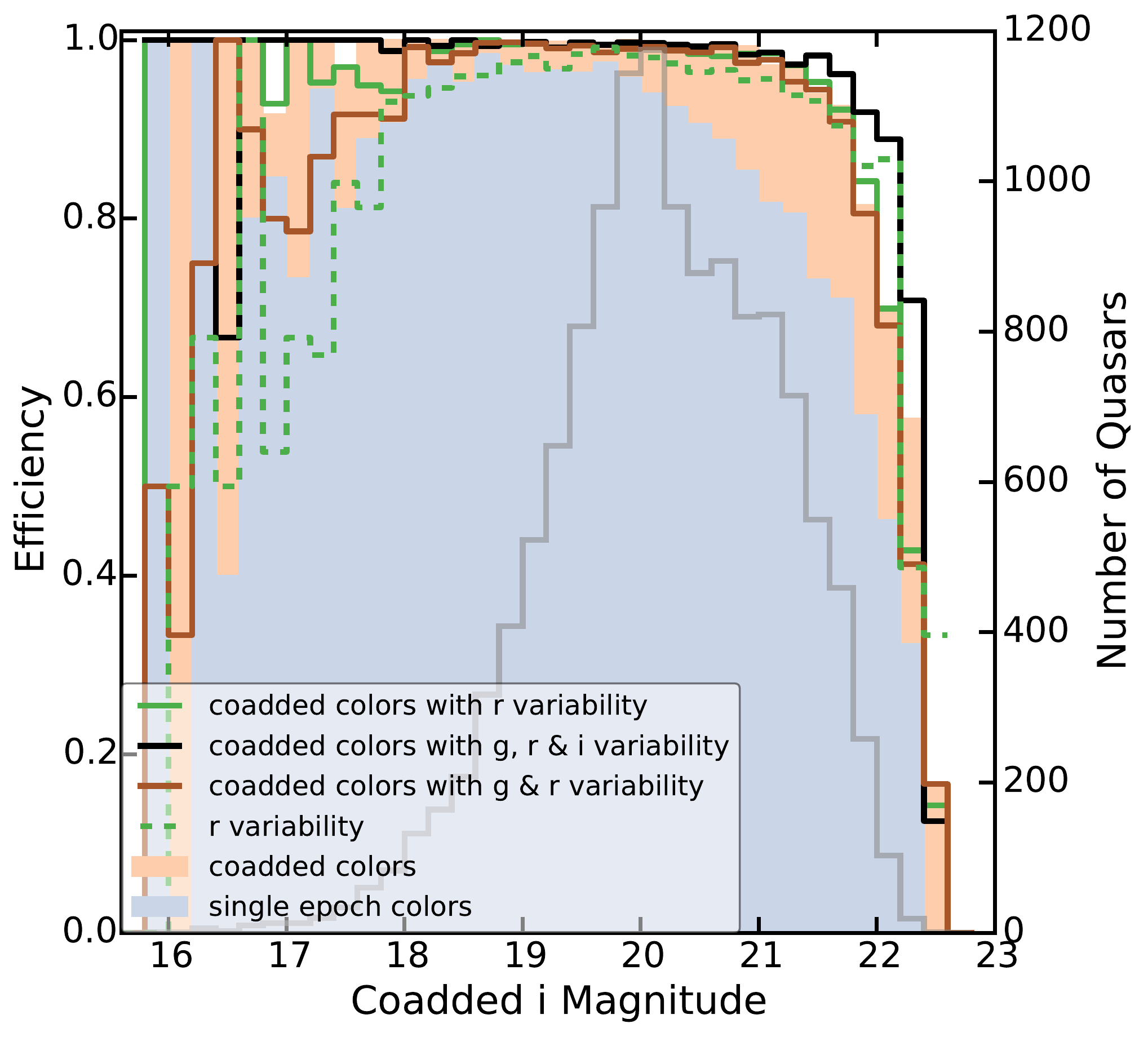}
\end{tabular}
\caption{Comparison of self tests using with different combinations of color and variability. These panels demonstrate that the combination of color and variability gives the best results for completeness and efficiency as a function of redshift and magnitude with more details in the text. Shown are the completeness (known quasars classified as quasars divided by known quasars) as a function of redshift ({\it left panel}), completeness as a function of coadded $i$-band magnitude ({\it center panel}), and efficiency (known quasars classified as quasars divided all objects classified as quasars) as a function of coadded $i$-band magnitude ({\it right panel}). The gray line shows the number of quasars in each bin (right axis).}
\label{fig:zhistogram_fraction_comparison}
\end{figure*}

We graphically summarize the results of the self tests in Figure~\ref{fig:zhistogram_fraction_comparison}. Quasar completeness as a function of redshift is shown in the left panel, quasar completeness as a function of $i$ magnitude in the center panel, and quasar efficiency as a function of $i$ magnitude in the right panel. For colors alone, both coadded and single epoch, there are regions of color space where the quasar training set and non-quasar training set overlap, resulting in redshift regions with poor completeness. Variability alone, as demonstrated by the $r$-band selection, does not have these redshift trends, but has a lower efficiency than coadded colors at all other redshifts. The addition of coadded colors to the $r$-band variability information helps to improve upon the colors alone at all redshifts, but in particular in the dips at $z \sim 2.7$ and $z \sim 3.5$.  Using coadded colors together with variability in multiple bands improves the classification even further (e.g., compare the solid green lines to the dotted green lines).  The left panel of Figure~\ref{fig:zhistogram_fraction_comparison} shows that adding the $i$-band variability makes things worse (possibly because the $i$-band has a lower signal-to-noise ratio than $g$ or $r$ given that quasars generally have blue spectral energy distributions), but note that there are relatively few high-redshift objects and the middle panel shows that the loss of completeness is coming from very faint objects. Moreover, the right panel shows that adding the $i$-band variability improves the efficiency. Table~\ref{table:Self Test Completeness and Efficiency} shows that while adding the $i$-band variability reduces the completeness by 1\%, it compensates by increasing the efficiency by 2\%.

These self tests of the quasar and non-quasar training sets validate our hypothesis that the most successful option is a combination of coadded colors and variability.  No combination of colors and variability was highest in both completeness and efficiency; however, the combination of coadded colors and both $g$ and $r$ variability parameters give the most robust selection with a combined product of completeness and efficiency of 93.88\% (see Table~\ref{table:Self Test Completeness and Efficiency}) and was consistent in completeness across all redshift values (see Figure~\ref{fig:zhistogram_fraction_coadded}).  As such, for our analysis of the test set in the next section, we have adopted coadded colors with both $g$ and $r$ variability parameters as our basis set.

\section{Building a Quasar Candidate Catalog}\label{sec:BuildingCatalog}
Now that the most efficient set of parameters are chosen, in Section~\ref{sec:ResultsTestSet} the algorithm is applied to the test set using the full quasar training set.  Finally, in Section~\ref{sec:ResultsBins} we test a process where the algorithm is used to perform simultaneous classification and redshift estimation. Specifically, the test set is classified using a series of quasar training sets that only contains quasars from limited redshift ranges.

\subsection{Classifying the Test Set} \label{sec:ResultsTestSet}
In the previous section we identified coadded colors combined with both $g$ and $r$ variability as producing the best classification for the training set objects. We now apply the selection to the test set. The NBC KDE algorithm was used to perform an 8\--D classification ($u-g$, $g-r$, $r-i$, $i-z$, $A_g$, $\gamma_g$, $A_r$, and $\gamma_r$), using the same bandwidths used during the self tests and an identical prior. The objects identified as quasar candidates, with $P(Q|d)>0.5$, are listed in the catalog (available online) which is described in more detail in Section~\ref{sec:Catalog}. 

\begin{figure*}
\capstart
\centering
\begin{tabular}{cc}
\includegraphics[width=3in,height=3in]{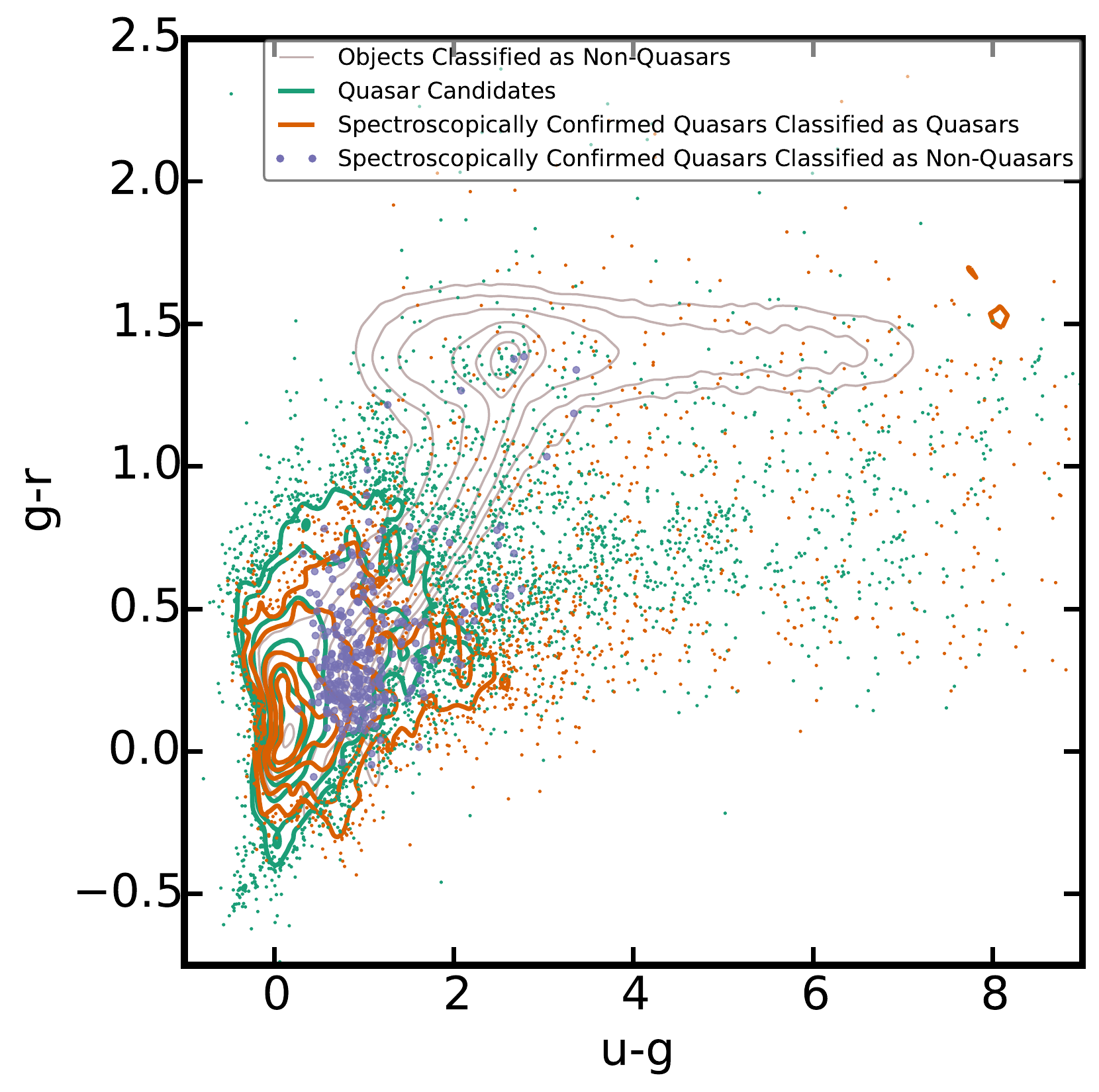} & 
\includegraphics[width=3in,height=3in]{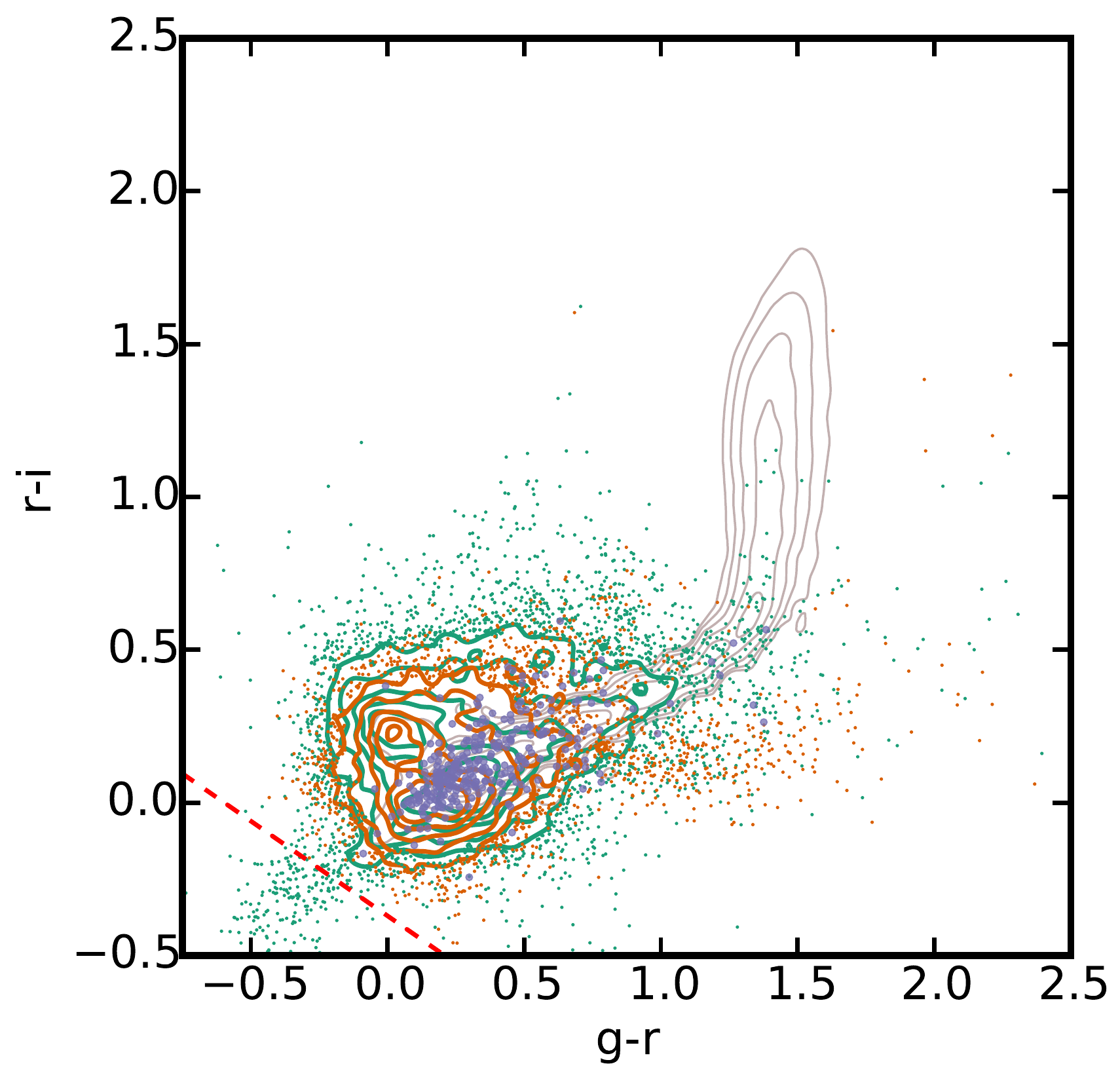} \\
\includegraphics[width=3in,height=3in]{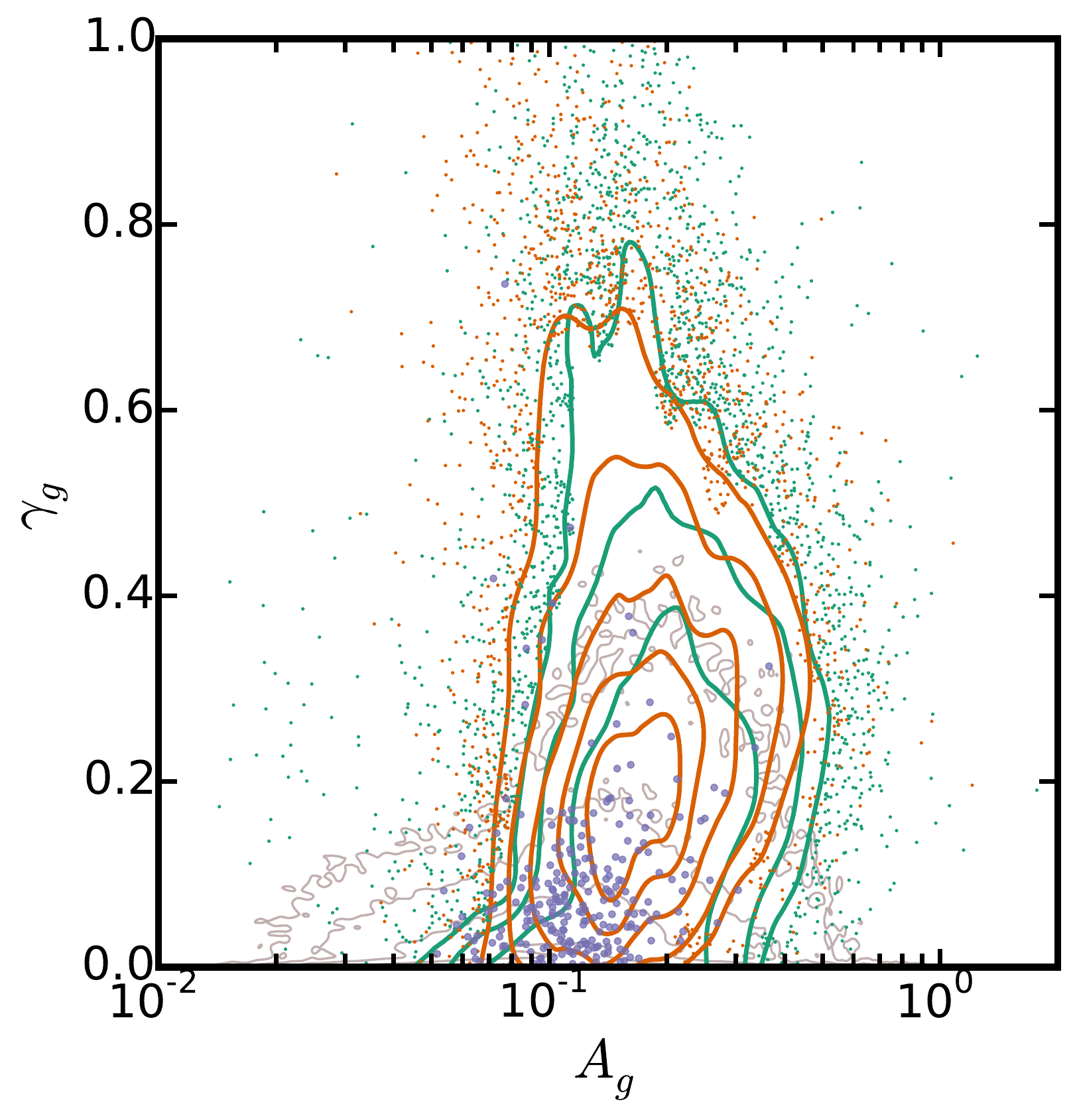} & 
\includegraphics[width=3in,height=3in]{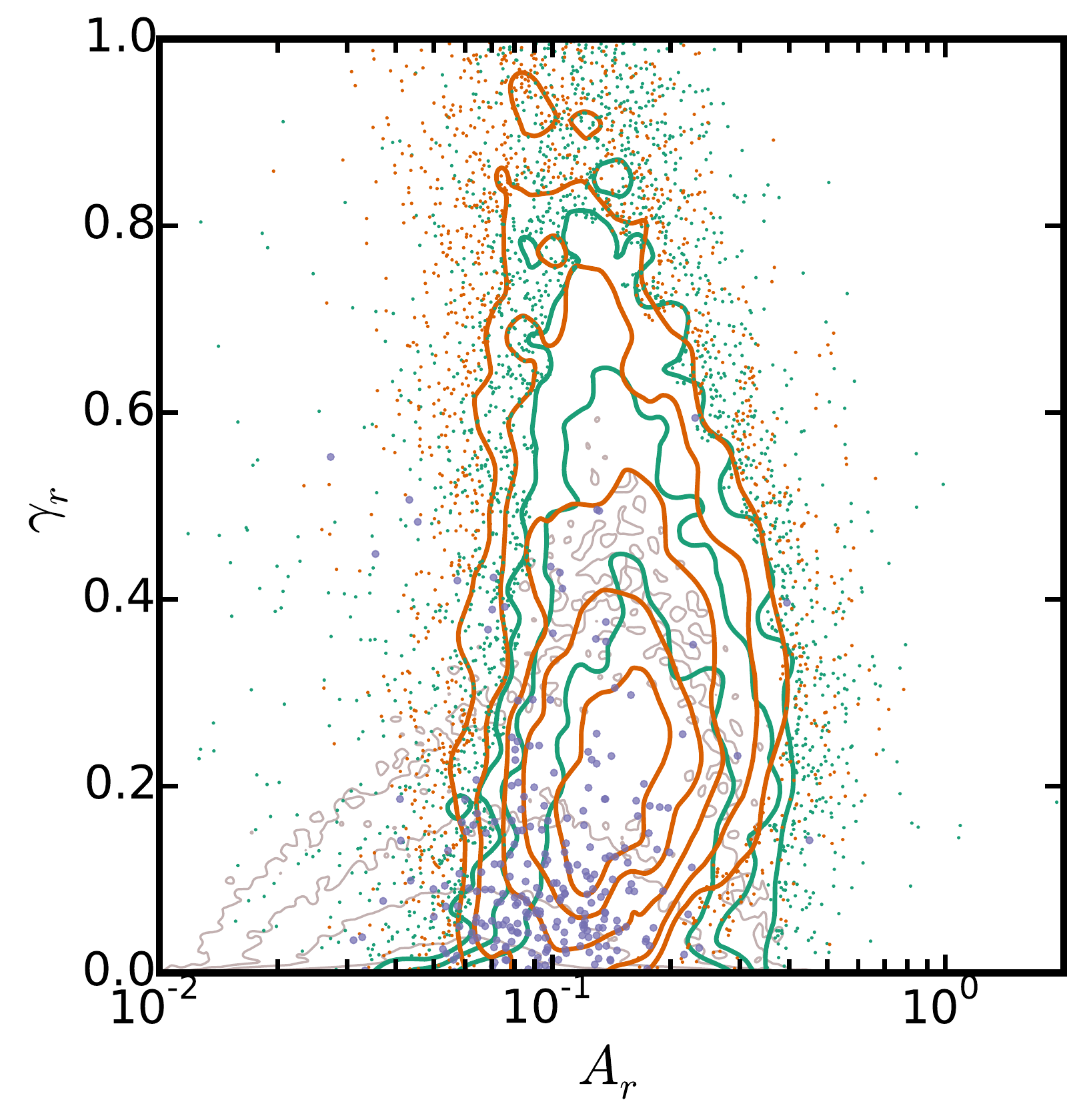}
\end{tabular}
\caption{Color and variability parameter space plots showing the results of test set classification using a single quasar training set covering the full quasar redshift range (Section~\ref{sec:ResultsTestSet}). These panels demonstrate that the incorrectly classified quasars lie in the area where quasars and non-quasars overlap in color and variability space and that the candidate quasars closely mirror the distribution of the known quasars and extend slightly beyond in the parameter space (including a region known to be inhabited by white dwarfs in the blue corner of the upper right panel). {\it Colors left panel}: $u-g$ color vs.\ $g-r$,  {\it colors right panel}:  $g-r$ vs.\ $r-i$,  {\it variability left panel}: $A_g$ vs.\ $\gamma_g$, and {\it variability right panel}: $A_r$ vs.\ $\gamma_r$. Objects in the test set classified as non-quasars are shown as gray contours\footnote{Levels for contours in Figures~\ref{fig:colorcolor_nobins_testsetresults} and \ref{fig:colorcolor_bins_testsetresults}: gray: colors - 95\%, 90\%, 80\%, 60\%, 40\%, 20\%, variability -  98\%, 95\%, 90\%, 80\%; green: colors - 90\%, 80\%, 60\%, 40\%, 20\%, variability - 90\%, 80\%, 60\%; orange: 90\%, 80\%, 60\%, 40\%, 20\%.}, quasar candidates that are not spectroscopically identified are shown as green contours and scatter points for outliers, spectroscopically identified quasars classified as quasars are shown as orange contours and scatter points for outliers, and spectroscopically identified quasars incorrectly classified as non-quasars are shown as purple dots. The red dashed line in the upper right panel is the white dwarf cut described in Eq.~\ref{eq:WDCut}.}
\label{fig:colorcolor_nobins_testsetresults}
\end{figure*}

The results of the classification are shown in Figure~\ref{fig:colorcolor_nobins_testsetresults}. We will discuss the new candidate quasars, their characteristics, and contaminants in Sections~\ref{sec:Catalog} and \ref{sec:Discussion}. In general, the candidate quasars (green contours) closely mirror the distribution of the known quasars (orange contours) and extend slightly beyond in the parameter space. The incorrectly classified quasars lie in the area where quasars and non-quasars overlap in color and variability space.  When comparing to the quasar distribution as a function of redshift shown in Figure~\ref{fig:colorcolor_redshift}, the candidate quasars extend beyond the known quasars into mid-redshift and high-redshift regions of color space. The candidate quasars have a higher density in the areas overlapping the non-quasars (gray contours), than the known quasars. This could be caused by the variability parameters selecting quasars that were missed by color selection because they are hidden in the stellar locus, or stellar contaminants in our selection.  There are also some new candidates in the bluest corner of $g-r$ vs.\ $r-i$ color space which are likely white dwarf contaminants that we will attempt to purge in Section~\ref{sec:Catalog}.

\subsection{Classification using Redshift Bins}\label{sec:ResultsBins}

\begin{figure}
\centering
\includegraphics[width=3in,height=3in]{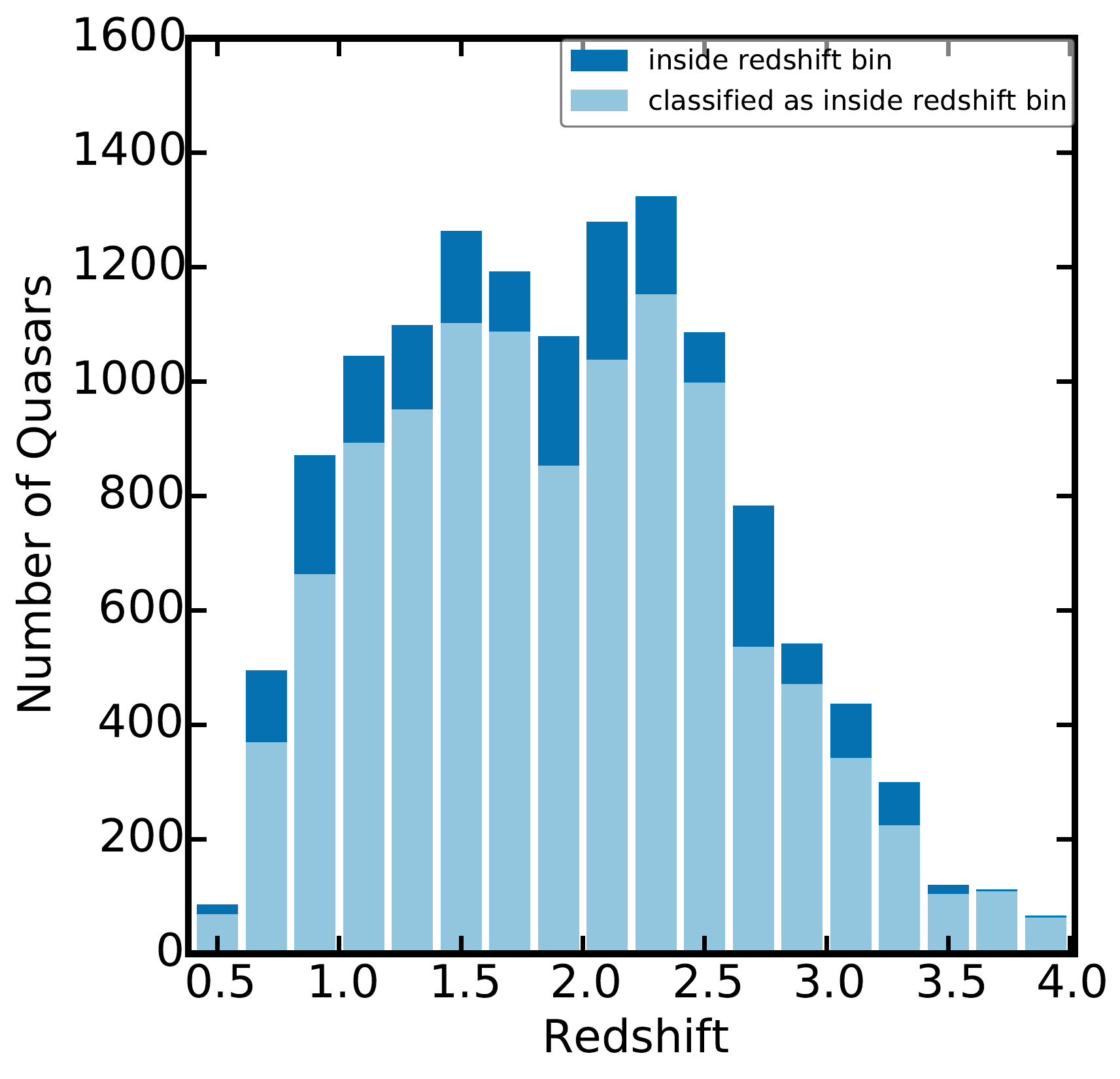}
\caption{Classification of a test set of quasars with known spectroscopic redshifts, using the training sets divided into redshift bins. Dark blue indicates all quasars in that bin, light blue indicates quasars classified with the correct redshift. The ratio of the two is the completeness of quasars inside the redshift bin.}
\label{fig:ClassificationofSpectroscopicallyConfirmedQuasars}
\end{figure}

\begin{figure*}
\capstart
\centering
\begin{tabular}{cc}
\includegraphics[width=3in,height=3in]{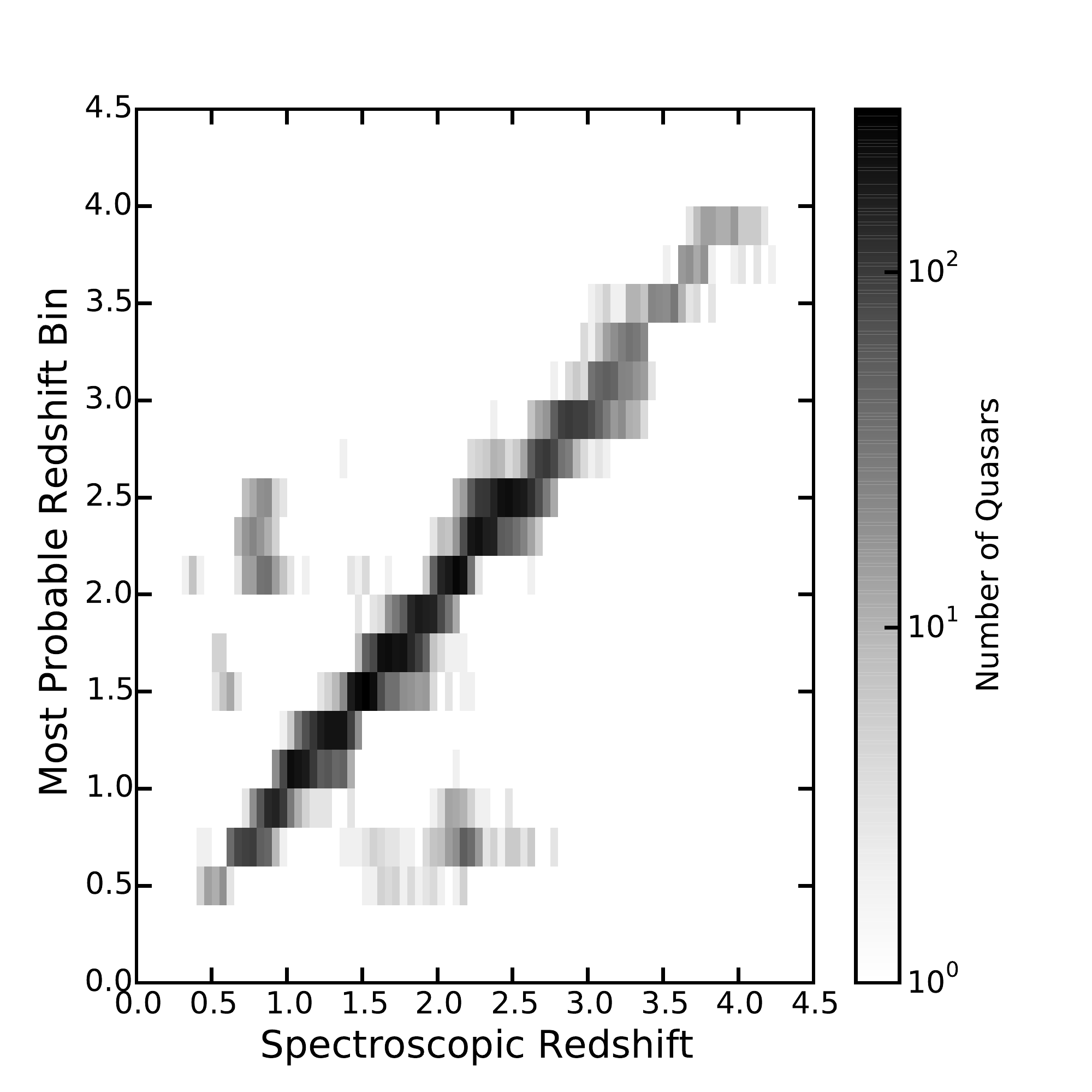} & 
\includegraphics[width=3in,height=3in]{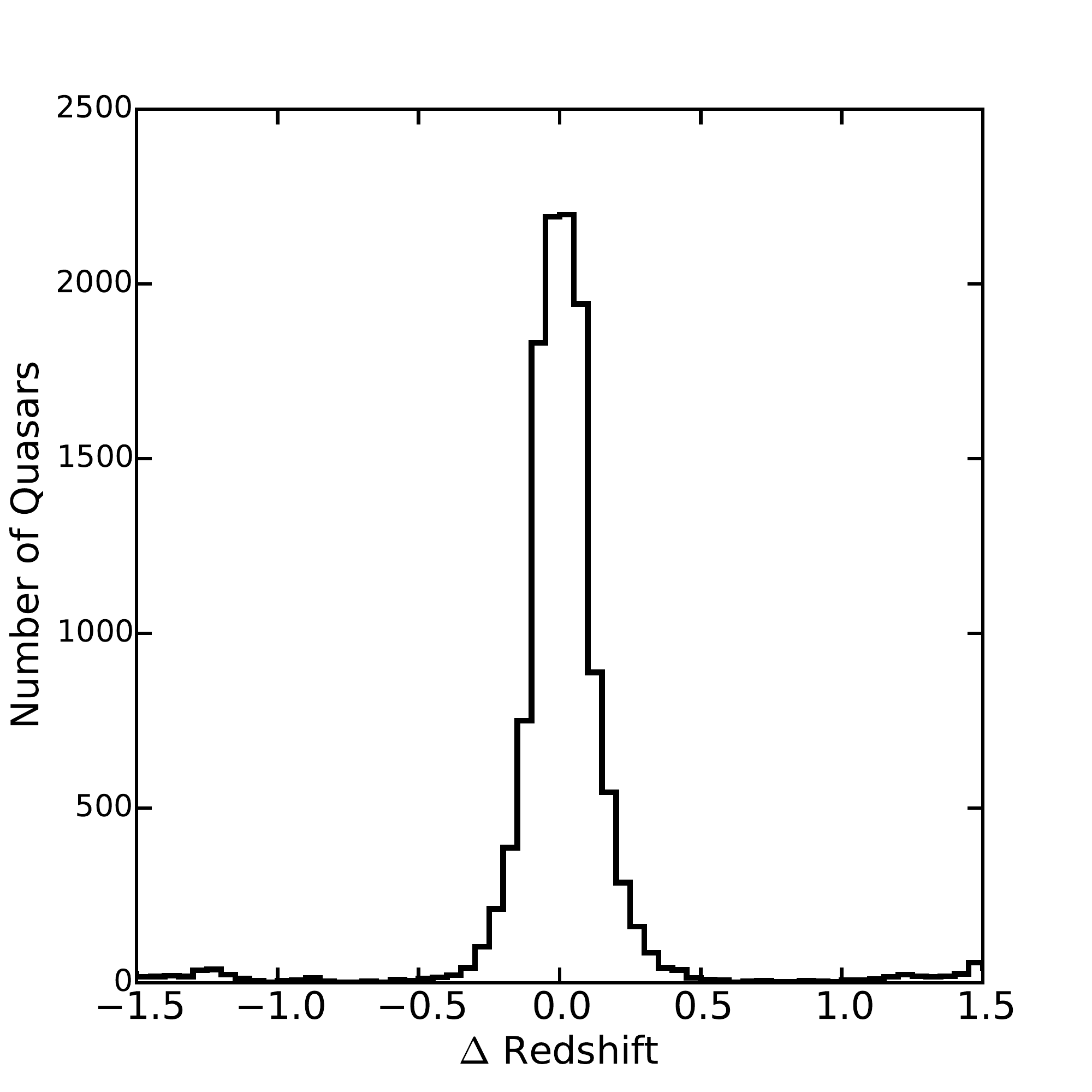}
\end{tabular}
\caption{Comparison of spectroscopic redshift to the bin into which known quasars were classified with the highest probability. {\it Left panel:} Spectroscopic redshift vs.\ the most probable redshift bin. {\it Right panel:} Histogram of $\Delta z$ (the most probable redshift bin minus the spectroscopic redshift). Only 5.6\% of the quasars have $|\Delta z| > 0.5$ \label{fig:Self_Test_Bins_zspec_vs_bin}}
\end{figure*}

Quasar colors depend on redshift as shown in Figure~\ref{fig:colorcolor_redshift}.   As such, it is possible to identify quasars while simultaneously estimating their redshifts (e.g., \citealt{Suchkov:2005}; \citealt{Bovy:2012}).  We test the extension of our method in a similar manner simply by limiting the quasar training set to a narrow redshift region.  By doing so, we are able to select quasars with colors similar to other quasars of that redshift, thereby simultaneously providing a rough estimate of the redshift.

To accomplish this, the full quasar training set (see Section~\ref{sec:Test Set and Training Sets}) was divided into 18 separate training sets by redshift: non-overlapping redshift bins from 0.4 to 4.0 with a bin width of 0.2.  The quasars outside each redshift bin were added to the {\em non-quasar} training set.  A handful of quasars that were significant outliers (5$\sigma$) from the modal color in each bin were removed from the quasar training set.  These outliers could be caused by errors in the photometry and/or heavy dust reddening. Including them caused us to find objects with those colors that are not really quasars or are quasars at a different redshift. 

As above, a self test was performed on the training sets for each redshift bin to find the optimal bandwidths.  Specifically, the redshift-bin training sets were used to classify the full quasar training set (13,221 quasars spanning the full redshift range). The results of these self tests are shown in Table~\ref{table:Classification of Spectroscopically Confirmed Quasars} and Figures~\ref{fig:ClassificationofSpectroscopicallyConfirmedQuasars} and \ref{fig:Self_Test_Bins_zspec_vs_bin}.  These show that the completeness of quasar classification (both identifying known quasars as quasars {\em and} also as being in the correct redshift bin) is generally better than 75\%. The contamination (here quasars from the wrong redshift bin being selected) is typically less than 10\%.

Of the 13,221 training set quasars, 12,535 were classified in at least one bin (94.8\% overall completeness). These objects are shown as a density plot in Figure~\ref{fig:Self_Test_Bins_zspec_vs_bin} in $\Delta z = 0.2$ photometric redshift bins. The regions of misclassification at spectroscopic redshifts $\sim0.75$ and $\sim2.1$ stem from degeneracies in color-redshift space.

\begin{figure*}
\capstart
\centering
\begin{tabular}{cc}
\includegraphics[width=3in,height=3in]{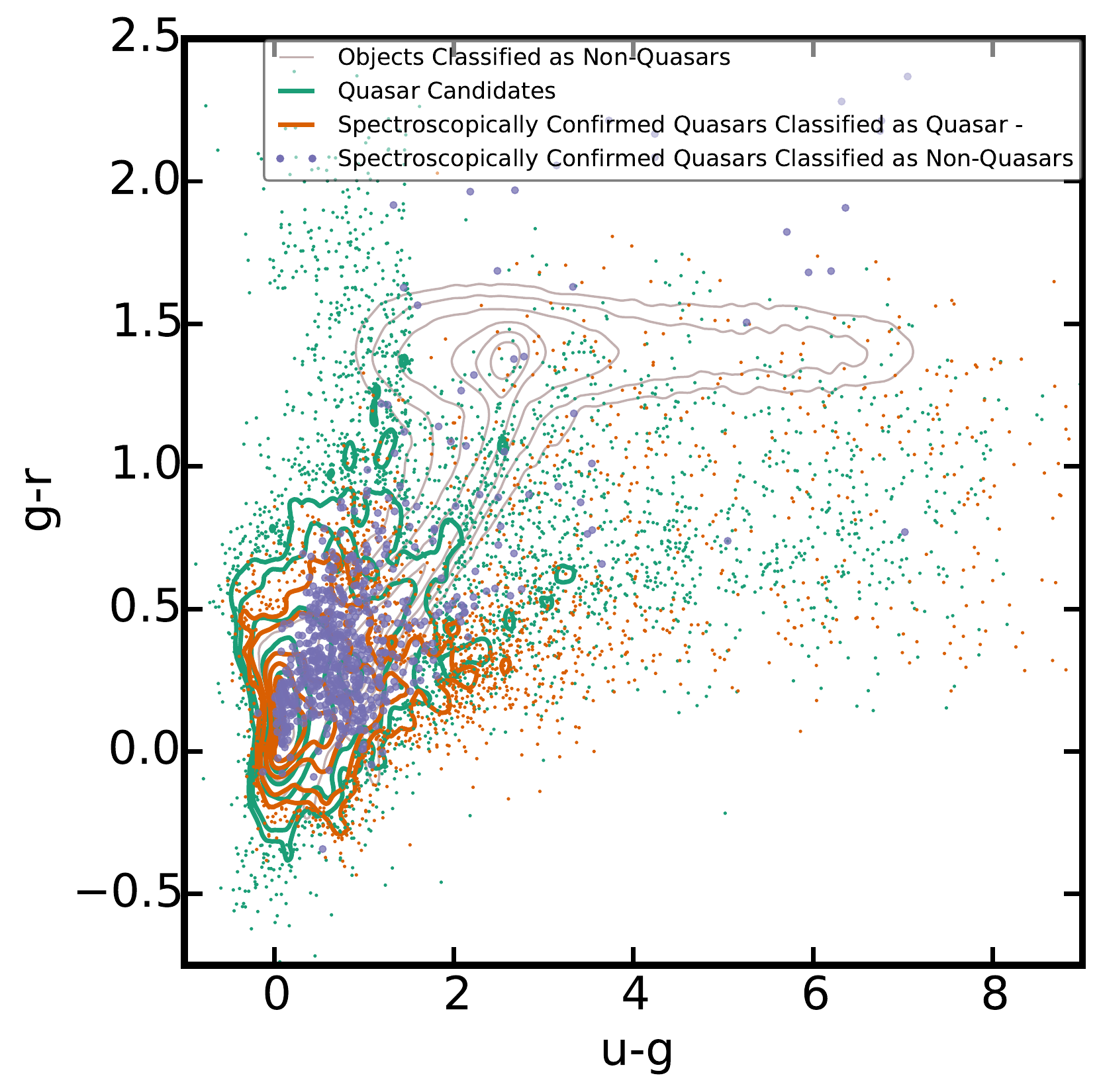} & 
\includegraphics[width=3in,height=3in]{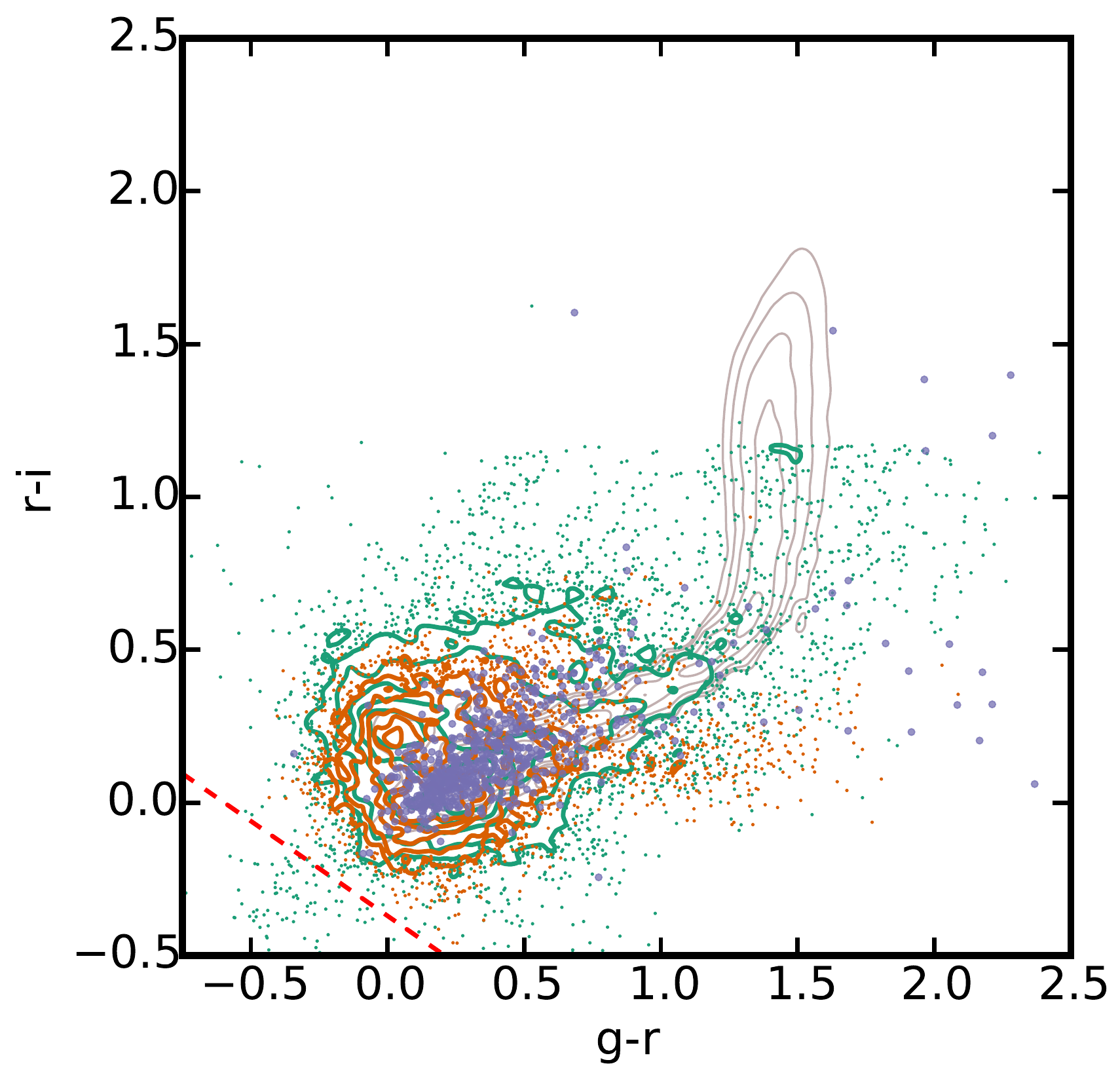} \\
\includegraphics[width=3in,height=3in]{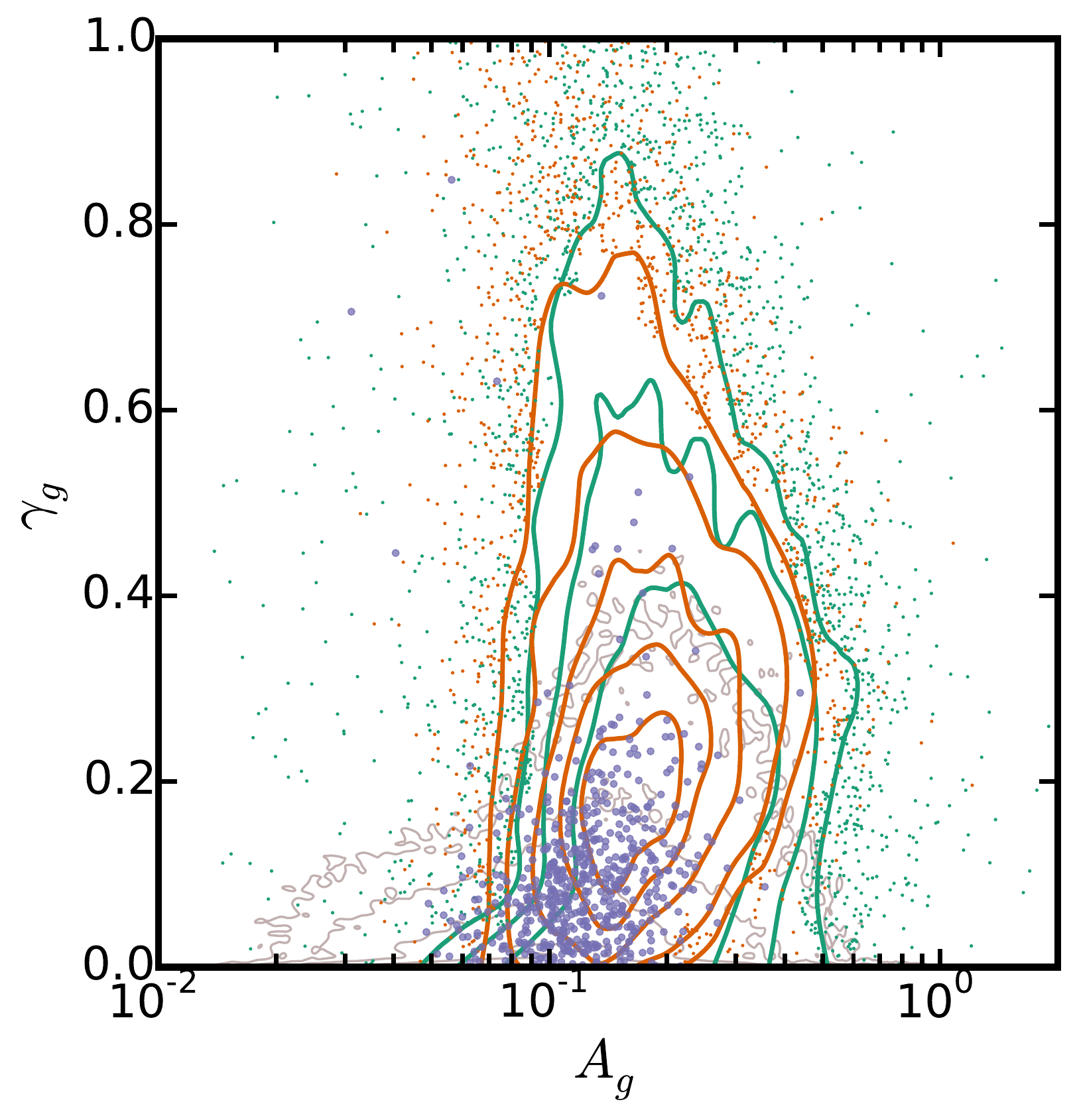} & 
\includegraphics[width=3in,height=3in]{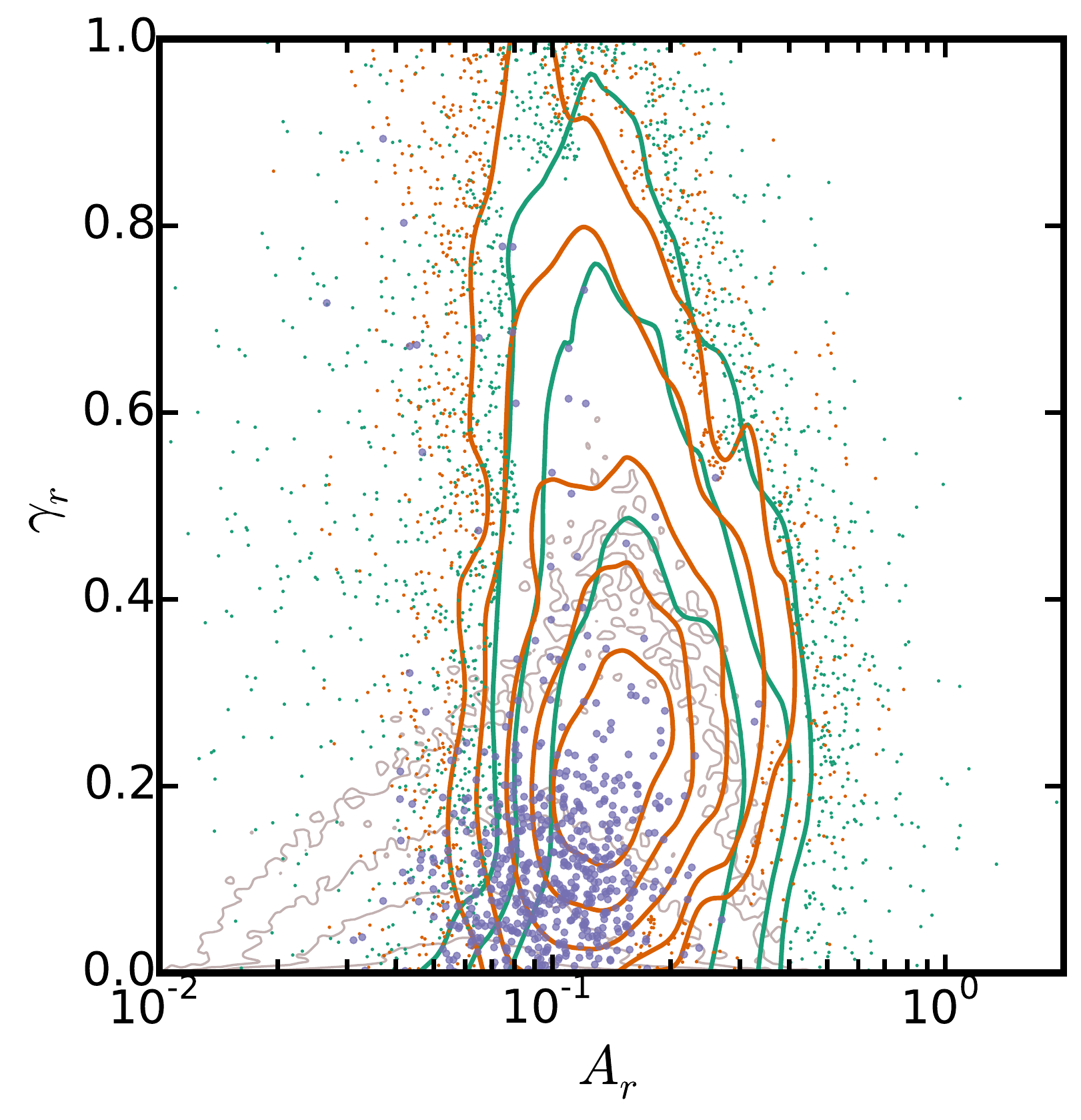}
\end{tabular}
\caption{As Figure~\ref{fig:colorcolor_nobins_testsetresults}, color and variability space plots showing the results of test set classification, but using redshift bins (described in Section~\ref{sec:ResultsBins}). In the bottom panels, we find that the selection in variability parameter space shows no noticeable difference to Figure~\ref{fig:colorcolor_nobins_testsetresults}, which is not surprising as $A_g$ vs.\ $\gamma_g$ and $A_r$ vs.\ $\gamma_r$ have no strong redshift trends. However, there are slight differences in color space  ({\em top panels}). We discuss these further in Section~\ref{sec:Catalog}.}
\label{fig:colorcolor_bins_testsetresults}
\end{figure*}

With the self test completed, we finally classify the test set described in Section \ref{sec:Test Set and Training Sets}, the same that was classified in Section \ref{sec:ResultsTestSet}. For each of the non-overlapping redshift bins from 0.4 to 4.0, each object in the test set is returned as either a quasar candidate or a non-quasar candidate. If it is found to be a quasar candidate, we calculate the quasar probability (in addition to the initial binary classification). Many objects were found to be quasar candidates in several bins and the classification probability in each bin was calculated. Results of the classification are given in Table~\ref{table:Classification with Bins}; Figure~\ref{fig:colorcolor_bins_testsetresults} shows the results of the classification in color and variability parameter space, as in Figure~\ref{fig:colorcolor_nobins_testsetresults}. We discuss the difference in this selection and the selection in Section~\ref{sec:ResultsTestSet} in Section~\ref{sec:Catalog}. An analysis of the quasar candidates is performed in Section~\ref{sec:Discussion}.

\begin{deluxetable*}{lrrrrrr}
\tablecolumns{7}
\tablewidth{0pc} 
\tabletypesize{\small}
\tablecaption{NBC KDE Results: Test Set Classification of Spectroscopically Confirmed Quasars \label{table:Classification of Spectroscopically Confirmed Quasars}}
\tablehead{ \colhead{redshift bin}  & \multicolumn{3}{c}{number inside redshift bin} & \multicolumn{3}{c}{number outside redshift bin}\\
\cline{2-3} \cline{4-5} \cline{6-7}
\colhead{} & correct &  total & fraction &  correct &  total & fraction}
\startdata
$0.4 < z \leq 0.6$ & 67 & 84 & 0.798 & 12788 & 13137 & 0.973 \\
$0.6 < z \leq 0.8$ & 368 & 494 & 0.745 & 11855 & 12727 & 0.932 \\
$0.8 < z \leq 1.0$ & 662 & 870 & 0.761 & 11704 & 12351 & 0.948 \\
$1.0 < z \leq 1.2$ & 891 & 1043 & 0.854 & 11368 & 12178 & 0.934 \\
$1.2 < z \leq 1.4$ & 949 & 1097 & 0.865 & 11307 & 12124 & 0.933 \\
$1.4 < z \leq 1.6$ & 1100 & 1262 & 0.872 & 11147 & 11959 & 0.932 \\
$1.6 < z \leq 1.8$ & 1085 & 1191 & 0.911 & 10766 & 12030 & 0.895 \\
$1.8 < z \leq 2.0$ & 851 & 1078 & 0.790 & 11343 & 12143 & 0.934 \\
$2.0 < z \leq 2.2$ & 1036 & 1278 & 0.811 & 11150 & 11943 & 0.934 \\
$2.2 < z \leq 2.4$ & 1151 & 1322 & 0.871 & 10349 & 11899 & 0.870 \\
$2.4 < z \leq 2.6$ & 996 & 1084 & 0.919 & 10572 & 12137 & 0.871 \\
$2.6 < z \leq 2.8$ & 535 & 782 & 0.684 & 11866 & 12439 & 0.954 \\
$2.8 < z \leq 3.0$ & 469 & 540 & 0.869 & 12093 & 12681 & 0.954 \\
$3.0 < z \leq 3.2$ & 340 & 435 & 0.782 & 12377 & 12786 & 0.968 \\
$3.2 < z \leq 3.4$ & 223 & 298 & 0.748 & 12587 & 12923 & 0.974 \\
$3.4 < z \leq 3.6$ & 103 & 119 & 0.866 & 12933 & 13102 & 0.987 \\
$3.6 < z \leq 3.8$ & 107 & 111 & 0.964 & 12966 & 13110 & 0.989 \\
$3.8 < z \leq 4.0$ & 61 & 65 & 0.939 & 13026 & 13156 & 0.990
\enddata
\tablecomments{Fraction of quasars inside the redshift bin correctly classified as inside the redshift bin and quasars outside the redshift bin correctly classified as outside the redshift bin from the leave-one-out cross-validation of the training sets, using the training sets divided into redshift bins.}
\end{deluxetable*}

\begin{deluxetable*}{lcc|rrrrrr}
\tablecolumns{7}
\tablewidth{0pc} 
\tabletypesize{\small}
\tablecaption{NBC KDE Results: Test Set Classification with Redshift Bins \label{table:Classification with Bins}}
\tablehead{ \colhead{redshift bin} & \multicolumn{2}{c}{QSO candidates} & \multicolumn{5}{c}{known QSOs returned} \\
\cline{2-3} \cline{4-8}
 & all & qso\_prob $>$ 0.8 & known QSOs & returned & fraction & qso\_prob $>$ 0.8 & fraction}
\startdata
$0.4 < z \leq 0.6$ & 2925 & 380 & 84 & 67 & 0.798 & 46 & 0.548 \\
$0.6 < z \leq 0.8$ & 3433 & 801 & 494 & 367 & 0.743 & 293 & 0.593 \\
$0.8 < z \leq 1.0$ & 3590 & 767 & 870 & 671 & 0.771 & 332 & 0.382 \\
$1.0 < z \leq 1.2$ & 4775 & 1920 & 1043 & 883 & 0.847 & 567 & 0.544 \\
$1.2 < z \leq 1.4$ & 6238 & 2981 & 1097 & 945 & 0.861 & 656 & 0.598 \\
$1.4 < z \leq 1.6$ & 5543 & 2237 & 1262 & 1097 & 0.869 & 754 & 0.598 \\
$1.6 < z \leq 1.8$ & 7838 & 3516 & 1191 & 1083 & 0.909 & 740 & 0.621 \\
$1.8 < z \leq 2.0$ & 5931 & 2585 & 1078 & 840 & 0.779 & 574 & 0.533 \\
$2.0 < z \leq 2.2$ & 5195 & 1948 & 1278 & 1034 & 0.809 & 582 & 0.455 \\
$2.2 < z \leq 2.4$ & 4162 & 2354 & 1322 & 1146 & 0.867 & 895 & 0.677 \\
$2.4 < z \leq 2.6$ & 4540 & 2477 & 1084 & 993 & 0.916 & 832 & 0.768 \\
$2.6 < z \leq 2.8$ & 3023 & 1028 & 782 & 524 & 0.670 & 327 & 0.418 \\
$2.8 < z \leq 3.0$ & 2246 & 1295 & 540 & 465 & 0.861 & 410 & 0.759 \\
$3.0 < z \leq 3.2$ & 1390 & 753 & 435 & 334 & 0.768 & 260 & 0.598 \\
$3.2 < z \leq 3.4$ & 1228 & 644 & 298 & 223 & 0.748 & 181 & 0.607 \\
$3.4 < z \leq 3.6$ & 1122 & 671 & 119 & 102 & 0.857 & 99 & 0.832 \\
$3.6 < z \leq 3.8$ & 596 & 399 & 111 & 106 & 0.955 & 106 & 0.955 \\
$3.8 < z \leq 4.0$ & 514 & 348 & 65 & 60 & 0.923 & 58 & 0.892 \\
\tableline
Total & 32108 &  20962 & 13153 & 10940 & 0.831 & 7712 & 0.586
\enddata
\tablecomments{Classification of the full test set of objects, using the training sets divided into redshift bins. Total will not be a sum of the above rows because many objects were classified in multiple bins.}
\end{deluxetable*}

\section{Redshift Estimation}\label{sec:ResultsPhotometricRedshifts}

In this section we will improve on the accurate, but not precise, redshift estimation of Section~\ref{sec:ResultsBins} and compute photometric redshifts for the quasar candidates. First, we will describe the astrometric information (Section~\ref{sec:Astrometry}) and near-infrared colors (Section~\ref{sec:VHS}), that will be used in addition to optical colors (Section~\ref{sec:Colors}). We combine these inputs to calculate photometric redshifts using the method described in \citealt{Weinstein:2004}. We compare the robustness of our different redshift estimates in Section~\ref{sec:PhotometricRedshifts}.

\subsection{Astrometry}\label{sec:Astrometry}
In addition to colors, our analysis will make use of {\em astrometric} measurements of quasars (\citealt{Kaczmarczik:2009}). Light rays from extraterrestrial sources are bent according to Snell's law as they enter the Earth's atmosphere from the vacuum of space. A celestial source observed from the Earth will appear higher in the sky than it actually is, unless it is at the zenith. The amount of this deflection depends on the index of refraction in the air and the photon's angle of incidence. Since the index of refraction of air is a function of wavelength, shorter wavelength photons are bent more than longer wavelength photons. This effect is known as differential chromatic refraction (DCR).

The automated corrections for the DCR effect to the SDSS astrometry are computed as a function of a broad-band flux ratio. The DCR for any given object depends on the effective wavelength of the bandpass (the convolution of the object's SED and the filter transmission curve) of the object within a given bandpass, which in turn depends upon the filter's transmission properties and on the distribution of the source's flux within the bandpass. A pure power-law (without emission lines) changes the effective wavelength in a correctable way, but the DCR corrections become anomalous when there are emission lines.  For example, adding an emission line on the blue side of the filter makes the effective wavelength bluer, while adding an emission line on the red side makes the effective wavelength redder. For emission line objects (like quasars), the effective wavelength can be very different from the assumed power law, changing by as much as 150\AA\ in the $u$-band (\citealt{Kaczmarczik:2009}). The difference between the expected and observed astrometric displacements due to DCR enables the distinction of quasars and non-quasars in addition to providing an additional source of information about the redshift of the object. We examine the differential DCR offset (along the parallactic angle; \citealt{Filippenko:1982}) in the $u$-band ($auPar$) and in the $g$-band ($agPar$);  the effect is too small to measure in $r$, $i$, and $z$ given the astrometric errors of our data and the smaller DCR at longer wavelengths.

\cite{Kaczmarczik:2009} reduced the statistical error in the astrometric offsets of individual objects by normalizing the DCR offsets at multiple epochs (each with different airmass) to some fiducial airmass.  Here we take a different approach that we find to be more robust.  To first order, differential refraction is linear in $\tan (Z)$, where Z is the zenith angle, with zero intercept (no DCR at airmass of one at the zenith). Thus, a plot of multiple epochs of noisy quasar DCR measurements should cluster around a line with a fixed slope (for a given bandpass and object redshift) with zero intercept.

In a manner similar to our structure function fitting above, we use minimization of a log likelihood function to calculate the astrometric parameters in the $u$ and $g$ band. We fit the data with a straight line that runs through the origin and parameterize the DCR simply by the slope of the line. The light curve is cleaned of outliers in the same way as was done for the variability parameter calculation. We require at least 10 good observations in each band and at least one observation with airmass in the $r$ band greater than 1.5, which is $\tan (Z) \sim 1.1$---contrary to the variability analysis above since here higher airmass means a larger DCR signal despite greater extinction. We weight each observation by the $r$-band airmass since higher airmass observations are more rare and should have greater discriminatory power.  Further work could be performed in the future to determine if this weighting scheme is indeed optimal.  

\begin{figure}
\capstart
\centering
\includegraphics[width=3in,height=3in]{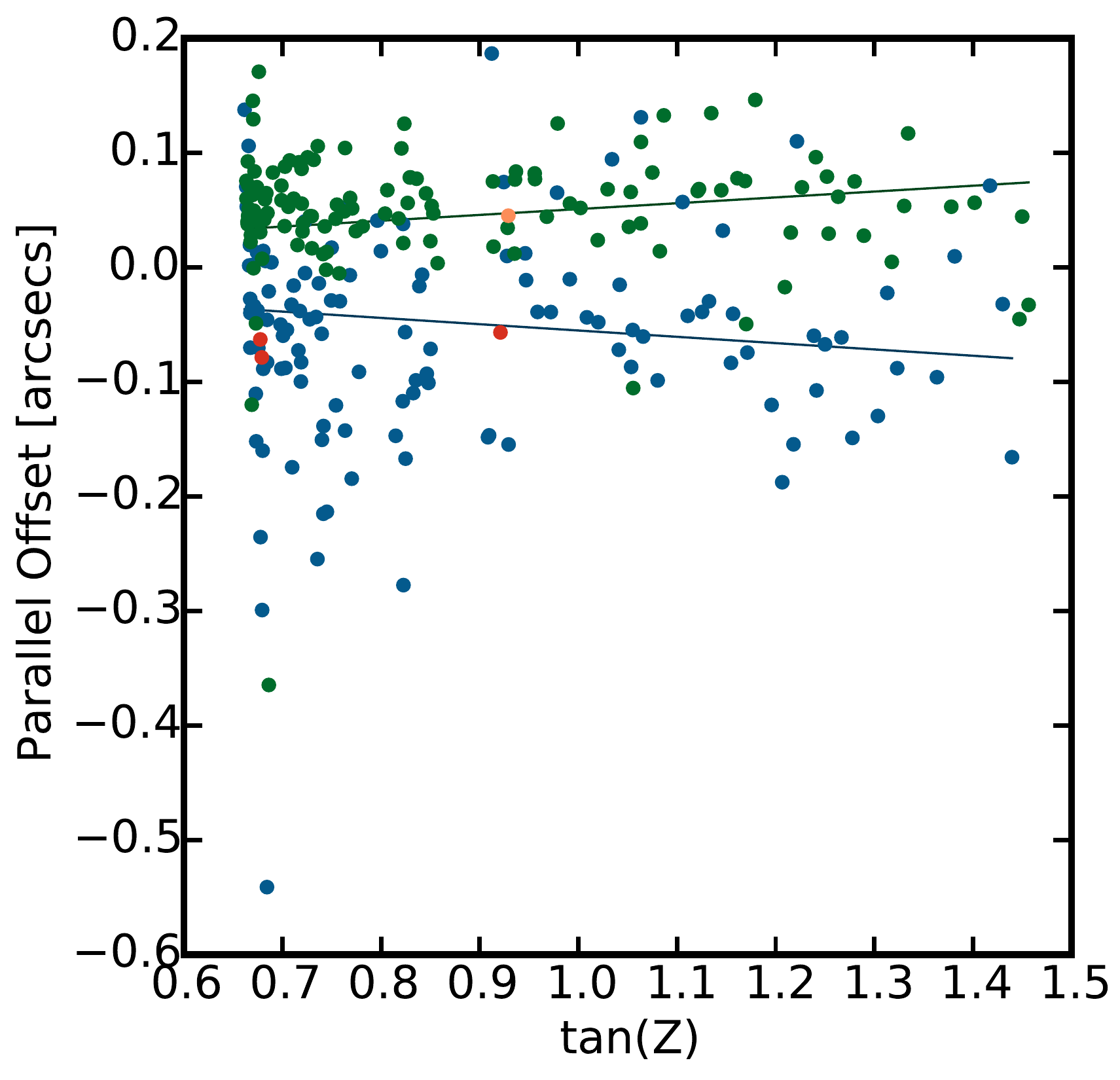}
\caption{The measured astrometric offset along the parallactic angle as a function of $\tan(Z)$. Shown is SDSS J013417.81-005036.2, a redshift 2.26 quasar from SDSS Stripe 82, the same object shown in Figure~\ref{fig:lightcurve}. This quasar is shown as an example representative of the data set. Each point refers to a different observation of this object, at a different airmass. The astrometric accuracy is $\sim 0.03$ arcsecs for $g< 20.0$, but up to 0.1 arcsecs for $g \sim 22.0$ \citep{Pier:2003}. $u$-band observations are shown in blue, with those points that were outliers removed from the light curve in Figure~\ref{fig:lightcurve} are shown in red. $g$-band observations are shown in green, with outliers removed from the light curve shown in orange. The fits, shown as solid blue and green lines, have an y-axis intercept of zero. For this quasar, the slope of the line (offset along the parallactic angle) in the $u$-band ($auPar$) is -0.055 and $g$-band ($agPar$) is 0.105. The astrometric redshift is found to be 2.57.}
\label{fig:DCRprocess}
\end{figure}

Figure~\ref{fig:DCRprocess} shows an example of this process for a single quasar with the $u$-band data in blue and the $g$-band data in green. These astrometric data can be used to constrain photometric redshifts for quasars in surveys where there are many observations and/or observations at high airmass that can provide constraints on the DCR slope. See Figure~7 of \citealt{Kaczmarczik:2009}. We will use the astrometric parameters $auPar$ and $agPar$ in Section~\ref{sec:PhotometricRedshifts} when calculating the photometric redshifts of the quasar candidates.

\begin{figure*}
\capstart
\centering
\begin{tabular}{cc}
\includegraphics[height=3in]{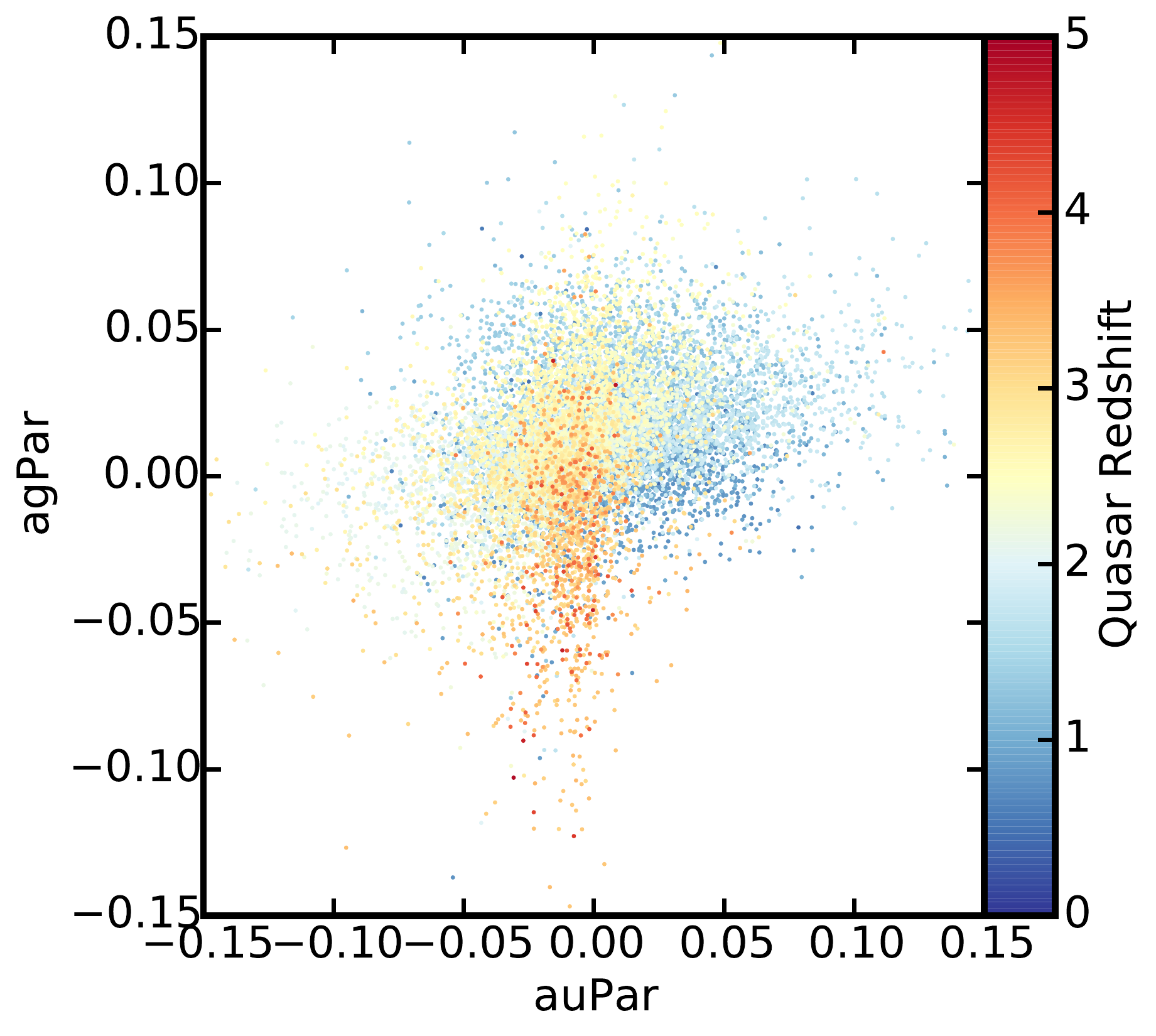} &
\includegraphics[height=3in]{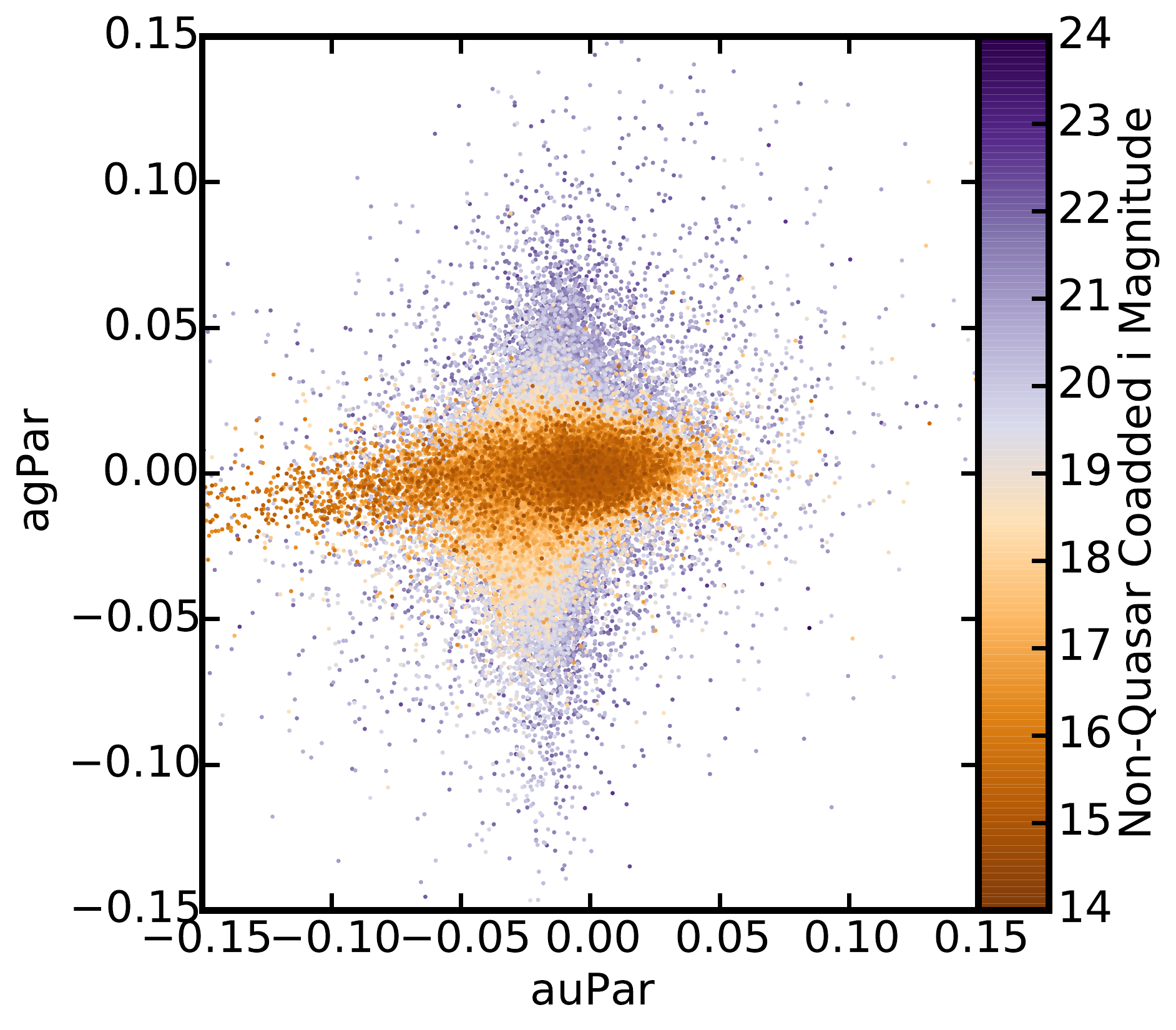}
\end{tabular}
\caption{Slope of the line (offset along the parallactic angle) with respect to redshift in the $u$-band ($auPar$) and $g$-band ($agPar$) as a function of redshift for the quasar sample ({\it left panel}) and as a function of magnitude for non-quasars ({\it right panel}). {\it Left panel:} While the changes in these astrometric parameters are not as strong as the changes in color with redshift, they provide another source of redshift information. {\it Right panel:} The differences between the distributions in the {\it left panel} and {\it right panel} can aid in the separation of quasars from non-quasars. See Section~\ref{sec:FutureWork}. For example, objects with large negative values of $auPar$ are more likely to be non-quasars than quasars.}
\label{fig:auParagPar}
\end{figure*}

In Figure~\ref{fig:auParagPar}, {\it left panel}, we plot all of the empirical DCR slopes for the quasar training set. The {\it right panel} of Figure~\ref{fig:auParagPar} shows that non-quasars and quasars have somewhat different signals in this parameter space.  We have only included point sources in this analysis, but the process should work for normal star forming galaxies too, as the 4000\AA\ break can produce significant astrometric shifts relative to the SED model assumed in the astrometric solution. In this pilot investigation, we have not used the DCR effect for classification; however, the information provided by DCR would add yet another piece of information that could be used to refine the classification probabilities of the objects in the test set. For example, objects with large negative values of $auPar$ are (empirically) more likely to be non-quasars than quasars.

\subsection{VISTA Hemisphere Survey}\label{sec:VHS}

While we select objects only using optical imaging data, we can make use of near-IR (NIR) photometry to improve our photometric redshift estimates. The VISTA Hemisphere Survey (\href{http://www.vista-vhs.org/}{VHS}) is a near-infrared survey with coverage in the southern hemisphere, including the full Stripe 82 footprint. The second VHS public data release (VHSDR2) was made available on the VISTA Science Archive (VSA)\footnote{http://horus.roe.ac.uk/vsa/index.html} in April 2014. These data include three bands $J$, $H$, and $Ks$, with (Vega) magnitude limits of $J = 20.2$, $H = 19.3$, and $Ks = 18.2$  (\citealt{McMahon:2013}). Using the Rayleigh criteria, the surveys were matched at 1.0$\arcsec$ (\citealt{Parejko:2008}): 48\% of the quasar candidates had matches in all three bands. It would be beneficial to calculate photo-z estimates for the remaining non-detections to put constraints on the quasar SED, but that is beyond the scope of this work.

\subsection{Photometric / Astrometric Redshifts}\label{sec:PhotometricRedshifts}

\begin{figure*}
\capstart
\centering
\includegraphics[width=7in]{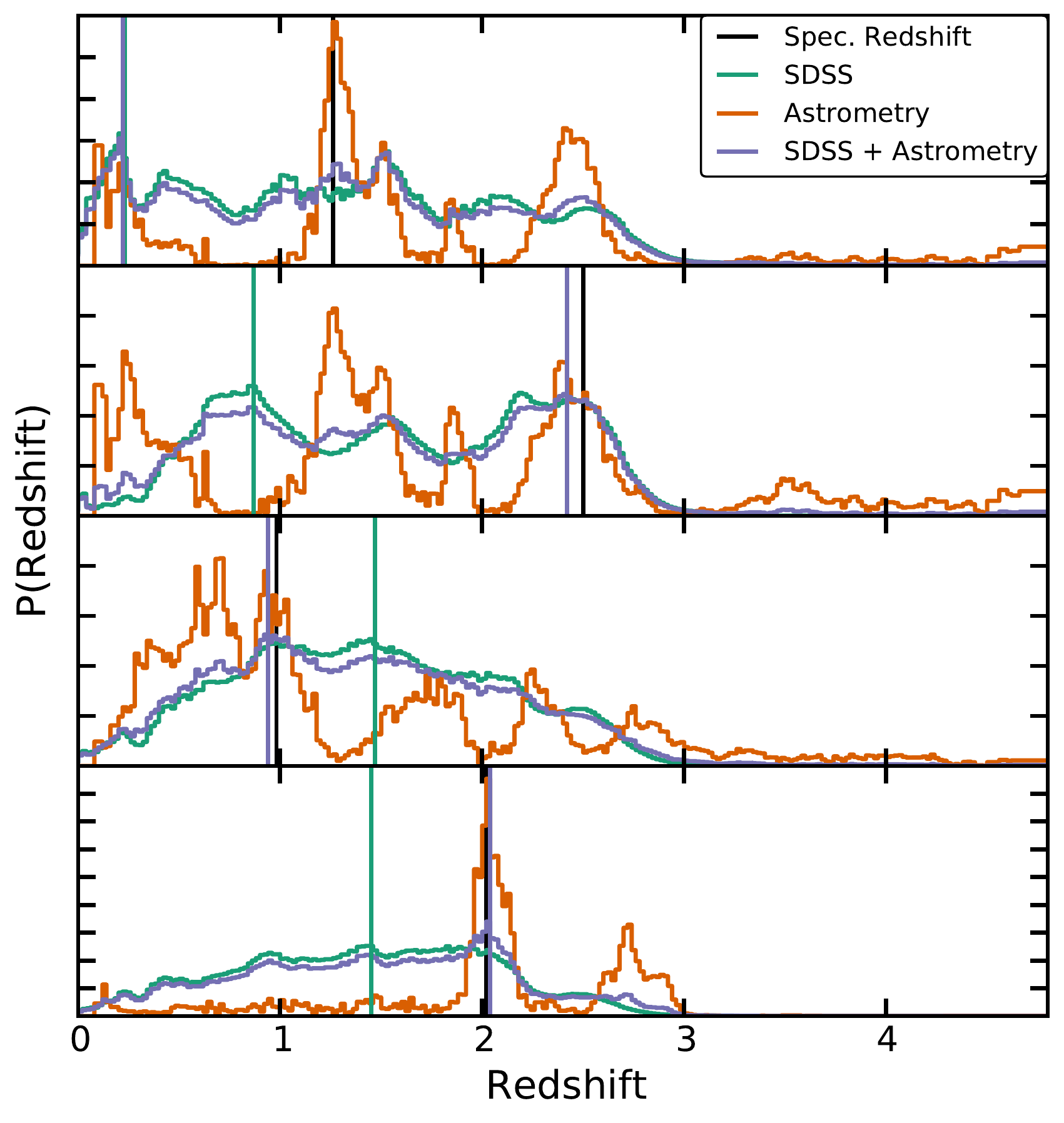}
\caption{Four selected example quasars demonstrating the photometric redshift probability function using the SDSS colors (green), astrometry (orange), and astro-photometric (purple). The expectation value of the SDSS colors PDF is shown as a vertical green line and the peak of the astro-photometric PDF is shown as a vertical purple line. The spectroscopic redshift is shown as a vertical black line.The colors curve is given five times the weight of the astrometry curve, then the two are added, and finally renormalized to create the purple curve. The top two panels demonstrate how, when the photometric redshifts returned by the colors, are inconsistent with the spectroscopic redshifts, the colors often return the spectroscopic redshift as one of the secondary peaks in the PDF. The astrometric PDF generally has several large peaks or an extended plateau. When the two PDFs are combined it often pulls out the correct peak in the colors PDF. In the top panel, the tertiary peak of the astro-photometric PDF correctly identifies the spectroscopic redshift for a low-redshift quasar where colors alone failed; a different weighting of the colors and astrometry PDFs might have picked up the correct peak. In the second panel, the primary peak of astro-photometric PDF identifies the spectroscopic redshift for a high-redshift quasar where colors alone failed. The third panel shows how the astrometry PDF helps to identify which peak in the colors PDF is correct. The bottom panel shows how a broad plateau in the colors PDF converges to the spectroscopic redshift by the addition of the astrometry PDF information.}
\label{fig:hist_photometric redshift}
\end{figure*}

Empirical photometric redshifts \citep{Richards:2001} were calculated for all of the objects that were found to be potential quasars in Sections \ref{sec:ResultsTestSet} or \ref{sec:ResultsBins}. The algorithm is described in detail in \cite{Weinstein:2004} and essentially involves least-squares fitting (without error weighting) between the candidate quasar colors and the mean (sigma clipped) colors of quasars as a function of redshift. The covariance matrix used in the process was calculated using the quasars with known spectroscopically determined redshifts. The quasars are binned by redshift in bins of width 0.02. The mean color-vector and the color covariance matrix is found for the quasars in each redshift bin; see Figure 4 of Richards et al.\ (2015, submitted). For each of the quasar candidates, we calculate how ``far" its colors are from these calculated mean colors and convert this information into a probability distribution as a function of redshift bin, as shown in Equation 5 of \cite{Weinstein:2004}. The peak of the probability distribution is reported as the photometric redshift and the confidence is calculated by integrating under the curve down to a threshold. A few examples of photometric redshift PDFs are shown in Figure~\ref{fig:hist_photometric redshift}.

\begin{figure*}
\capstart
\centering
\begin{tabular}{cc}
\includegraphics[width=3in,height=3in]{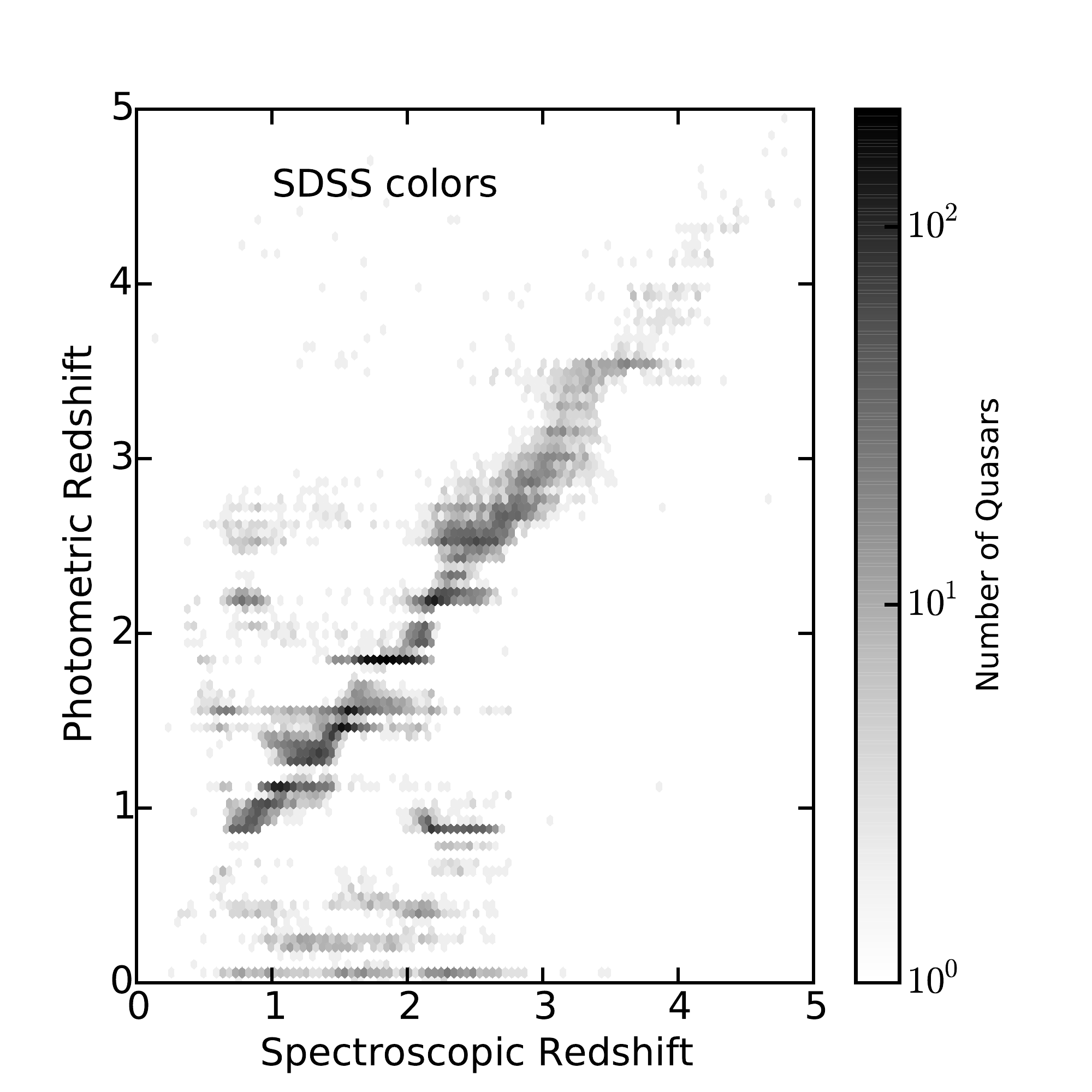} & 
\includegraphics[width=3in,height=3in]{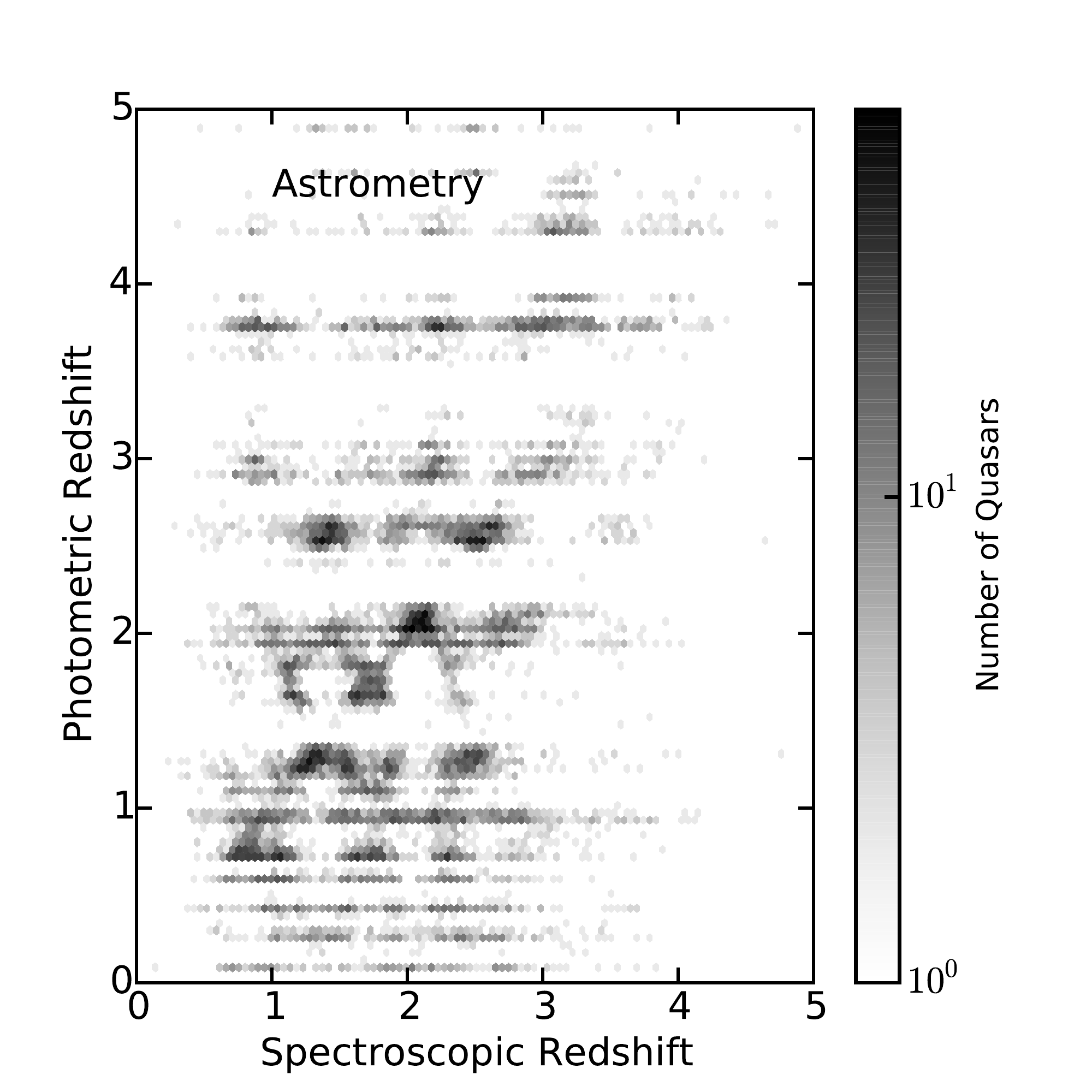} \\
\includegraphics[width=3in,height=3in]{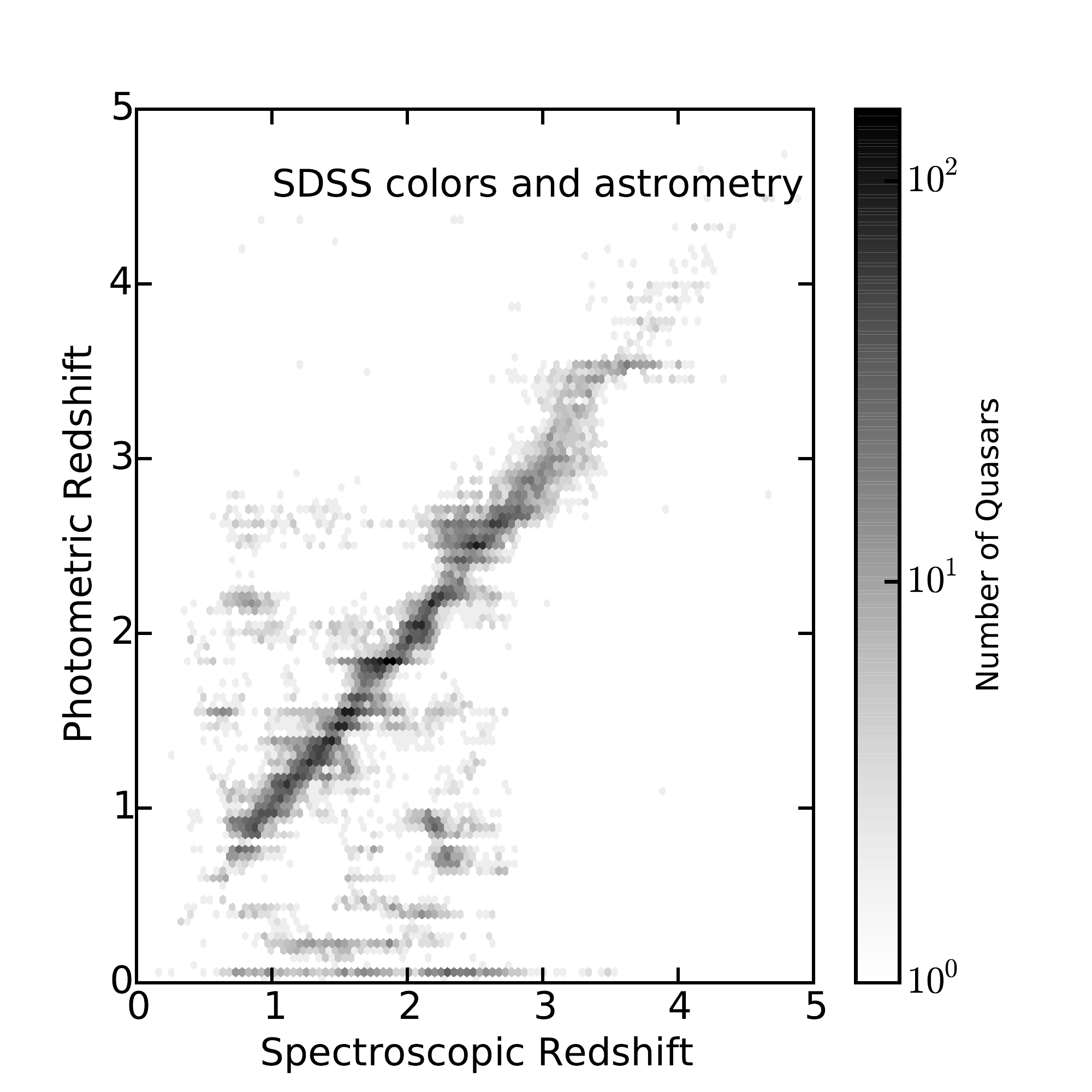} & 
\includegraphics[width=3in,height=3in]{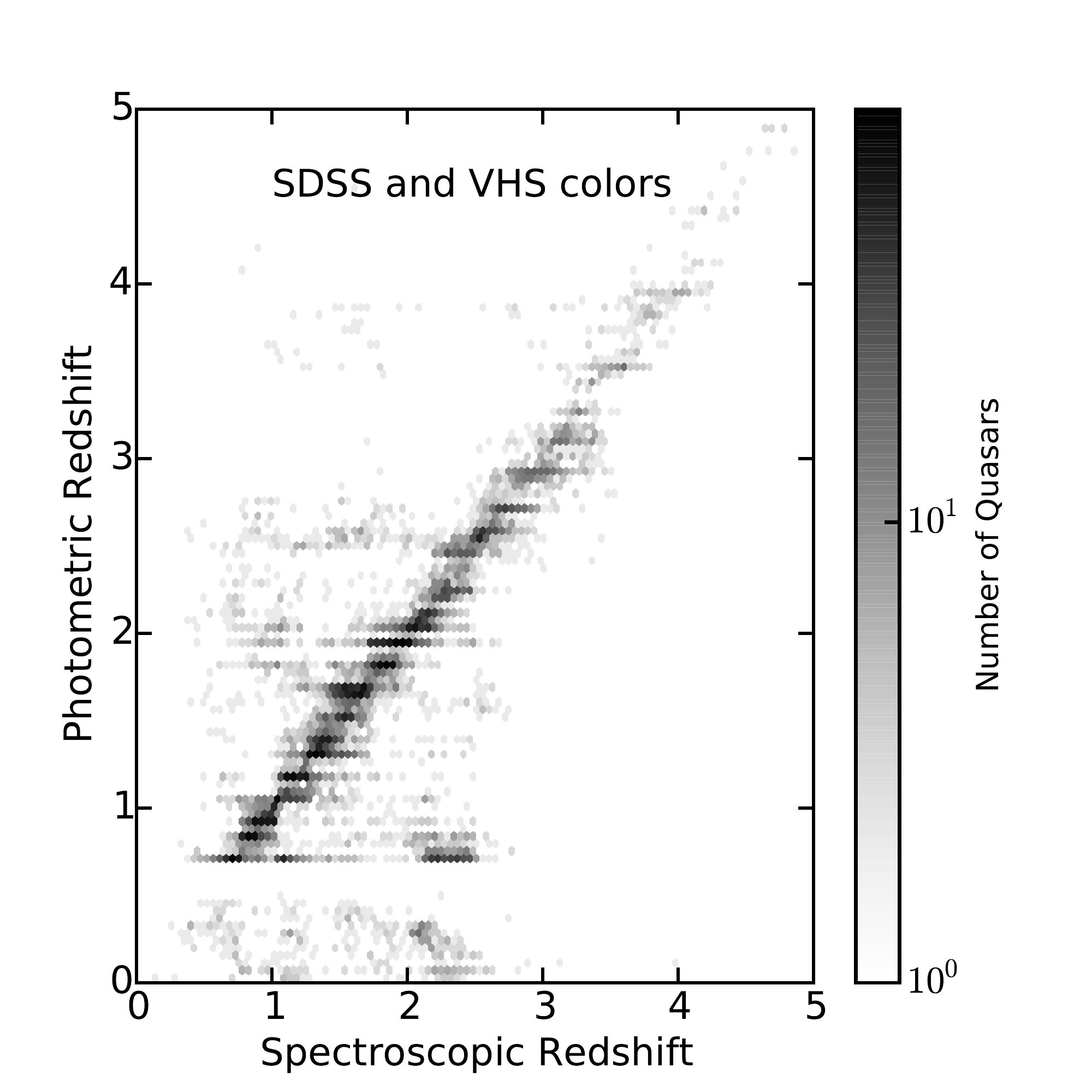}
\end{tabular}
\caption{Spectroscopic redshift vs.\ photometric redshift in hex bins with logarithmic gray scale, using ({\it top left panel}) SDSS colors (both single epoch and coadded, when available), ({\it top right panel}) astrometry, ({\it bottom left panel}) SDSS and astrometry PDFs combined, and ({\it bottom right panel}) SDSS and VHS adjacent colors. This illustrates those redshifts where the algorithm has the largest error rate (either due to degeneracy between distinct redshifts or smearing of nearby redshifts). The {\it bottom left panel} demonstrates that when the photometric redshifts returned by the SDSS colors are inconsistent with the spectroscopic redshifts, the addition of the astrometry PDF often pulls out one of the secondary peaks in the SDSS PDF as the spectroscopic redshift. The {\it bottom right panel} demonstrates how optical+IR magnitudes can similarly improve the photometric redshift accuracy. However, with the addition astrometry we can surpass the improvements due to IR data.}
\label{fig:zspec_photometric redshift}
\end{figure*}

\begin{figure}
\capstart
\centering
\includegraphics[width=3in,height=3in]{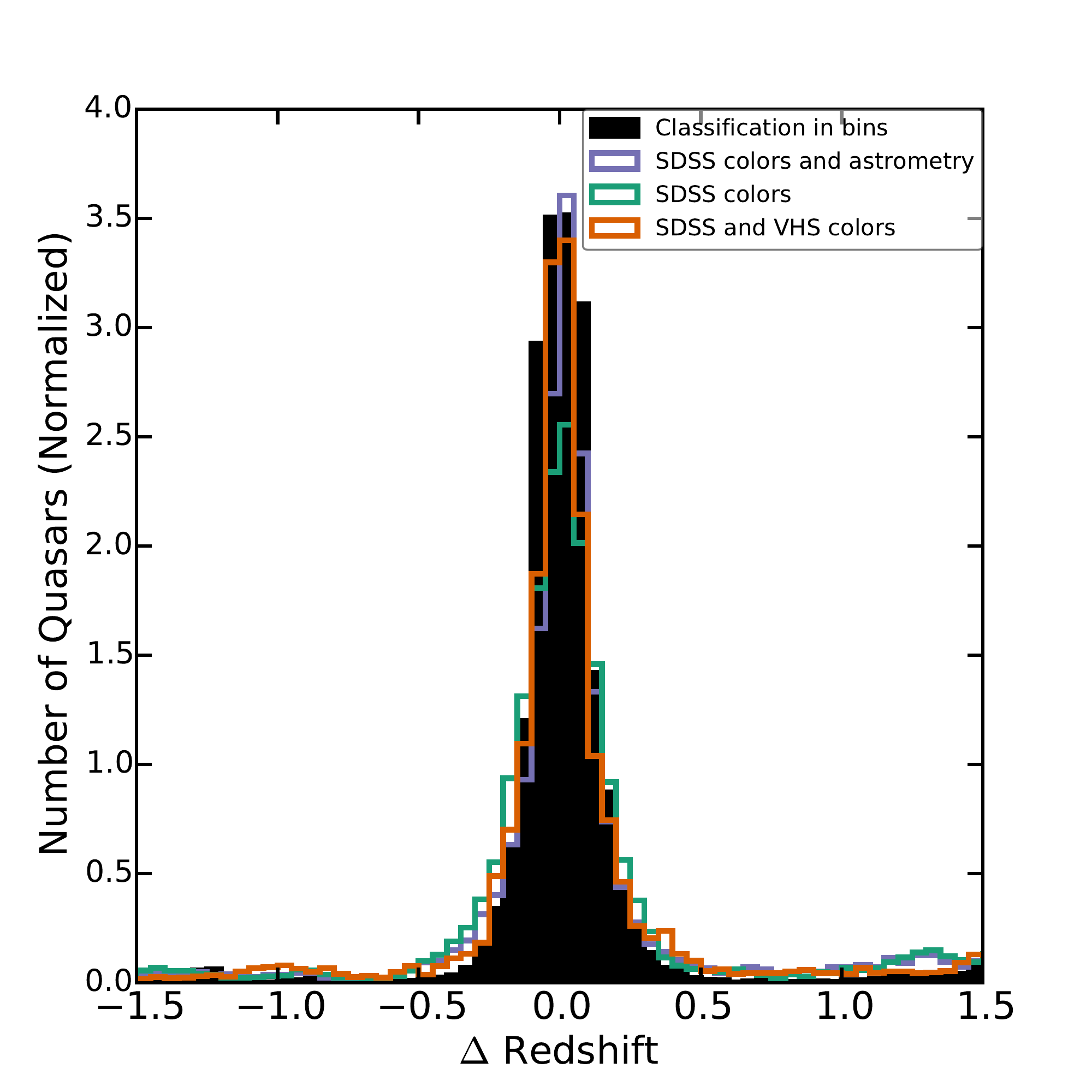}
\caption{Normalized histogram of the difference between spectroscopic redshift and photometric redshift for quasars. Note how the distribution tightens toward $\Delta \mathrm{z} = 0.0$ from the SDSS color photometric redshifts to the astro-photometric redshifts. Shown are SDSS colors (green), SDSS colors and astrometry (purple), and SDSS and VHS colors (orange). Shown in solid black is the histogram of classification in redshift bins from Figure~\ref{fig:Self_Test_Bins_zspec_vs_bin}.}
\label{fig:hist_photometricredshift}
\end{figure}

First, the photometric redshift was calculated using SDSS adjacent colors ($u-g$, $g-r$, $r-i$, $i-z$).  The mean colors were calculated using all MQC objects with known spectroscopic redshifts (i.e., not just the Stripe 82 quasars) using coadded photometry when available. We did this to improve the constraints on the photometry for high-redshift quasars. Those objects without coadded photometry have larger photometric errors, but the increase in the number of objects overcomes the noise. The color-based photo-z PDF of 4 representative objects is shown in green in Figure~\ref{fig:hist_photometric redshift}. The 13,419 quasars on Stripe 82 with spectroscopic redshifts are shown in Figure~\ref{fig:zspec_photometric redshift} ({\it top left panel}). Of these objects, 5,843 (43.5\%) have a calculated photometric redshift within 0.1 of the spectroscopic redshift and 10,201 (76.0\%) are within 0.3, as seen in Figure~\ref{fig:hist_photometricredshift}. The quasars around redshift 0.8 and 2.2 have particularly poor photometric redshifts because of a color-redshift degeneracy. This is described in detail in Section 4.2.3 of \cite{Weinstein:2004}.

Next, a redshift based on the astrometric data (the {\em astrometric} redshift) was calculated using the parameters described in Section~\ref{sec:Astrometry}. The mean vector and the covariance matrix were calculated using $auPar$ and $agPar$, using the same method as for the SDSS adjacent colors. The astrometric redshift PDF is shown in orange in Figure~\ref{fig:hist_photometric redshift}.  The 13,028 quasars on Stripe 82 with spectroscopic redshifts and for which we were able to calculate astrometric redshifts are shown in Figure~\ref{fig:zspec_photometric redshift} ({\it top right panel}). This process gives poorer redshift estimates than the SDSS photometric redshifts, but the purpose is to break degeneracies in the photometric redshifts by combining photometric and astrometric information.  That is, the astrometric redshift serves as an informative prior.

Next, the astrometric redshift PDFs and the photometric redshift PDFs are combined using weighted averages in a similar manner as \cite{CarrascoKind:2014} (Section~3.1.2 and Equation~7) to make {\em astro-photometric} redshifts.  Specifically, we have combined the PDFs by adding rather than multiplying in order to enable a relative weighting of the two PDFs.  In future work, we will consider a multiplicative joining of the data with smoothing to provide relative weighting. The colors curve is given five times the weight of the astrometry curve chosen based on empirical experiments with different weights. The resulting curve is shown in Figure~\ref{fig:hist_photometric redshift} in purple.  When the photometric redshifts returned by the colors alone are inconsistent with the spectroscopic redshifts, the correct redshift is generally one of the secondary peaks in the color-based PDF. The astrometric-redshift PDF generally has a plateau at one end of the redshift range or several large peaks. When the two PDFs are combined, it pulls out the correct peak in the color-based PDF as the best estimate of the redshift. The 13,028 training set quasars in Stripe 82 with spectroscopic redshifts and astrometric values are shown in Figure~\ref{fig:zspec_photometric redshift} ({\it bottom left panel}). Of these objects, 6,717 (51.6\%) have a calculated astro-photometric redshift within 0.1 and 10,010 (76.8\%) are within 0.3, as seen in Figure~\ref{fig:hist_photometricredshift}. 

Finally, for the 17,321 quasar candidates with matches to the VHS catalog (about 48\%) (see Section~\ref{sec:VHS}) the photometric redshift was calculated using the SDSS and VHS adjacent colors ($u-g$, $g-r$, $r-i$, $i-z$, $z-J$, $J-H$, $H-K$). The 9,244 quasars on Stripe 82 with spectroscopic redshifts and matches to VHS data are shown in Figure~\ref{fig:zspec_photometric redshift} ({\it bottom right panel}). Of these objects, 4,951 (53.6\%) have a calculated photometric redshift within 0.1 of the spectroscopic redshift and 7,250 (78.4\%) are within 0.3, as seen in Figure~\ref{fig:hist_photometricredshift}.

Figure~\ref{fig:hist_photometricredshift} demonstrates that adding either NIR colors or astrometric information significantly improves the redshift estimates over using only optical colors.   Comparison of the continuously-determined redshifts versus the discrete redshift binning from Section~\ref{sec:ResultsBins}, suggests that the binning method is somewhat more accurate (in terms of having fewer outliers), but not as precise as the astro-photometric redshifts or optical+NIR photometric redshift.

\begin{figure*}
\capstart
\centering
\includegraphics[width=7in]{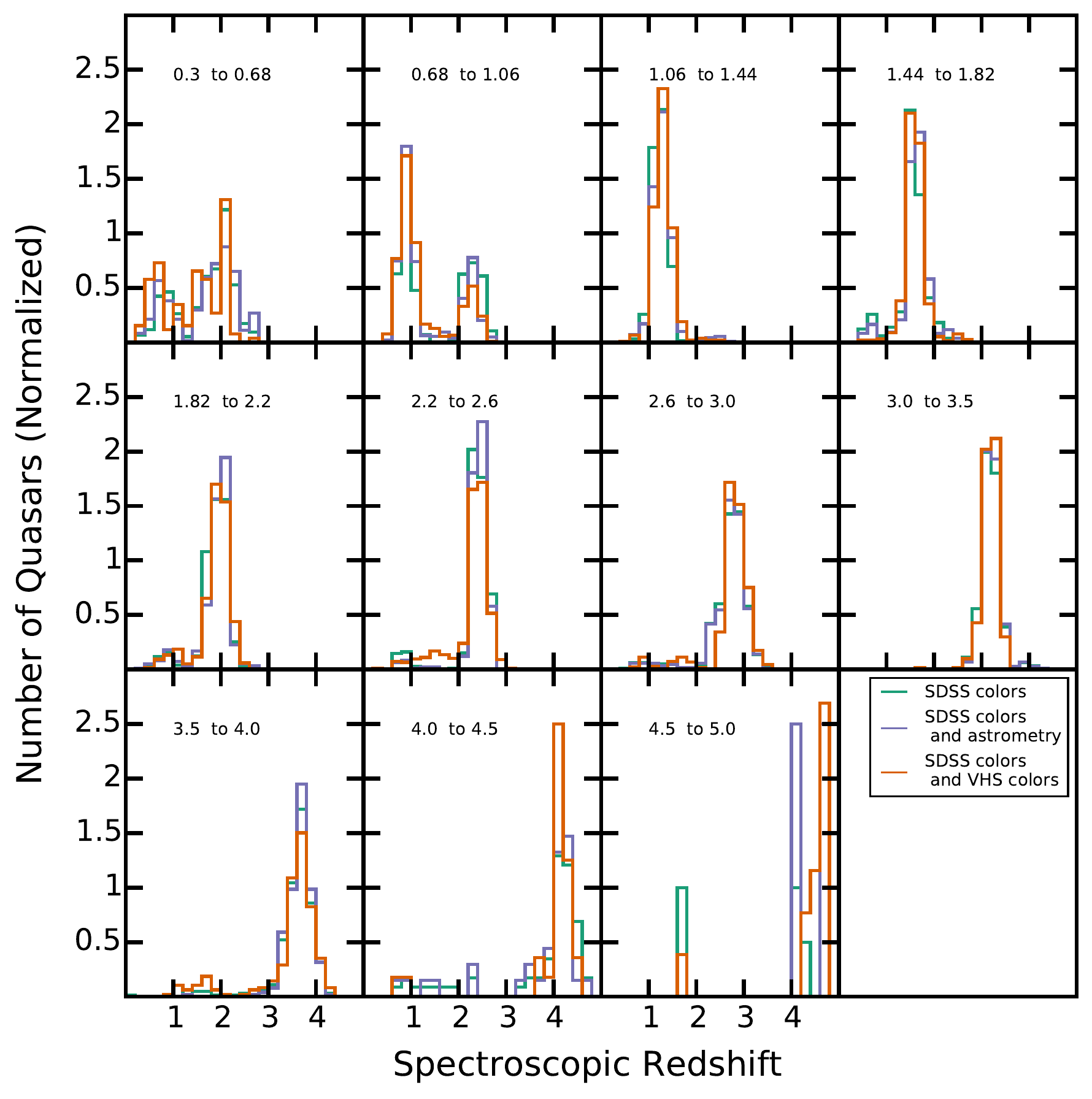}
\caption{Normalized histogram of spectroscopic redshift in panels based on bins of photometric redshift from 0.3 to 5.0 in the same bins as the luminosity function in Section~\ref{sec:DiscussionNumberCountsLuminosityFunction}. These panels demonstrate which photometric redshift ranges are most unreliable and most reliable. Photometric redshifts were calculated using SDSS colors (green), SDSS colors and astrometry (purple), and SDSS and VHS colors (orange). In particular, note the bimodal distribution at $0.68<z_{\rm phot}<1.06$ compared to the precision at $1.06<z_{\rm phot}<1.44$ and $3.0<z_{\rm phot}<3.5$. This bimodality is caused by degeneracies in color-redshift space. We correct for photometric redshift errors when calculating the quasar luminosity function in Section~\ref{sec:DiscussionNumberCountsLuminosityFunction}.}
\label{fig:hist_photometricredshift_inbins}
\end{figure*}

We graphically summarize the quality of the photometric redshifts in Figure~\ref{fig:hist_photometricredshift_inbins} by showing the distribution of true redshifts within a given photometric redshift bin. The photometric redshift bins were chosen to match those of the \cite{Richards:2006} quasar luminosity function. It will be necessary to correct for such photometric redshift errors before determining the quasar luminosity function in Section~\ref{sec:DiscussionNumberCountsLuminosityFunction}. We find that objects with photometric redshifts of $z\sim1.25$ and $z\sim3.25$ are particularly robust, whereas the $z\sim0.85$ objects are often mistaken for $z\sim2.2$.  This is caused by degeneracies in color-redshift space. As shown in Figure~1 of \cite{Richards:2001}, the colors of particular quasars can fall within the $1 \sigma$ distribution of the color-redshift relation at many redshifts. Using all four SDSS colors decreases the areas of degeneracy and adding IR colors or astrometry decreases them still further. The degeneracies found in this work are similar to those described in Section~3.4 of \cite{Richards:2001}.

Overall, we find that optical+NIR magnitudes can improve the photometric redshift accuracy; however, with astro-photometric redshifts we can surpass the improvements due to NIR data alone.

\vfill

\section{Catalog}\label{sec:Catalog}

From the classification test set, described in Section~\ref{sec:Test Set and Training Sets}, we present a FITS catalog of the 36,569 objects classified as quasars in either Section~\ref{sec:ResultsTestSet} or \ref{sec:ResultsBins}.  The number of objects and their origin (\ref{sec:ResultsTestSet} or \ref{sec:ResultsBins}) is summarized in Table~\ref{table:QuasarCandidates} and a description of the columns in the binary FITS catalog table are provided for reference in Appendix~\ref{sec:CatalogColumns}. The catalog is available online.

Another Bayesian selection method using optical and mid-infrared (MIR) colors (Richards et al. 2015, submitted) was able to clean out contaminating bright stars using some simple color cuts. We similarly use MIR color cuts to clean bright stars out of our final candidate list. To do so, we matched the quasar candidate catalog to the \href{http://wise2.ipac.caltech.edu/docs/release/allwise/}{{\it WISE} ALLWISE data release}\footnote{wise2.ipac.caltech.edu/docs/release/allwise/}. Of our candidates, 19,720 (53.9\%) had matches in both W1 and W2 (AB magnitudes). For these objects, we made the following cuts:
\begin{equation} \label{eq:WISECut1}
i < 19.5
\end{equation}
\begin{equation} \label{eq:WISECut2}
i < (-5.5(W1 - W2) + 19.5)
\end{equation}
following Richards et al.\ (2015, submitted) and using the coadded $i$ magnitude.  This process identified 573 candidates that are flagged as likely stellar contaminants in the catalog as noted in Table~\ref{table:ColumnNames}. The majority of these objects have colors that are consistent with the stellar locus and have a mean $i$ magnitude of 16.8.

Most white dwarf contaminants are below {\it WISE} detection thresholds. Thus, to eliminate these contaminants we made the following optical color cut, guided by the SDSS white dwarf catalog of \cite{Fusillo:2015}:
\begin{equation} \label{eq:WDCut}
(r - i) < (-0.62(g - r) - 0.37).  
\end{equation}
We used the coadded magnitudes and confirmed that this cut would remove none of the spectroscopically confirmed quasars from our training set. It removes 48\% of the known white dwarfs in \cite{Fusillo:2015} and identified 178 quasar candidates as possible white dwarfs. These candidates are flagged as likely white dwarf contaminants in the catalog as noted in Table~\ref{table:ColumnNames}. These possible white dwarfs are all in the bluest corner of $g-r$ vs.\ $r-i$ color space and have a mean $i$ magnitude of 21.7. 

All together, after the ALLWISE and white dwarf cuts, there are a total of 35,820 ``good" quasar candidates in Stripe 82. (Perform the following query to retrieve these objects from the catalog: {\tt WISEcut\_label}  == 0 \& {\tt WDcut\_label}  == 0 \& {\tt candidate\_label} == 1.) These candidates are used in the analysis that follows.

\begin{deluxetable*}{l cc | cc | l rl rl}
\tablecolumns{9} 
\tablewidth{0pc} 
\tabletypesize{\small}
\tablecaption{Quasar Candidates \label{table:QuasarCandidates}}
\tablehead{\colhead{Data Set} & \multicolumn{2}{c}{Candidate Quasars} & \multicolumn{2}{c}{w/ spectra} & \multicolumn{5}{c}{w/o spectra} \\
\colhead{ } & \colhead{Total} & \colhead{Fraction} & \colhead{Total} & \colhead{Completeness} & \colhead{Total} & \multicolumn{2}{c}{$i <19.9$} &\multicolumn{2}{c}{$i > 19.9$}}
\startdata
All Candidates & 36569 & 0.040 & 12953 & 0.980 & 23616 & 1570 & 0.066 &  22046 & 0.934 \\
Whole Redshift Range & 33673 & 0.037 & 12898 & 0.976 & 20775 & 1048 & 0.050 & 19727 & 0.950 \\
Redshift Bins & 32108 & 0.035 & 12511 & 0.946 & 19597 & 1282 & 0.065 & 18315 & 0.935 \\
Both Methods & 29212 & 0.032 & 12456 & 0.942 & 16756 & 760 & 0.045 & 15996 & 0.955 \\
After WISE and WD Cut & 35820 & 0.039 & 12953 & 0.980 & 22867 & 991 & 0.043 & 21876 & 0.957
\enddata
\end{deluxetable*}

Classification over the whole redshift range (as described in Section~\ref{sec:ResultsTestSet}) returned 33,240 quasar candidates, or 3.63\%, of the 916,587 objects in the test set---roughly consistent with the prior of 5\%. Of the 13,221 spectroscopically confirmed quasars that could have been returned, we found 12,898 (97.6\% completeness). Classification in redshift bins (Section~\ref{sec:ResultsBins}) returned 31,600 objects as potential quasars. Of the 13,221 spectroscopically confirmed quasars that could have been returned, we found 12,511 (94.6\% completeness).  Thus, our attempts at simultaneous classification and redshift estimation are somewhat less complete than our efforts to classify quasars regardless of redshift.  Using either method, of the 13,221 spectroscopically confirmed quasars that could have been returned, we found 12,953 (98.0\% completeness).

Of the candidates, 29,020 (81.0\%) were identified by both methods. As shown in Figures~\ref{fig:colorcolor_nobins_testsetresults} and \ref{fig:colorcolor_bins_testsetresults}, the quasars selected using these two methods show similar distributions.  In the bottom panels, we find that the selection in variability parameter space shows no noticeable difference, which is not surprising as $A_g$ vs.\ $\gamma_g$ and $A_r$ vs.\ $\gamma_r$ have no strong redshift trends. However, there are slight differences in color space  ({\em top panels}). Using the quasar training set in redshift bins we select more $g-r > 1.0$, $u-g < 2.0$, and $z_{\rm phot} > 3.4$ quasar candidates, many of them potential contaminant stars. Using the full redshift range we select more objects in the bluest corner of $g-r$ vs. $r-i$ space, many of them flagged as potential white dwarf contaminants.

\begin{figure}
\capstart
\centering
\includegraphics[width=3in,height=3in]{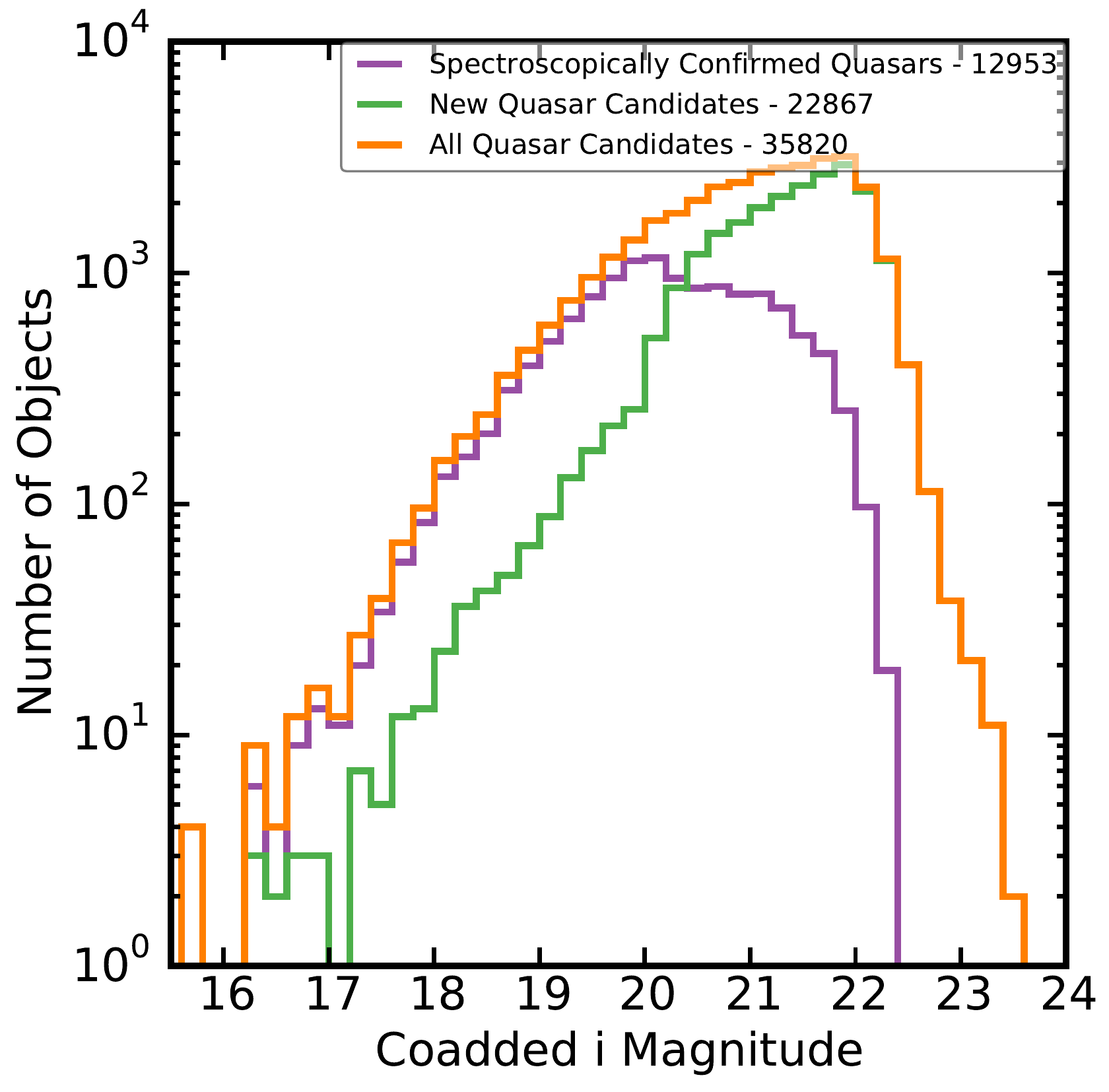}
\caption{Histogram of coadded $i$ band magnitude for known Stripe 82 quasars and new quasar candidates. In purple are the previously known, spectroscopically confirmed quasars returned by the selection. The quasar candidates returned by the selection are shown in orange and the new quasar candidates are shown in green.}
\label{fig:testset_coadd_gr_imag}
\end{figure}

As described in Section~\ref{sec:MQC}, the SDSS I/II quasars were primarily color-selected to $i<19.1$ for low-redshift and to $i<20.2$ for high-redshift \citep{Richards:2002}, but the target selection on Stripe 82 was deeper, initially going to $i=19.9$ for low-redshift  and $i=20.4$ for high-redshift; later to $i=20.2$ for low-redshift sources and $i=20.65$ for radio sources; and later to $i<21.0$ for variable sources \citep{Adelman-McCarthy:2006}. As such, when we consider the completeness of previous spectroscopic observations on Stripe 82, it is important to consider the magnitude of the objects.  The ``good" quasar candidates are shown in Figure~\ref{fig:testset_coadd_gr_imag}.  Note the change in character of the new quasar candidates at $i \sim 20.0$.

According to \cite{Vanden-Berk:2005}, the completeness of the SDSS quasar selection algorithm for sources with $i<19.1$ is $C_q = 94.9^{+2.6}_{-3.8}$\% at the 90\% confidence level. We will consider the completeness of existing quasar spectroscopy on Stripe 82 both brighter and fainter than this limit. Our region of selection extends beyond the region of uniform spectroscopic follow-up by SDSS: $-10\degree < RA < 50\degree$, therefore in order to do this comparison, we must limit our examination to this region. This includes 12,107 of the 22,867 ``good" quasar candidates in the catalog that are not spectroscopically confirmed. There are 1,090 (3,183) spectroscopically confirmed quasars brighter than a coadded $i$-band magnitude of 19.1 (19.9) and we find 61 (192) additional quasar candidates. Assuming that all of our new ``good" candidates are real, this completeness of 94.7\% (94.3\%) agrees well with \cite{Vanden-Berk:2005}. However, we might have expected it to be higher given the additional spectra taken on Stripe 82 since 2005 as part of the BOSS program.

Fainter than this limit, it could be that quasars are not being targeted or that there simply have not been enough fibers devoted to quasar candidates to find all of the objects that we consider to be valid quasar candidates. There are 4,591 spectroscopically-confirmed quasars dimmer than a coadded $i$-band magnitude of 19.9 and with a redshift $z < 3.0$. To this we add 9,536 quasar candidates with astro-photometric redshift $z < 3.0$. There are 561 spectroscopically-confirmed quasars dimmer than a coadded $i$-band magnitude of 19.9 and with a redshift $z > 3.0$. To this we add 576 quasar candidates with astro-photometric redshift $z > 3.0$.

\begin{figure*}
\capstart
\centering
\begin{tabular}{cc}
\includegraphics[width=3in,height=3in]{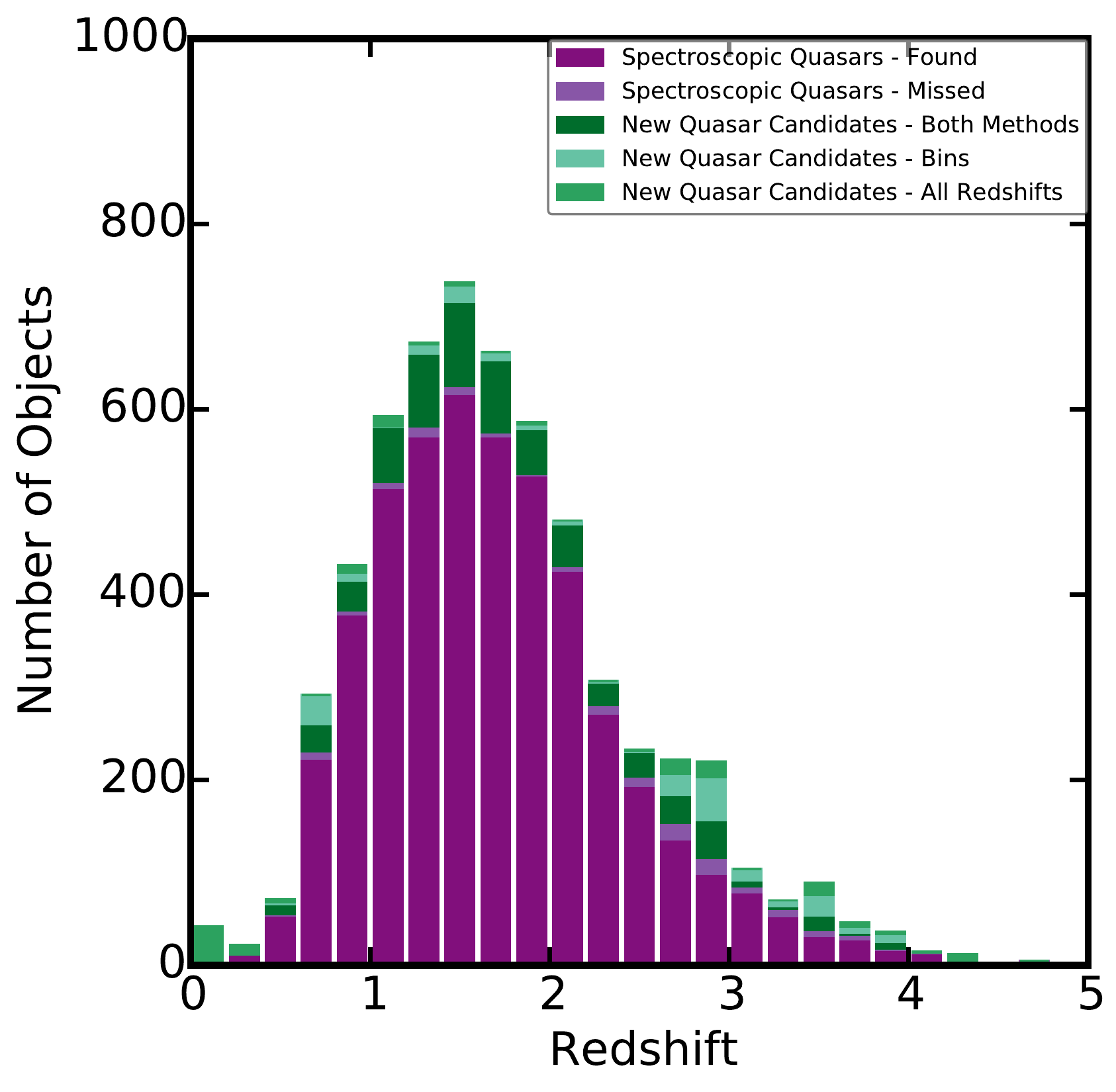} & 
\includegraphics[width=3in,height=3in]{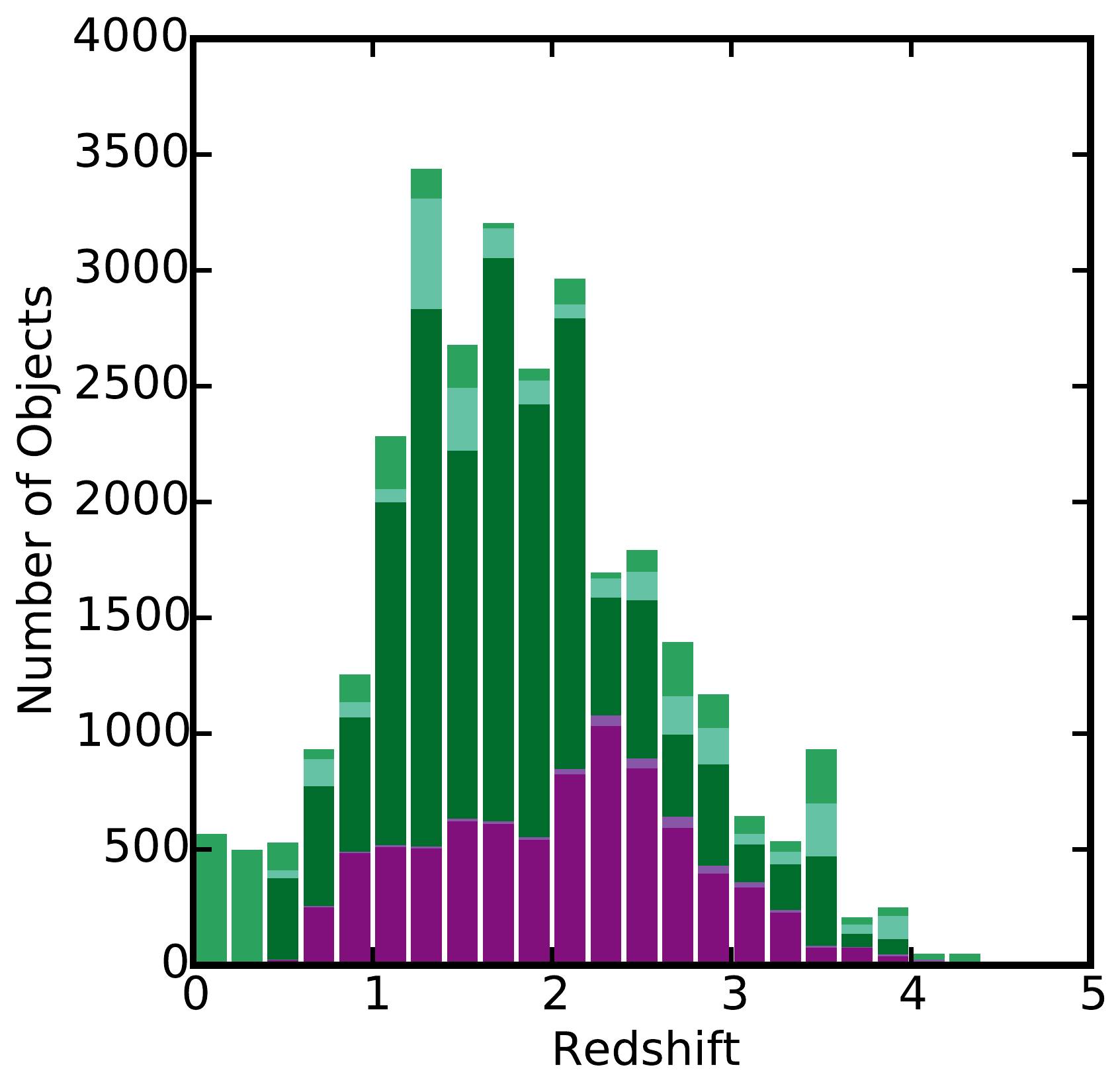}
\end{tabular}
\caption{Stacked histogram of redshift for known Stripe 82 quasars and new quasar candidates. {\it Left panel} shows the quasars and candidates $i<19.9$ and {\it right panel} shows $i>19.9$. Spectroscopic quasars found as quasar candidates and spectroscopic quasars missed are both binned by spectroscopic redshift. Quasar candidates found by both methods and quasar candidates found only using a binned quasar training set are both binned by where the candidate was classified with the highest probability. These bins only span $0.4 < z < 4.0$. Quasar candidates found only using a quasar training set over the full redshift range are binned by the astro-photometric redshift.}
\label{fig:ClassificationwithBins}
\end{figure*}

Figure~\ref{fig:ClassificationwithBins} shows the completeness and new quasar selection as a function of redshift. The {\it left panel} shows the quasars and candidates for $i<19.9$ and {\it right panel} shows $i>19.9$. In short, we have shown that current methods, (only colors, only variability, and other techniques used for Stripe 82 target selection) still are incomplete.  Next-generation surveys like LSST will have to adopt more sophisticated methods, of which ours is just a pilot example, to fully exploit the data.

While we find new quasars in Stripe 82, the catalog also includes 466 objects that were {\em not} selected by our algorithms as quasar candidates, but that are spectroscopically confirmed quasars. This incompleteness demonstrates where there is room for improvement beyond our pilot project. For the sake of completeness, to illustrate where we may be less sensitive, and to make it easier to compute the completeness corrections for our catalog without needing another data source, these quasars are included in our catalog.  They are indicated by {\tt candidate\_label} $ == 0$. In general, they are in the densest part of the stellar locus and have very small $\gamma_{g}$ and $\gamma_{r}$ values. More than 50\% are between redshifts $2.2 < z < 3.2$ and more than are third are $i > 21.5$ compared to 5\% and 9\% of the quasar training set as a whole, making these objects particularly difficult to distinguish as quasars. 

\section{Discussion} \label{sec:Discussion}
We will now explore the quality of the quasar catalog by comparing to other cuts and catalogs, in addition to evaluating it for remaining contaminants. In Section~\ref{sec:DiscussionSchmidtCuts} we use the quasar variability selection box from \cite{Schmidt:2010} as a comparison in time domain classification. In Section~\ref{sec:DiscussionBOSSQuasarSelection}, we evaluate how well our algorithm recovers quasars from BOSS DR10 and DR12 quasar catalogs. Finally, in Section~\ref{sec:DiscussionNumberCountsLuminosityFunction}, we evaluate completeness and contamination of the candidate quasars using number counts and luminosity function analysis.

\subsection{Comparison to Other Variability Based Selection}\label{sec:DiscussionSchmidtCuts}
\begin{figure*}
\capstart
\centering
\begin{tabular}{cc}
\includegraphics[width=3in,height=3in]{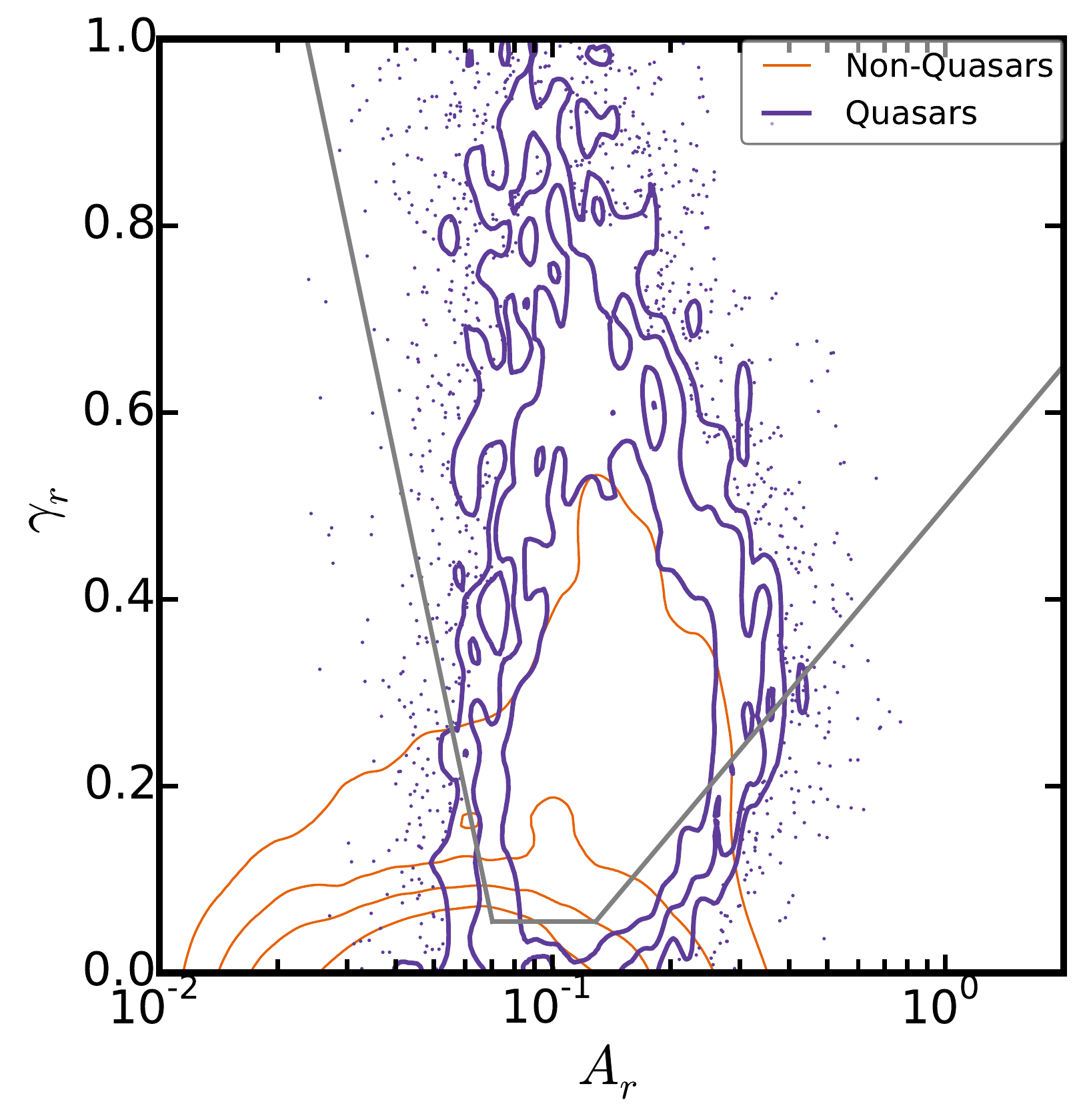} & 
\includegraphics[width=3in,height=3in]{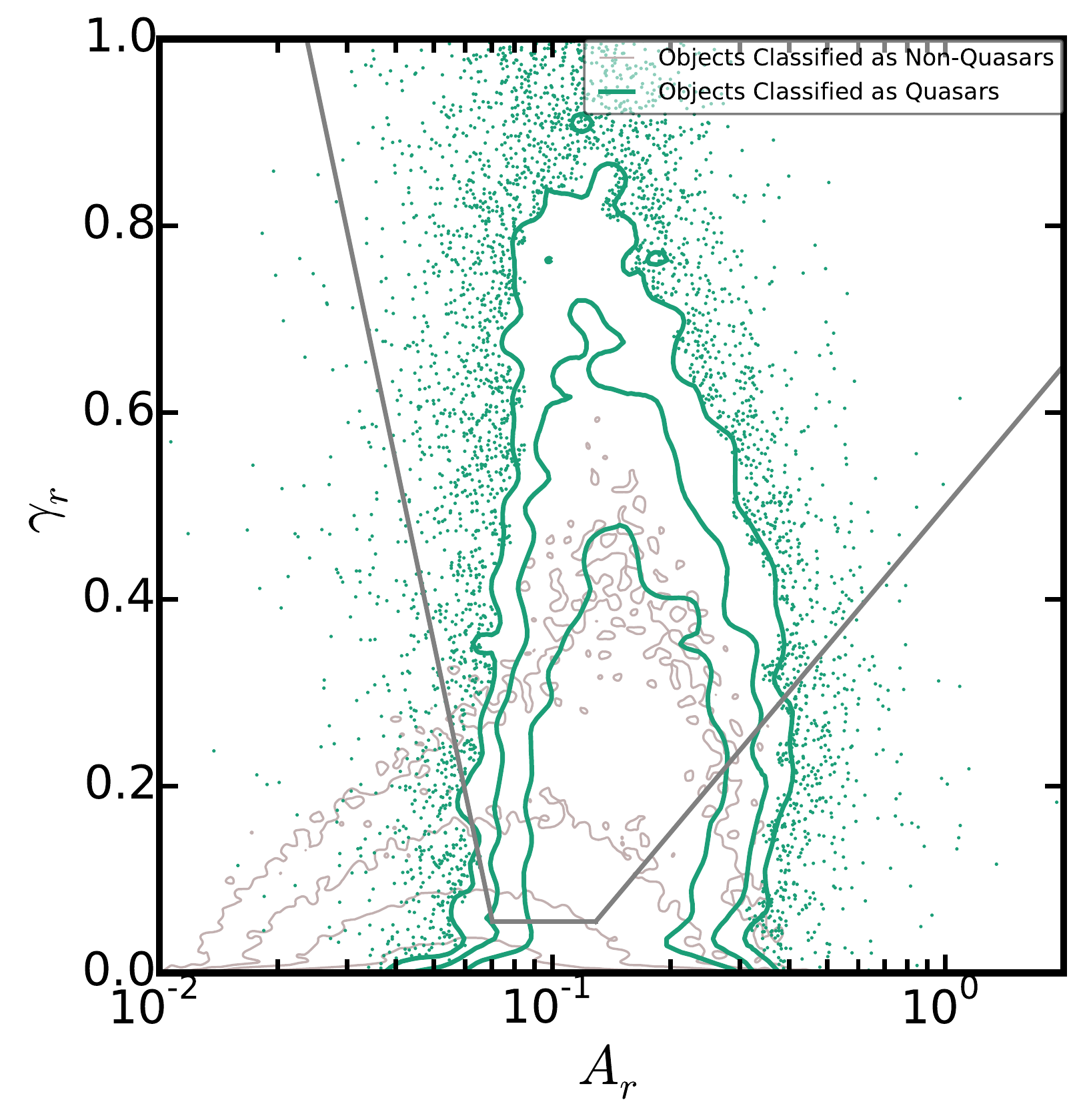}
\end{tabular}
\caption{$A_r$ vs.\ $\gamma_r$ for the training sets ({\it left panel}) and test set ({\it right panel}) shown with the \cite{Schmidt:2010} variability selection cuts (Equations \ref{eq:SchmidtCutRight} - \ref{eq:SchmidtCutMiddle}) as gray lines. {\it Left panel:}  Orange contours show the non-quasar training set  and purple contours and scatter points show the quasar training set. {\it Right panel:}  Gray contours show all objects in the test set classified as non-quasars and green contours and scatter points show all objects in the test set classified as quasars.}
\label{fig:SchmidtCuts}
\end{figure*}

First we compare our results to the performance of the (variability-based) quasar selection box (in $A$ and $\gamma$ space) defined in Equations 7 - 9 of \cite{Schmidt:2010}:
\begin{equation} \label{eq:SchmidtCutRight}
\gamma_r = 0.5 \log (A_r) + 0.50
\end{equation}
\begin{equation} \label{eq:SchmidtCutLeft}
\gamma_r = -2.0 \log (A_r) - 2.25
\end{equation}
\begin{equation} \label{eq:SchmidtCutMiddle}
\gamma_r = 0.055 .
\end{equation}
Using Stripe 82 data, \cite{Schmidt:2010} achieve a completeness of 90\% and an efficiency of 96\% with this box.  Applying the same cuts to our own training sets, as shown in Figure~\ref{fig:SchmidtCuts} {\it left panel}, results in 87\% completeness and 74\% efficiency. We achieved very different results because we have very different quasar and non-quasar data sets. \cite{Schmidt:2010} used quasars with $15.4 < i < 22.0$ with a mean of 19.5 and only 5000 bright F/G-star colored objects with $0.2 < g - r < 0.48$ and $14.0 < g < 20.2$. We used quasars with $15.9 < i < 22.7$ with a mean of 20.2 and 72,680 non-quasars (not just F/G stars) with $14.8 < g < 25.5$ and a mean of 20.6.

Applying these cuts instead to our full test set, as shown in Figure~\ref{fig:SchmidtCuts} {\it right panel}, gives 49,649 quasar candidates.  Of these, 23\% are spectroscopically confirmed quasars and another 27\% are objects that we identified as quasar candidates in either Section~\ref{sec:ResultsTestSet} or \ref{sec:ResultsBins} (with the remaining being previously-unidentified potential new candidates).  If all of our previously identified candidates were actually quasars and the remaining objects identified by these cuts were instead contaminants, then the efficiency of this variability quasar selection box would be 50\% and the completeness would be 69\%. The majority of the quasar candidates outside the box are dimmer than a coadded {\it i}-band magnitude of 20, where most variability is below the noise level. 

This comparison suggests that selection by variability alone, while working well to discriminate between relatively bright F/G stars and quasars, is incomplete when using a realistic sample of non-quasar contaminants, and that our hybrid approach of combining color and variability will yield better results for future surveys.

In \cite{Graham:2014} they compare the performance of variability selection using a power law fit to the structure function (SF), a DRW fit, and Slepian wavelet variance (SWV). Using the power law fit to the SF (SWV) to classify quasars on Stripe 82 they achieve 92\% (86\%) completeness and 93\% (92\%) efficiency.

\subsection{BOSS Quasar Selection}\label{sec:DiscussionBOSSQuasarSelection}

As described in Section~\ref{sec:MQC}, in addition to color selection, some of the BOSS quasars on Stripe 82 were targeted using an algorithm based on the same parameterization of variability used herein. We matched our candidate catalog to the SDSS-III/BOSS Data Release 10 Quasar Catalog (DR10Q; \citealt{Paris:2014}) to see how well we recovered these quasars. These quasars are indicated by {\tt DR10\_label} $ == 1$. There are 9,590 quasars on Stripe 82 in DR10Q and 7,241 were point sources that met the quality cuts to be included in our test set. Of these 7,241 known quasars, we recovered 7,034 (97.1\% completeness) as candidate quasars. The quasars we missed have $i < 22.0$ with a mean of $i = 20.0$ and have $\gamma_{g} < 0.25$: much less variable than the quasar training set on average.

We found 6,562 quasar candidates in the BOSS redshift range ($2.2 < z < 3.5$) based on astro-photometric redshifts. Of these, 49\% are training set quasars with spectroscopic redshifts $2.2 < z < 3.5$ ($i < 22.7$ with mean $i = 20.7$) and another 3\% are known quasars with spectroscopic redshifts outside this range. Of the remaining 48\% (3,157 quasar candidates), 1,614 are high probability candidates with {\tt qso\_prob > 0.8}. These are the objects that are highly likely to be quasars that BOSS has missed, which is consistent with the known incompleteness of BOSS \citep{Ross:2012}. Our high probability candidates have $i < 23.0$ with a mean of $i = 21.4$, suggesting that we are able to extend our selection to less luminous objects using the combined color and variability approach.

Since our test set was built, the twelfth data release quasar catalog of SDSS-III was made public (DR12Q; P\^{a}ris et al. 2015, in prep). Since DR10Q, additional spectroscopic plates were taken on Stripe 82, resulting in 2,054 DR12Q quasars on Stripe 82 that are not in the quasar training set, 1,162 were point sources that met the quality cuts to be included in our test set. We matched our candidate catalog to DR12Q to see how well we recovered these new quasars. These objects are indicated by {\tt DR12\_label} $ == 1$. Of the quasars new in DR12Q, we recovered 1,141 (98.2\% completeness). The objects that were missed have $i < 22.1$ with a mean of $i = 21.3$ and have $\gamma_{g} < 0.33$. Again, they are much less variable than the quasar training set on average.

\subsection{Number Counts and the Luminosity Function}\label{sec:DiscussionNumberCountsLuminosityFunction}
\begin{figure*}
\capstart
\centering
\begin{tabular}{cc}
\includegraphics[width=3in,height=3in]{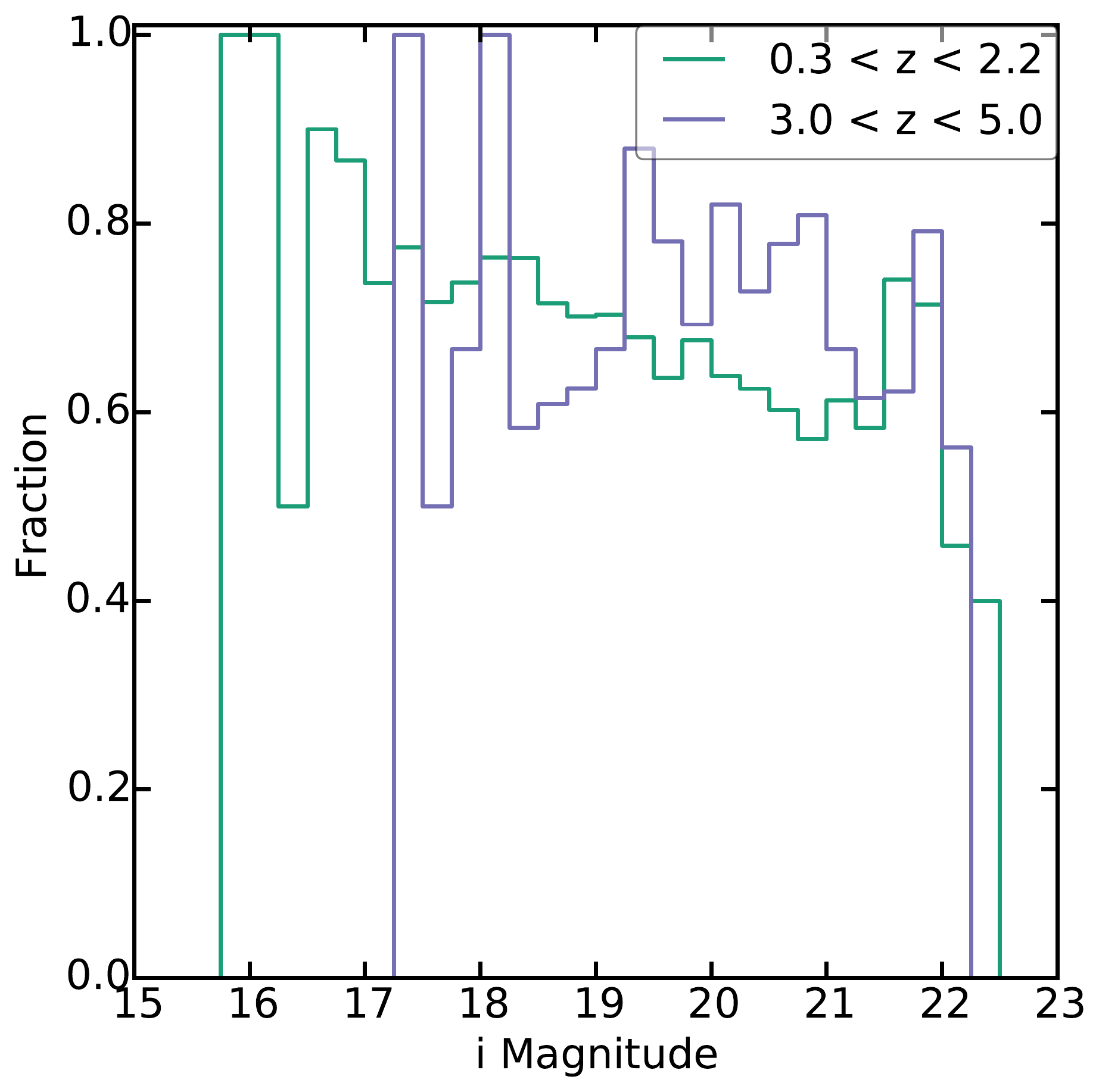} & 
\includegraphics[width=3in,height=3in]{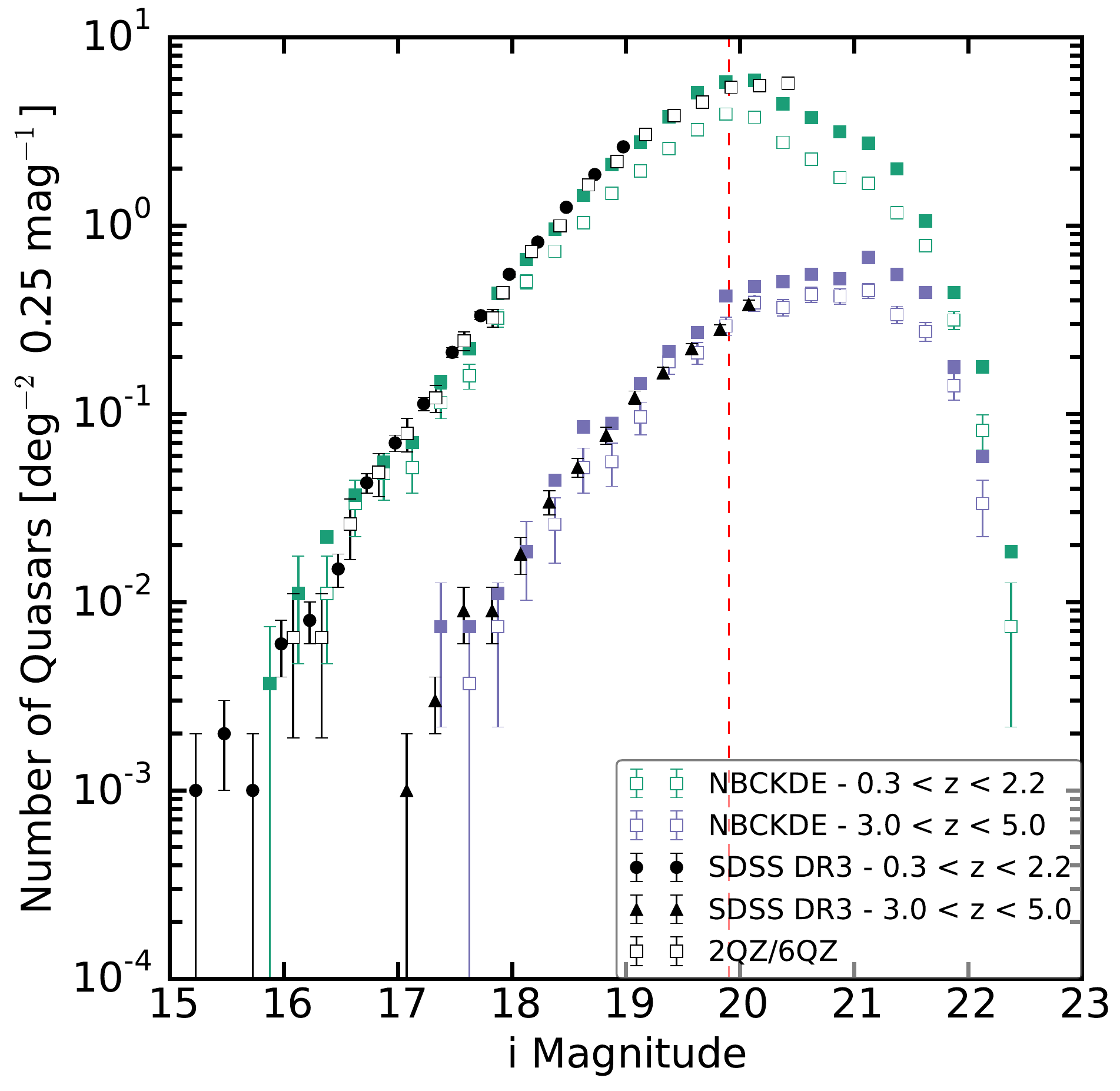}
\end{tabular}
\caption{{\it Left panel:} Ratio of quasars in the MQC on Stripe 82 to those MQC quasars returned by classification using a training set over the full redshift range. This allows us to correct for objects with too few observations to calculate variability parameters, the exclusion of extended sources, and incompleteness in the selection algorithm. The fraction is given as a function of coadded {\it i}-band magnitude for two redshift ranges. {\it Right panel:} Quasar number counts as a function of redshift and {\it i}-band magnitude. Black points give the spectroscopic number counts reported in \cite{Richards:2009_DR6catalog}; circles for $z < 2.2$ and triangles for $3 < z < 5$. The open purple and green squares give the raw number counts (with Poisson error bars) for the candidates reported here. The filled colored squares give the number counts corrected using the {\it left panel}. The vertical dashed red line at $i=19.9$ indicates the target selection depth for low-redshift on Stripe 82.}
\label{fig:NumberCounts}
\end{figure*}

In Figure~\ref{fig:NumberCounts} we reproduce the number counts analysis shown in Figure~9 of \cite{Richards:2009_DR6catalog}, using our candidate quasars. The counts have been corrected for incompleteness as given by the fraction of MQC quasars recovered as shown in Figure~\ref{fig:NumberCounts} {\it left panel}.  In short, the correction is the ratio of known quasars to quasar candidates. This process corrects for: objects with too few observations to calculate variability parameters, the exclusion of extended sources, and incompleteness in the selection algorithm. The {\it right panel} shows the number of quasars per deg$^{2}$ and per 0.25 mag as a function of coadded {\it i}-band magnitude. Open points represent the raw number counts, while the closed points give the completeness-corrected number counts. The turnover at $i = 19.9$ is due to the incompleteness of the spectroscopic sample. This analysis suggests that our selection algorithm is neither heavily contaminated (e.g., as might be evidenced by a large excess of bright objects versus known quasars), nor very incomplete---since the corrected counts agree well with the spectroscopic quasar distribution.

Next, we calculate the quasar luminosity function (QLF) for our candidate quasars. This QLF calculation was not intended to be a scientific result of this pilot project, as we expected more incompleteness and contamination than shown in Figure~\ref{fig:NumberCounts}.  However, the result does suggest that accurate determination of the QLF will be possible with photometric selection from LSST and other next-generation surveys.

In order to compare space densities at different redshifts, we must correct our photometry for the effects of redshift on the portion of the spectrum sampled by a given filter. We do this by using a mean K-correction for $z=2$ in the $i$-band as described in \citet[][Section 5]{Richards:2006}.

\begin{figure*}
\capstart
\centering
\begin{tabular}{ccc}
\includegraphics[width=2.3in]{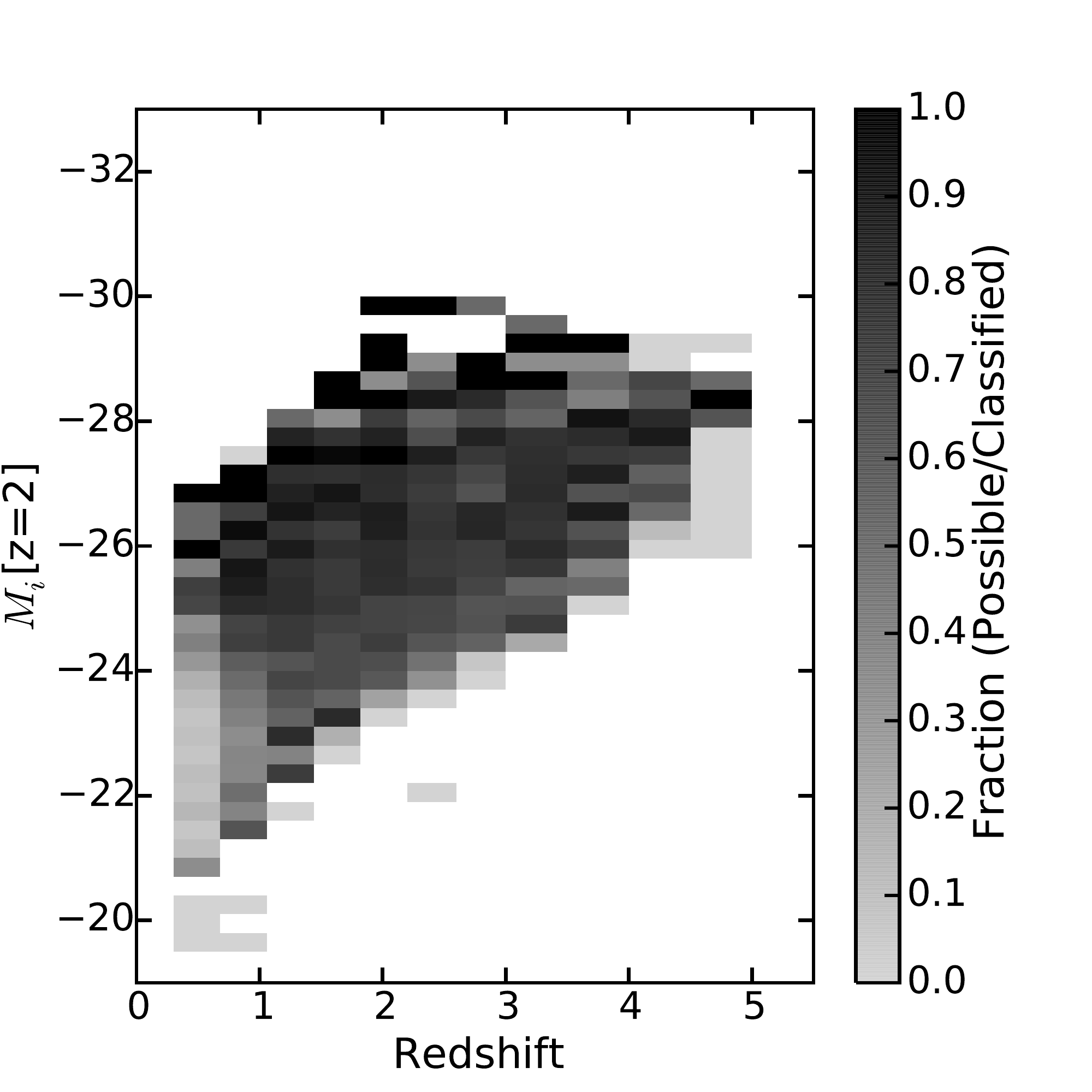} & 
\includegraphics[width=2.3in]{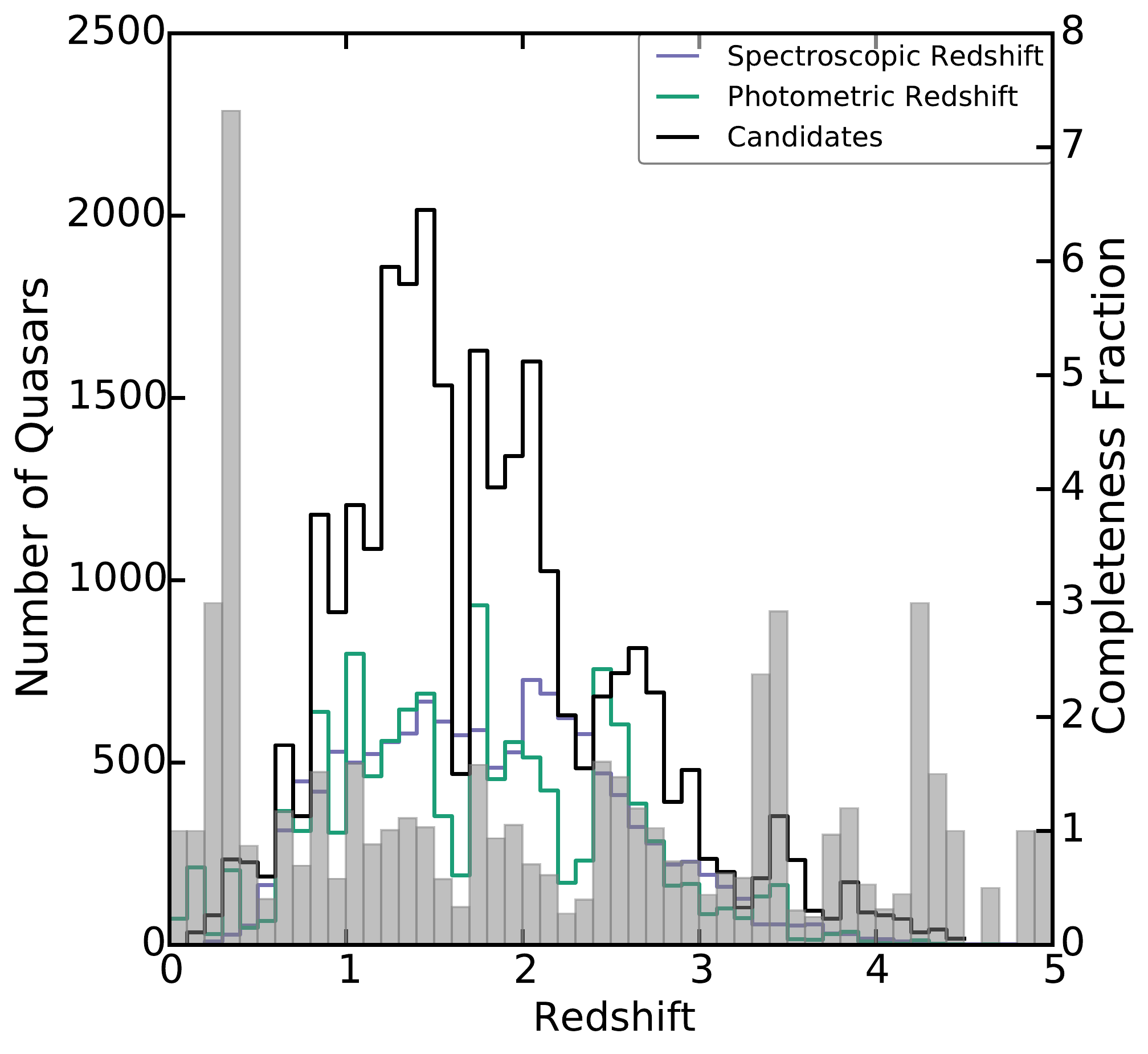} & 
\includegraphics[width=2.3in]{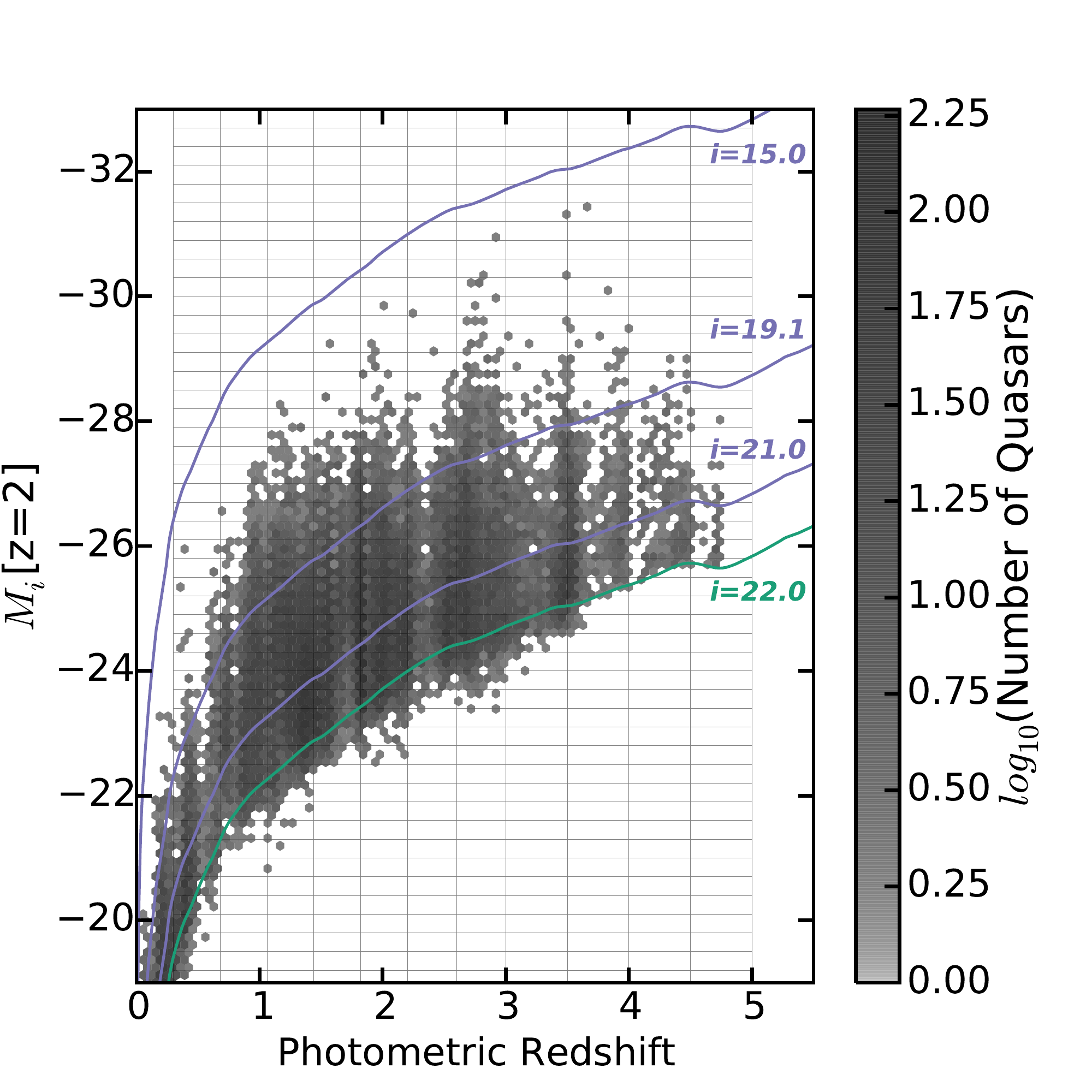}
\end{tabular}
\caption{Corrections and cuts used in the QLF in Figure~\ref{fig:candidatequasars_LuminosityFunction}.  {\it Left panel:} Completeness fraction, in bins of redshift and absolute magnitude, $M_i[z = 2]$, for candidate selection. Similar to Figure~\ref{fig:NumberCounts} {\it left panel}, but in two dimensions.  The number of quasars with spectroscopic redshifts on Stripe 82, even if they were excluded from our training set and test set, was divided by all quasars with spectroscopic redshifts that were recovered as candidate quasars. This is to correct for incompleteness from too few observations to calculate variability parameters, the exclusion of extended sources, and incompleteness in the classification algorithm. {\it Center panel:} Completeness fraction for astro-photometric redshifts. All of the training set quasars are binned by spectroscopic redshift (purple) and astro-photometric redshifts (green). The ratio of the two is shown in grey (right axis). The astro-photometric redshifts of the candidate quasars, after being corrected by the completeness fraction and assuming that objects without spectroscopic redshifts have the same astro-photometric redshift errors as those with spectroscopic redshifts, are shown in black. {\it Right panel:} Astro-photometric redshift vs.\ absolute magnitude, $M_i[z = 2]$, of all quasar candidates. The green line shows the brightness limit for bins that are used in computing the luminosity function. Purple curves show the $i$ = 15.0, 19.1, and 20.2 magnitude limits for SDSS spectroscopic follow-up.}
\label{fig:candidatequasars_LuminosityFunction_corrections}
\end{figure*}

As we have seen, and as discussed in \citet[][Section 3.4.1]{Ross:2013}, variability selection is less biased than color selection, but we cannot assume variability selected samples are complete and unbiased. Just as with the number counts above, the candidate object QLF must be corrected for the completeness fraction and, additionally, for systematic errors in astro-photometric redshifts. For the QLF, we need to correct for incompleteness in two dimensions: redshift and absolute magnitude (luminosity).  The gray-scale $M-z$ bins in the {\em left panel} of Figure~\ref{fig:candidatequasars_LuminosityFunction_corrections} gives the fraction of MQC quasars recovered.  This includes quasars that were not included in our test set so as to correct to the true number of quasars, not just those that met our test set criteria.

Since catastrophic errors in astro-photometric redshifts can distort the QLF, corrections were determined as follows: using bins of $\Delta z$ = 0.1, the number of quasars with astro-photometric redshift in that bin was divided by the number of quasars with spectroscopic redshifts in that bin. The resulting ratio is the correction that needs to be applied to objects in each astro-photometric redshift bin to statistically account for errors in the astro-photometric redshift distribution (as opposed to correcting individual values) and is shown in Figure~\ref{fig:candidatequasars_LuminosityFunction_corrections} {\it center panel}. The two corrections are multiplied together and used as a weight for the objects in the QLF.

\begin{figure*}
\capstart
\centering
\includegraphics[width=7in]{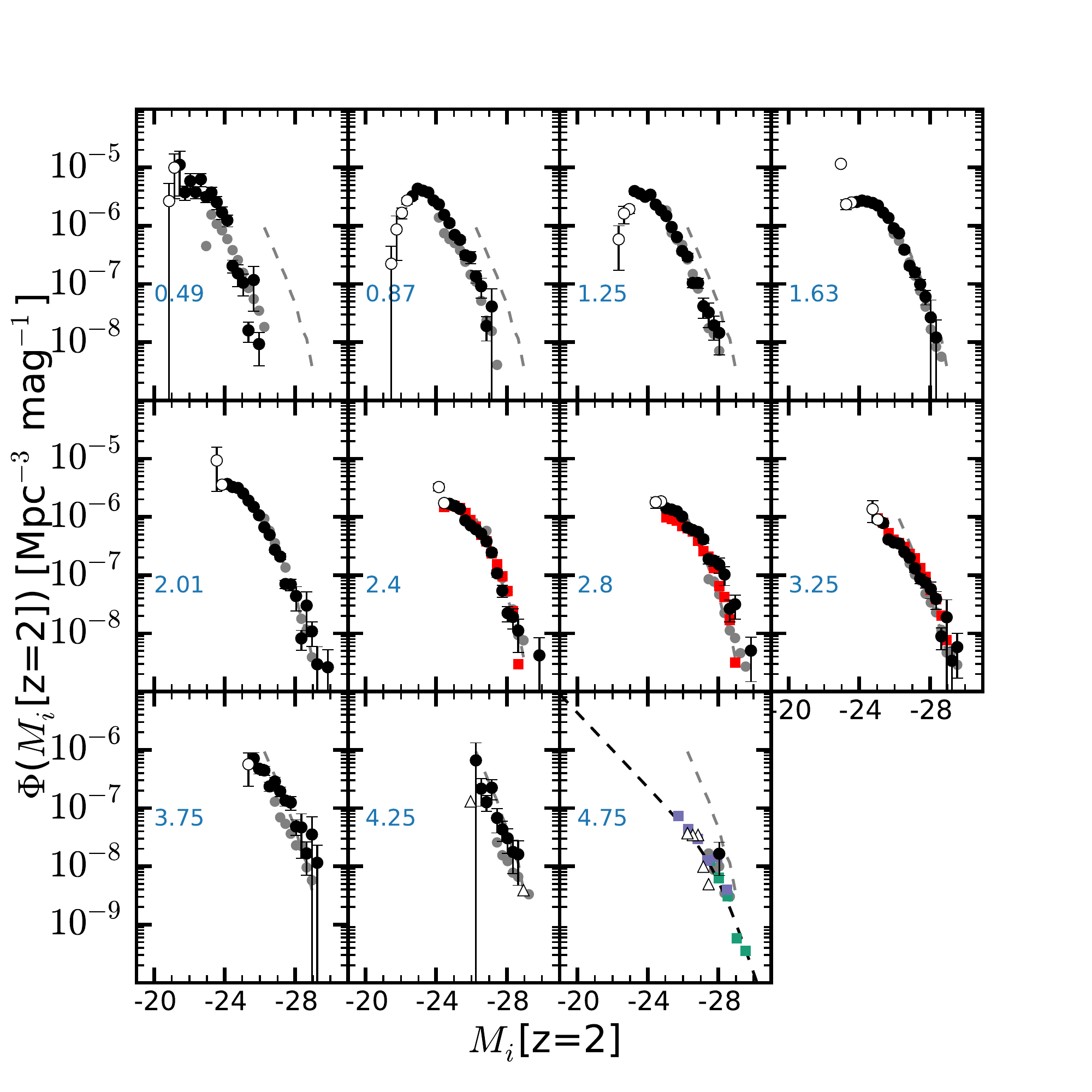}
\caption{$M_i[z = 2]$ binned luminosity function of the sample with astro-photometric redshifts using the method from \cite{Page:2000} (with Poisson error bars). The mean redshift of each slice is given in each panel. Black filled circles are complete bins, empty triangles indicate the lower limit for complete bins where the completeness fraction (shown in Figure~\ref{fig:candidatequasars_LuminosityFunction_corrections}  {\it left panel}) is 0, and empty circles are partial bins (a portion of the bin is dimmer than $i=22$). The grey circles show the binned luminosity function and the grey dashed line shows the $z = 2.01$ curve both from \citet[Figure~18]{Richards:2006} for comparison. In the $z =$ 2.4, 2.8, and 3.25 panels, the red squares show the binned luminosity function for BOSS quasars from DR9 from \citet[Figure~11]{Ross:2013}. In the 4.75 panel, the green squares, purple squares, and dashed black line show the binned luminosity function at $z = 4.9$ for Stripe 82, DR7, and double power law fits from the maximum likelihood analysis from \citet[Figure~12 and Figure~13]{McGreer:2013}.}
\label{fig:candidatequasars_LuminosityFunction}
\end{figure*}

We compute the QLF by binning the quasars in redshift and absolute magnitude, using the method from \cite{Page:2000}. Figure~\ref{fig:candidatequasars_LuminosityFunction_corrections} {\it right panel} shows absolute magnitude as a function of astro-photometric redshift for all quasar candidates. The grid shows the bins within which the QLF is calculated. The edges of the redshift bins are 0.30, 0.68, 1.06, 1.44, 1.82, 2.20, 2.6, 3.0, 3.5, 4.0, 4.5, and 5.0. The $M_i$ bins are in increments of 0.3 mag. The adopted limiting magnitude of $i = 22.0$, is shown as a green line. The resulting $i$-band QLF is shown as black dots with Poisson error bars in Figure~\ref{fig:candidatequasars_LuminosityFunction}.

As with the number counts, the QLF analysis shows relatively close agreement with the space density of known quasars. There is evidence for both incompleteness and contamination in the lowest redshift bin. This is perhaps not surprising given the effects the host galaxy has on quasar colors and apparent variability and the fact that we only include point sources. We show the \citet[Figure~18]{Richards:2006} and \citet[Figure~11]{Ross:2013} SDSS spectroscopic QLFs in the $z=$ 2.4, 2.8, and 3.25 bins. This comparison reveals that our QLF agrees better with the \cite{Ross:2013} QLF. The \cite{Ross:2013} QLF has the smaller corrections of the two spectroscopic QLFs, which suggests that the \cite{Richards:2006} QLF was undercorrected. In the three highest redshift bins our QLF suggests a higher space density than the \cite{Richards:2006} QLF. This could be a sign of contamination in our catalog, though it could also be true to some extent, given the relatively large completeness fraction for candidate selection needed for the smaller spectroscopic sample from which the \cite{Richards:2006} QLF was derived. Most importantly, given the lack of contamination and the dependability of the completeness corrections, this analysis bodes well for our future ability to determine the QLF for faint populations in post-SDSS sky surveys.

\section{Future Work}\label{sec:FutureWork}

The purpose of this investigation was to demonstrate that using a combination of optical colors and variability parameters improves quasar classification efficiency and completeness over the use of colors alone. This is one step toward finding an optimal strategy for photometric quasar selection. 

In future, we hope to use a data set that includes both point sources and extended sources, thus incorporating the variable nucleus with the steady host galaxy. Additionally, we plan to explore alternative parameterizations of quasar variability. The underlying mechanism and most appropriate model for quasar variability remain open questions and there are more sophisticated models (e.g. \citealt{Kelly:2009}; \citealt{MacLeod:2010}; \citealt{Kasliwal:2015}) that merit exploration. Given the large quantity of data expected in future surveys such as LSST, a more computationally efficient approach than the structure function may become important; e.g., the \cite{Kelly:2009}, \cite{Kozlowski:2010}, and \cite{MacLeod:2010} approaches require only O(N) rather than O(N$^2$) operations to determine the model parameters for a light curve with N data points. As described in Section~\ref{sec:Variability}, the likelihood method is biased and more robust approaches such as those described in the appendices of \cite{Kozlowski:2010} or \cite{Hernitschek:2015} should be investigated. Currently, we use variability data from each band separately. We hope to explore the various methods for merging bands together, even with non-simultaneous observations as will be the case with LSST.

This work relies on KDE for classification and it is important to explore other methods to see if they will be more successful. In the future, we hope to make use of other types of algorithms (e.g. Feigelson et al. \citeyear{Feigelson:2003}; \citeyear{Feigelson:2012}; \citealt{Chakraborty:2013}), such as random forests (e.g. \citealt{Gao:2009}; \citealt{Richards:2011}; \citealt{CarrascoKind:2013}), gradient boosting machines \citep{Hastie:2001}, and Bayesian classification with hash tables \citep{Gupta:2014}. Additionally, in the future our catalogs will not have binary classifications, but will simply give probabilities for all objects. 

We have used the combination of optical and mid-infrared (MIR) colors for quasar selection in another paper (Richards et al. 2015, submitted). In the future we will combine optical, IR, and variability data to produce the most complete and efficient catalog possible.

In order to improve the astro-photometric redshift estimations, we will multiply smoothed PDFs instead of adding by weights. Additionally, we will incorporate the clustering redshift estimation of \cite{Menard:2013} and \cite{Rahman:2015} and explore photometric and astro-photometric redshift accuracy without $u$ band observations (to mimic DES and Pan-STARRS observations). Finally, we will explore how simultaneous color and variability classification performs using other time-domain surveys, including DES \citep{DES:2005}, Pan-STARRS \citep{Kaiser:2010}, and LSST simulated data \citep{Connolly:2014}.

\section{Conclusions}\label{sec:Conclusions}

Using the Non-parametric Bayesian Classification Kernel Density Estimation (NBC KDE) quasar selection algorithm, we demonstrated that using a combination of optical colors and variability parameters improves quasar classification efficiency and completeness over the use of colors alone. For classification using colors alone, there are redshift ranges with poor completeness where the quasar and non-quasar training sets overlap in color space. Variability alone does not have these redshift trends, but it has a lower efficiency than coadded colors at all other redshifts. The addition of coadded colors to variability information improves classification at all redshifts. Using variability alone, colors alone, and combining variability and colors we achieve we achieve 91\%, 93\%, and 97\% quasar completeness and 98\%, 98\%, and 97\% efficiency respectively, with particular improvement in the selection of quasars at $2.7<z<3.5$, as shown in Figure~\ref{fig:zhistogram_fraction_comparison}.

We classified quasars and estimated their redshifts simultaneously by limiting the training set to non-overlapping redshift bins from 0.4 to 4.0 with a bin width of 0.2. We successfully classified known quasars into the correct redshift bins with 75\% or higher completeness, depending on the redshift bin, as shown in Figure~\ref{fig:ClassificationofSpectroscopicallyConfirmedQuasars}.

Overall, we identified 35,820 type 1 quasar candidates in the SDSS Stripe 82 field using the combination of optical photometry and variability either over the full redshift range or within one of the redshift bins. Of the 13,221 spectroscopically confirmed quasars that could have been returned, we found 12,953 (98.0\% completeness). Of the 22,867 quasar candidates that are not spectroscopically confirmed, 21,876 (95.7\%) are dimmer than a coadded i-band magnitude of 19.9. Figure~\ref{fig:testset_coadd_gr_imag} shows the magnitude distribution of the candidate quasars.

Photometric redshift estimates of these candidates using optical photometry and astrometric parameters are accurate to within $|\Delta z| < 0.1$ for 51.6\% of quasars and within 0.3 for 76.8\% of quasars. The combination of optical photometry and astrometry makes the photometric redshifts more accurate when colors alone returns the correct redshift as one of the secondary peaks in the PDF. The astrometric PDF pulls out the correct peak in the color PDF, as shown in Figure~\ref{fig:hist_photometric redshift}. We find that objects with photometric redshifts of $z\sim1.25$ and $z\sim3.25$ are particularly robust.

In Figure~\ref{fig:SchmidtCuts}, our color and variability selection was compared to other cuts in variability space that have been used on Stripe 82. We demonstrated that variability alone is incomplete and that our hybrid approach will yield better results for future surveys. Additionally, we have shown that our selection recovered 97\% of the quasars in the DR12 quasar catalog and we selected additional candidates in the BOSS redshift range with high confidence (and at even higher redshift).

We used MIR color cuts to remove a small number of bright star contaminants from our final candidate list. Our number counts and quasar luminosity function analyses, shown in Figures~\ref{fig:NumberCounts} and \ref{fig:candidatequasars_LuminosityFunction}, show there is little contamination remaining and that there is relatively close agreement with the space density of known quasars.

From the NBC KDE classification test set, we present a catalog of known quasars and candidate quasars on Stripe 82. The catalog is available as a FITS file online. Future work along these lines will be needed to capitalize on the imaging data produced by Pan-STARRS, DES, and LSST.

\acknowledgments
This work was supported in part by NASA-ADAP grant NNX12AI49G and NSF grant 1411773. This research made use of \href{http://www.astropy.org/}{Astropy}\footnote{astropy.org} (\citealt{astropy}), \href{http://www.star.bristol.ac.uk/~mbt/topcat/}{TOPCAT}\footnote{starlink.ac.uk/topcat}, (\citealt{TOPCAT}), and \href{http://www.star.bristol.ac.uk/~mbt/stilts/}{STILTS}\footnote{starlink.ac.uk/stilts} (\citealt{STILTS}). All figures made use of \href{http://matplotlib.org/}{matplotlib}\footnote{matplotlib.org} (\citealt{matplotlib}). Figures~\ref{fig:colorcolor_redshift}, \ref{fig:A_gamma}, \ref{fig:colorcolor_nobins_testsetresults}, \ref{fig:colorcolor_bins_testsetresults}, and \ref{fig:SchmidtCuts} were made using \href{https://github.com/CKrawczyk/densityplot}{densityplot}\footnote{github.com/CKrawczyk/densityplot} (\citealt{densityplot}). Special thanks to Alex Gray for developing the code for the NBC KDE algorithm and to Erica Smith for carefully reviewing the manuscript.

Funding for the SDSS and SDSS-II has been provided by the Alfred P. Sloan Foundation, the Participating Institutions, the National Science Foundation, the U.S. Department of Energy, the National Aeronautics and Space Administration, the Japanese Monbukagakusho, the Max Planck Society, and the Higher Education Funding Council for England. The SDSS Web Site is http://www.sdss.org/.

Funding for SDSS-III has been provided by the Alfred P. Sloan Foundation, the Participating Institutions, the National Science Foundation, and the U.S. Department of Energy Office of Science. The SDSS-III web site is http://www.sdss3.org/.

\appendix

\section{Catalog Columns}\label{sec:CatalogColumns}
In Section~\ref{sec:Catalog} we created a catalog of quasar candidates. All of the columns in the catalog are described in Table~\ref{table:ColumnNames}, but a few columns need some extra explanation.

Columns 4 to 13 are the single epoch magnitudes and magnitude errors used for the single epoch classification. We used a randomly chosen epoch from the observations for each object. The single epoch magnitudes are asinh magnitudes from \cite{Lupton:1999}.  Columns 14 to 18 are the coadded magnitudes and magnitude errors. The coadded magnitudes are from \cite{Annis:2014}. Columns 19 to 24 are the VHS magnitudes and magnitude errors in Vega. Columns 25 to 28 are the WISE magnitudes and magnitude errors in AB.

Columns 30 to 34 are labels. Specifically, column 30 is the candidate label: if the object was classified as a quasar in either Section~\ref{sec:ResultsTestSet} (over the whole redshift range) or \ref{sec:ResultsBins} (in redshift bins) the value is 1, otherwise it is 0. Column 31 is the MQC label: if the object is in the catalog the value is 1, otherwise it is 0. Column 32 is the DR10Q label: if the object is in the catalog the value is 1, otherwise it is 0. Column 33 is the DR12Q label: if the object is in the catalog the value is 1, otherwise it is 0. Column 34 is the WISE cut label: if the object is cut by Eqs. \ref{eq:WISECut1} and \ref{eq:WISECut2} the value is 1, otherwise it is 0. Column 35 is the white dwarf cut label: if the object is cut by Eq. \ref{eq:WDCut} the value is 1, otherwise it is 0. To retrieve the quasar candidates that pass these cuts (the ``good" quasar candidates) perform this query on the catalog: {\tt WISEcut\_label}  == 0 \& {\tt WDcut\_label}  == 0 \& {\tt candidate\_label} == 1. To limit to the new candidates (not spectroscopically confirmed quasars) add:  \& {\tt zspec} $< 0$.

Columns 48 to 55 are the various classification results. Specifically, columns 48 to 50 are the results of classifying the test set over the full redshift range as described in Section~\ref{sec:ResultsTestSet}. Column 48 is the probability of being a quasar, column 49 is the star density from the KDE ($P(D|M)$), and column 50 is the quasar density from the KDE. If the object was not found to be a candidate over the full redshift range the value is -9999. Columns 51 to 55 are the results of classification using redshift bins as described in Section~\ref{sec:ResultsBins}. Column 51 is the probability of being a quasar, column 52 is the star density, and column 53 is the quasar density. Each is a vector with 18 cells, one for each redshift bin from 0.4 to 4.0. If the object was not found to be a candidate in any bin, all cell values are -9999. Column 54 is the maximum value of column 51, and column 55 is the center of the redshift bin corresponding to that maximum probability. If the object was not found to be a candidate in any bin these columns will be -9999. 

Columns 56 to 73 are the various redshift estimation results. If we were unable to calculate a redshift estimate for the object the value will be -9999. Column 74 indicates whether the object's $g-i$ color is within 1$\sigma$ (0.68), 2$\sigma$ (0.95), or 3$\sigma$ (0.99) of the mean color for quasars at the astro-photometric redshift. Outliers are an indication of either bad estimated redshifts or non-quasar contaminants. Columns 75 to 82 are the details of matching to all spectra taken on SDSS Stripe 82.

\begin{deluxetable*}{l l l}
\tablecolumns{3} 
\tablewidth{0pc}
\tabletypesize{\small}
\tablecaption{Column Names \label{table:ColumnNames}}
\tablehead{\colhead{index} & \colhead{Name} & \colhead{Description}}
\startdata
1 & id & SDSS Coadded ParentID \\
2 - 3  & ra, dec & coadded right ascension, declination \\
4 - 8 & u, g, r, i, z & single epoch magnitude \\
9 - 13 & uErr, gErr, rErr, iErr, zErr & single epoch magnitude error\\
14 - 18 & coadd\_u, coadd\_g... & coadded  magnitude \\
19 - 21 & J, H, KS & VHS magnitude \\
22 - 24 & JErr, HErr, KSErr & VHS magnitude error \\
25 - 26 & W1, W2 & WISE magnitude \\
27 - 28 & W1Err, W2Err & WISE magnitude error \\
29 & zspec & spectroscopic redshift, if none, the value is -9999.\\
30 & candidate\_label & 1 if selected as a quasar candidate, 0 otherwise.\\
31 & MQC\_label & 1 if in MQC, 0 otherwise. \\
32 & DR10Q\_label & 1 if in DR10Q, 0 otherwise. \\
33 & DR12Q\_label & 1 if in DR12Q, 0 otherwise. \\
34 & WISEcut\_label & 1 if cut by Eqs. \ref{eq:WISECut1} and \ref{eq:WISECut2}, 0 otherwise. \\
35 & WDcut\_label & 1 if cut by Eq. \ref{eq:WDCut}, 0 otherwise. \\
36 - 45 & A\_u, gamma\_u, A\_g, gamma\_g... & variability parameters, if none, the value is -9999. \\
46 - 47 & auPar, agPar & astrometry parameters, if none, the value is -9999.\\
48 & qso\_prob & probability of being a quasar \\
49 & star\_dens & star density \\
50 & qso\_dens & quasar density \\
51 & qso\_prob\_bins & vector - probability of being a quasar \\
52 & star\_dens\_bins & vector - star density \\
53 & qso\_dens\_bins & vector - quasar density\\
54 & qso\_prob\_max & maximum value of {\em qso\_prob\_bins} vector \\
55 & qso\_prob\_max\_bin & redshift bin of maximum value of {\em qso\_prob\_bins} vector\\
56 & photoz\_ugriz\_pdf & vector - full photo-z PDF, SDSS colors\\
57 & photoz\_ugriz\_low & low redshift end of the peak in photo-z PDF, SDSS colors \\
58 & photoz\_ugriz\_best & peak of photo-z PDF, SDSS colors \\
59 & photoz\_ugriz\_high & high redshift end of the peak in photo-z PDF, SDSS colors \\
60 & photoz\_ugriz\_prob & probability of photo-z, SDSS colors\\
61 & photoz\_astrometry\_pdf & vector - full photo-z PDF, astrometry\\
62 & photoz\_astrometry\_low & low redshift end of the peak in photo-z PDF  astrometry\\
63 & photoz\_astrometry\_best & peak of photo-z PDF astrometry \\
64 & photoz\_astrometry\_high & high redshift end of the peak in photo-z PDF astrometry \\
65 & photoz\_astrometry\_prob & probability of photo-z astrometry\\
66 & photoz\_added\_pdf & vector - full photo-z PDF, SDSS colors and astrometry \\
67 & photoz\_added\_max & max of photo-z PDF, SDSS colors and astrometry\\
68 & photoz\_added\_max\_bin & redshift bin of maximum value of photo-z PDF, SDSS colors and astrometry \\
69 & photoz\_ugrizJHK\_pdf & vector - full photo-z PDF, SDSS and JHK colors\\
70 & photoz\_ugrizJHK\_low & low redshift end of the peak in photo-z PDF, SDSS and JHK colors \\
71 & photoz\_ugrizJHK\_best & peak of photo-z PDF, SDSS and JHK colors \\
72 & photoz\_ugrizJHK\_high & high redshift end of the peak in photo-z PDF, SDSS and JHK colors \\
73 & photoz\_ugrizJHK\_prob & probability of photo-z, SDSS and JHK colors \\
74 & gi\_sigma & g-i color offset from the mean color\\
75 & SDSSSPECMATCH & 1 if the object had a spectrum from the original SDSS, 0 otherwise \\
76 & BOSSSPECMATCH & 1 if the object had a spectrum from BOSS, 0 otherwise\\
77 & DR12QSOMATCH & 1 if the object is visually inspected as a quasars in the DR12Q, 0 otherwise\\
78 & ZSDSS & pipeline redshift from SDSS\\
79 & CLASSSDSS  & pipeline classification from SDSS\\
80 & ZBOSS & pipeline redshift from BOSS\\
81 & CLASSBOSS & pipeline classification from BOSS\\
82 & DR12QSO\_Z\_VI & redshift of the quasars if included in the DR12Q
\enddata
\end{deluxetable*}

\pagebreak

\bibliography{Quasar_Classification.bib}

\end{document}